\documentclass[11pt]{article}
\pdfoutput=1

\usepackage{jheppub}
\usepackage[T1]{fontenc} 

\usepackage{amssymb}
\usepackage{amsfonts}
\usepackage{amsbsy}
\usepackage{amsmath}
\usepackage{amsthm}
\usepackage{graphicx}
\usepackage{epstopdf}
\usepackage[vcentermath]{youngtab}
\usepackage{multirow}
\usepackage{latexsym} 
\usepackage{array}
\usepackage{color}

\definecolor{Red}{rgb}{1.00, 0.00, 0.00}
\definecolor{Blue}{rgb}{0.00, 0.00, 1.00}
\definecolor{Purple}{cmyk}{0.45,0.86,0,0}

\input cyracc.def 
\newfont{\twelvecyr}{wncyr10 at 12pt}

\newcommand{\fl}[1]{ \lfloor #1 \rfloor }

\def\F{\mathbb{F}}

\def\P{\mathbb{P}}

\def\n3a{t}

\def\tr{{\mathrm{tr}}}

\def\age{{\mathfrak{e}}}
\def\agso{{\mathfrak{so}}}
\def\agsu{{\mathfrak{su}}}
\def\agsp{{\mathfrak{sp}}}
\def\agf{{\mathfrak{f}}}
\def\aggg{{\mathfrak{g}}}

\def\gso{\rm SO}
\def\gsu{\rm SU}
\def\gsp{\rm Sp}

\newcommand{\beq}{\begin{equation}}
\newcommand{\eeq}{\end{equation}}
\newcommand{\ba}{\begin{array}}
\newcommand{\ea}{\end{array}}
\newcommand{\bea}{\begin{eqnarray}}
\newcommand{\eea}{\end{eqnarray}}
\newcommand{\bean}{\begin{eqnarray*}}
\newcommand{\eean}{\end{eqnarray*}}
\newcommand{\eref}[1]{(\ref{#1})}

\newcommand{\comment}[1]{}

\newcommand{\expec}[1]{\langle #1 \rangle}
\newcommand{\Bigparen}[1]{\Big(#1\Big)}
\newcommand{\Biggparen}[1]{\Bigg(#1\Bigg)}
\newcommand{\Biggsquare}[1]{\Bigg[#1\Bigg]}
\newcommand{\deform}[1]{\tilde{#1}}
\newcommand{\suthree}[0]{\gsu(3)}

\newcommand{\rseven}[0]{r}
\newcommand{\reight}[0]{r}
\newcommand{\rsixexp}[0]{r_{6}}
\newcommand{\rsevenexp}[0]{r_7}
\newcommand{\reightexp}[0]{r_{8}}

\title{\boldmath Matter in transition \\
}

\author[1]{Lara B. Anderson}
\author[1]{James Gray}
\author[2]{Nikhil Raghuram}
\author[]{and}
\author[2]{Washington Taylor}

\affiliation[1]{Department of Physics\\
Robeson Hall, 0435\\
Virginia Tech \\
850 West Campus Drive \\
Blacksburg, VA 24061, USA}
\affiliation[2]{Center for Theoretical Physics\\
Department of Physics\\
Massachusetts Institute of Technology\\
77 Massachusetts Avenue\\
Cambridge, MA 02139, USA}

\emailAdd{{\tt lara.anderson} {\rm at} {\tt vt.edu}}
\emailAdd{{\tt jamesgray} {\rm at} {\tt vt.edu}}
\emailAdd{{\tt nikhilr} {\rm at} {\tt mit.edu}}
\emailAdd{{\tt wati} {\rm at} {\tt mit.edu}}

\preprint{\today \hspace*{0.1in} MIT-CTP-4747}

\abstract{
We explore a novel type of transition in certain 6D and 4D quantum
field theories, in which the matter content of the theory changes
while the gauge group and other parts of the spectrum remain
invariant.  Such transitions can occur, for example, for SU(6) and
SU(7) gauge groups, where matter fields in a three-index antisymmetric
representation and the fundamental representation are exchanged in the
transition for matter in the two-index antisymmetric representation.
These matter transitions are realized by passing through
superconformal theories at the transition point.  We explore these
transitions in dual F-theory and heterotic descriptions, where a
number of novel features arise.  For example, in the heterotic
description the relevant 6D SU(7) theories are described by bundles on
K3 surfaces where the geometry of the K3 is constrained in addition to
the bundle structure.
On
the F-theory side, non-standard representations such as the
three-index antisymmetric representation of SU($N$) require
Weierstrass models that cannot be realized from the standard SU($N$)
Tate form.  We also briefly describe some other situations, with
groups such as Sp(3), SO(12), and SU(3), where analogous matter
transitions can occur between different representations.  For SU(3),
in particular, we find a matter transition between adjoint matter and
matter in the symmetric representation, giving an explicit Weierstrass
model for the F-theory description of the symmetric representation
that complements another recent analogous construction.  }
  
\begin{document}
\maketitle

\flushbottom

\section{Introduction}

A variety of different types of transitions can occur in physical
theories in which the massless or light spectrum of the theory
changes.  
For certain types of transitions, 6D supergravity forms a clear
framework in which to classify and analyze the possible changes of
spectrum; similar transitions occur in 4D supergravity theories,
though the detailed description can involve more subtle issues.
For 6D theories coupled to gravity, the different types of
transitions can be characterized by the massless spectrum of the
low-energy theory.  

The most dramatic of these transitions in 6D theories are the
\emph{tensionless string}
or \emph{small instanton} transitions
\cite{Seiberg-Witten, Morrison-Vafa-II}, which
involve a change in the number of
tensor multiplets in the theory, accompanied by a corresponding change
in the number of uncharged scalar hypermultiplets.
These transitions are described in F-theory
by blowing up or down points on the base manifold used for the
F-theory compactification,  and in the heterotic theory by shrinking
an instanton to a point.
In both pictures the resulting transition is
fundamentally nonperturbative in nature, and in the low-energy theory
it involves passing through a superconformal fixed point.
Higgsing/unHiggsing transitions, on the other hand, leave the number
of tensor multiplets unchanged but modify the gauge group of the
theory and generally change the number of vector multiplets
in addition to modifying the matter spectrum.  
Higgsing/unHiggsing type transitions have a simple description in both F-theory
and heterotic pictures, in terms of a tuning of Weierstrass moduli on
the one hand and tuning of bundle moduli on the other, and have a
perturbative description in the low-energy theory.

In this paper we describe another type of transition, in which both
the tensor and vector multiplet spectra remain unchanged, and only the
representation content of the matter fields is modified.  
While the possibility of such transitions has been noted in the
literature \cite{Kachru-Silverstein, Ovrut-Pantev-Park,
  mt-singularities, Grassi-Morrison-2}
these kinds of
pure matter
transitions have not been studied in  depth, and we identify a
number of new interesting transitions in this class here.
We describe
these transitions both from the F-theory point of view and in a dual
heterotic picture.  Because of the matter representations involved
(frequently involving symmetry enhancement of the singular fibers
associated to exceptional groups, {\it etc.}) these transitions are
not accessible in a perturbative Type IIB description and can only be
explored in F-theory. 
In the F-theory geometry these transitions can be realized by tuning
Weierstrass moduli so that certain codimension two singularities
coincide and then split into a distinct geometry. On the heterotic
side these transitions arise when an instanton is shrunk and moved
into a separate component of the bundle structure group in the same
$E_8$ component.  In both cases the transition can be described by moving 
along a one-parameter family of theories that passes
through
a strongly coupled superconformal fixed point, but does not move onto
the tensor branch. In both the heterotic bundle and the resolved F-theory geometry these transitions are realizable as geometric transitions (i.e. topology changing transitions).
Our description of these transitions in both F-theory and the
heterotic theory is for the most part quite general, but for
comparison of these perspectives we focus in particular on cases where
the F-theory geometry is compactified on a $K3$ fibration over a base
$B$ and the heterotic geometry describes an elliptic fibration over
the same base $B$.

While we primarily focus on 
compactifications to 6D to make the analysis completely concrete and
precise, the transitions we study are local phenomena that will
also arise in field theory without the supergravity coupling; these
transitions should also arise in a similar fashion in 4D theories.
Though some of the technical details and issues involved will be more
subtle in 4D due to the presence of a superpotential and additional
complexity in the theories with reduced supersymmetry, many aspects of
the analysis carried out here, including the general forms of F-theory
Weierstrass models and heterotic bundles, will hold in a large class
of dual geometries for 4D compactifications.
Only the details of the anomaly analysis and some specific features of
the heterotic constructions on specific geometries will depend upon
the dimensionality of the construction.

We begin in \S\ref{sec:SUGRA} with a low-energy description of matter
transitions in 6D theories with SU(6), {\rm SU}(7), and SU(8) gauge groups.
The strong constraints of anomaly cancellation dictate the transitions
that can occur in matter content without a change in the gauge group
of the theory.  In \S\ref{sec:F-theory} we describe these transitions
in F-theory using Weierstrass models, some of which do not have the
standard Tate form for SU($N$).  In \S\ref{sec:heterotic} we describe
the transitions from the heterotic point of view, where the transitions are manifested by instantons moving
between factors in the structure group of a bundle
within one $E_8$ factor.  In \S\ref{sec:matching} we relate the
F-theory and heterotic pictures using the spectral cover construction of the heterotic gauge bundle and explore novel forms of the stable degeneration limit of the F-theory compactification.  In \S\ref{sec:other} we briefly describe some  examples
of matter transitions in other groups, and \S\ref{sec:conclusions}
contains some concluding comments. A variety of useful technical results are provided in the Appendices.

\section{SU($N$) Matter transitions in 6D supergravity}
\label{sec:SUGRA}

\subsection{Anomaly-equivalent matter representations}

In 6D supergravity, anomaly cancellation conditions strongly constrain
the spectrum of massless matter fields that can be charged under a given gauge
group. In some cases, however, there are multiple distinct types of
matter  that give equivalent solutions to the anomaly equations.

One of the simplest examples occurs for SU(6) and higher
SU($N$) gauge
groups, with charged matter that transforms under the
three-index
antisymmetric representation, as described in
\cite{mt-singularities, Grassi-Morrison-2}. In 6D, the anomaly cancellation conditions
for a matter spectrum containing $x_R$ fields transforming in each
representation $R$ of an SU($N$) gauge group are
 (\cite{gs-west, Sagnotti}, as described in \cite{KMT-II})
\begin{align}
-a \cdot b & =   - \frac{1}{6}  \left( A_{\rm adj} - \sum_R
x_R A_R\right)  \label{eq:ab-condition}\\
0  & =
 B^{i}_{\rm adj}  -  \sum_R x_{R} B_{R} \label{eq:f4-condition}\\
b\cdot b & = -\frac{1}{3} \left( C_{\rm adj} - \sum_R x_R
C_R  \right) 
\label{eq:bij-condition}
\end{align}
where $a, b$ are the coefficients of $B R^2$ and 
$B F^2$ Green-Schwarz
terms, and are associated with
vectors in the
signature $(1, T)$ anomaly lattice of the 6D supergravity theory, so that
$a \cdot b,  b \cdot b$ are integers.  The coefficients $A_R, B_R,
C_R$ are numerical constants associated with each representation
computed from
\begin{align}
\tr_R F^2 & = A_R  \tr F^2 \\
\tr_R F^4 & = B_R \tr F^4+C_R (\tr F^2)^2 \;.
\end{align}

When multiple gauge factors $G_i =$ SU($N_i$)
are involved, each gauge factor has an associated
vector $b_i$ in the anomaly lattice, and we have the further condition
\begin{equation}
b_i \cdot b_j = \sum_{R, S}x^{ij}_{R S} A^i_R A^j_S\,,
\label{eq:intersection-condition}
\end{equation}
where $x^{ij}_{R S}$ is the number of fields in the representation $R
\otimes S$ of $G_i \times G_j$.  In the simplest
cases 
we are interested in here,
these are  bifundamental representations.

There is also a constraint that arises from the purely gravitational
anomaly cancellation condition
\begin{equation}
n_H-n_V = 273-29 n_T\,,
\end{equation}
where
$ n_H, n_V, n_T$ are the numbers of matter hypermultiplets, vector multiplets,
and tensor multiplets in the theory respectively.

When the only types of matter that arise are in $k$-index
antisymmetric ($\Lambda^k$) representations of SU($N$), we have, for some integer
$n$,
\begin{equation}
 b \cdot b = n, \;\;\;\;\; -a \cdot b =n+2 \,.
\end{equation}
In the F-theory picture, these configurations come from
gauge groups wrapped on rational (genus zero) curves of
self-intersection $n$.

The dimension and coefficients $A_R$-$C_R$ for the 
adjoint,
fundamental, symmetric,
antisymmetric, and
three-and four-index antisymmetric representations of SU($N$) are
listed in Table~\ref{t:coefficients}
along with the ``genus'' contribution of each representation
(\cite{Erler, kpt, mt-singularities}).

\begin{table}
\centering
\begin{tabular}{|c|c|c|c|c|c|c|}
\hline
 Rep. & $N$ & Dimension & $A_R$ & $B_R$ & $C_R$  & g\\
\hline
 Adj. & $N$ & $N^2-1$ & $2N$ & $2N$ & 6 & 1 \\
& 6, 7, 8 &  {\bf 35}, {\bf 48}, {\bf 63} &
12, 14, 16 & 12, 14, 16 & 6 & 1\\
\hline
 ${\tiny\yng(1)}$ &$N$ & $N$ & 1 & 1 & 0 & 0\\
\hline
 ${\tiny\yng(2)}$ & $N$ &$ \frac{N(N+1)}{2} $ & $ N+2 $ & $ N+8 $ & 3 & 1
\\[0.07in]
&6, 7, 8& {\bf  21}, {\bf  28}, {\bf  36} & 8, 9, 10 & 14, 15, 16 & 3 & 1\\
\hline
 ${\tiny\yng(1,1)}$ & $N$ &$ \frac{N(N-1)}{2} $ & $ N-2 $ & $ N-8 $ & 3 & 0
\\[0.07in]
&6, 7, 8& {\bf 15}, {\bf 21}, {\bf 28} & 4, 5, 6 & -2, -1, 0 & 3 & 0\\
\hline
& & & & & &\\
 ${\tiny\yng(1,1,1)}$ & $N$ & $ \frac{N(N-1)(N-2)}{6} $& $
 \frac{N^2-5N+6}{2} $&$ \frac{N^2-17N+54}{2} $ & $ 3N-12 $ & 0\\[0.07in]
& 6, 7, 8 & {\bf 20}[{10}], {\bf 35}, {\bf 56} &
6[3], 10, 15 & -6[-3], -8, -9 & 6, 9, 12 & 0\\
\hline
& & & & & &\\
 ${\tiny\yng(1,1,1,1)}$ & $N$ & $ \frac{N(N-1)(N-2)(N-3)}{24} $&
 $ \frac{(N-2)(N-3)(N-4)}{6} $ & $\frac{(N-4)(N^2-23N+96)}{6}$ & 
$\frac{3 (N^2 -9 N + 20)}{2}$  &  0\\[0.07in]
 & 8 & {\bf 70}[{35}] & 20[10] & -16[-8] & 18[9] & 0\\
\hline
\end{tabular}
\caption{Values of the group-theoretic coefficients $A_R, B_R, C_R$,
  dimension and genus
for some representations of SU($N$), $N \geq 4$, with specific values
computed for convenience for $N = 6, 7, 8$.  Values in brackets refer
to half-hypermultiplets for self-conjugate representations.}
\label{t:coefficients}
\end{table}

The following
combinations of SU(6) matter fields give equivalent contributions to each of
the anomaly cancellation conditions (including the purely
gravitational anomaly condition $n_H - n_V= 273-29 n_T$).
\begin{equation}
\frac{1}{2} {\bf 20} \;\left(\;\frac{1}{2}{\tiny\yng(1,1,1)}\;\right)+
{\bf  6}\left({\tiny\yng(1)}\right)\; \; \leftrightarrow \;
{\bf  15}\left(\;{\tiny\yng(1,1)}\;\right)
+{\bf  1} \,.
 \label{eq:equivalence-6}
\end{equation}
We refer to these combinations of matter fields as
\emph{anomaly equivalent} \cite{Grassi-Morrison-2}.
Note that we can have a half-hypermultiplet for the {\bf  20}, since
this is a self-conjugate (pseudoreal) representation of SU(6).
These combinations of representations can be seen to be equivalent by
checking that the contribution to each of the terms $\sum_{R}x_R A_R,
\sum_{R}x_R B_R, \sum_{R}x_R C_R$ are the same on both sides of (\ref{eq:equivalence-6}).
The equivalence of these representations under the anomaly conditions
suggests that there is no obstruction to a transition between
SU(6) theories
with the different matter representations.  
Explicit local models from F-theory realizing transitions between these
matter representations were identified in \cite{mt-singularities}, as
described in more detail in the following  section.
The main focus of this paper is the detailed analysis of this and
related types of matter transitions in global models from the dual
F-theory and heterotic perspectives.

For SU(7),  there is a similar type of anomaly equivalence and
associated transition
\begin{equation}
 {\bf   35} \;\left(\;{\tiny\yng(1,1,1)}\;\right)+
5 \times{\bf  7}\left( {\tiny\yng(1)}\right)\; \; \leftrightarrow \;
3 \times {\bf   21}\left(\;{\tiny\yng(1,1)}\;\right)
+7 \times{\bf  1} \,.
\label{eq:equivalence-7}
\end{equation}

For SU(8), there are anomaly equivalent matter representations

\begin{equation}
 {\bf   56} \;\left(\;{\tiny\yng(1,1,1)}\;\right)+
9 \times{\bf   8}\left({\tiny\yng(1)}\right)\; \; \leftrightarrow \;
4 \times {\bf   28}\left(\;{\tiny\yng(1,1)}\;\right)
+16 \times{\bf  1} \,.
\label{eq:equivalence-81}
\end{equation}

and

\begin{equation}
\frac{1}{2} {\bf   70} \;\left(\frac{1}{2}\;{\tiny\yng(1,1,1,1)}\;\right)+
8 \times{\bf   8}\left({\tiny\yng(1)}\right)\; \; \leftrightarrow \;
3 \times {\bf  28}\left(\;{\tiny\yng(1,1)}\;\right)
+15 \times{\bf  1} \,.
\label{eq:equivalence-82}
\end{equation}

In the following sections, we describe the extent to which these
equivalences correspond to transitions that have realizations in
global F-theory and heterotic models.  There is a similar anomaly
equivalence for the 3-index antisymmetric ($\Lambda^3$) representation
of SU(9), which we discuss further below.  Similar equivalences would
seem at first to be possible for $\Lambda^3$ representations of SU(10)
and higher SU($N$) and for $\Lambda^4$ representations of SU(9) and
above, {\it etc.}, but global considerations (discussed below) seem to rule
out such models for all values of $T$.

\subsection{SU($N$) blocks in 6D supergravity}
\label{sec:blocks}

The anomaly conditions constrain the total matter content that can be
charged under a given component SU($N$) of the full 6D gauge group.
For the most generic 6D SU($N$) models with specific values of
$b \cdot b = n$, a ``genus'' $g = 1-(a \cdot b + b \cdot b)/2$, and $N \geq 4$,
the matter content contains only adjoint, fundamental, and
antisymmetric representations, and is given by
\begin{equation}
g \times ({\rm adjoint}) + \left[ 16 (1-g) + (8-N) n \right]
\times {\tiny\yng(1)}
+\left( n + 2-2g \right) \times {\tiny\yng(1,1)} \,.
\label{eq:model-general}
\end{equation}
The sense in which this matter content is the most generic is that it
corresponds to the theory with the largest number of uncharged scalar fields.  Each of
the anomaly-equivalent combinations involving a triple or
quadruple-antisymmetric representation described above removes some
number of scalar matter fields, corresponding to a more refined
 ``tuning'' of the
low-energy field theory model.

For genus $g = 0$ models, we have the following spectra for SU(6),
SU(7), {\rm SU}(8), where $b \cdot b = n$

\begin{equation}
SU(6): (16 + 2n + r)  \times {\bf 6} ({\tiny\yng(1)})
+ (n + 2 -r)\times {\bf 15} \left({\tiny\yng(1,1)}\right)
+r \times \frac{1}{2}{\bf 20} \left(\frac{1}{2}{\tiny\yng(1,1,1)}\right) \,.
\label{eq:model-6}
\end{equation}

\begin{equation}
SU(7):  (16+ n+ 5r) \times {\bf 7} ({\tiny\yng(1)})
+(n + 2-3r) \times {\bf 21} \left({\tiny\yng(1,1)}\right)
+r \times {\bf 35} \left({\tiny\yng(1,1,1)}\right) \,.
\label{eq:model-7}
\end{equation}

\begin{equation}
SU(8): (16 + 9r + 8r')  \times {\bf 8} ({\tiny\yng(1)})
+(n + 2-4r-3r') \times {\bf 28} \left({\tiny\yng(1,1)}\right)
+r \times {\bf 56} \left({\tiny\yng(1,1,1)}\right) 
+ r' \times \frac{1}{2}{\bf 70} \left(\frac{1}{2}{\tiny\yng(1,1,1,1)}\right) \,.
\label{eq:model-8}
\end{equation}

The possible matter spectra for  SU(6) factors was described in
\cite{Bershadsky-all} and there also related to F-theory and heterotic theory.
As we see in the later sections, both in F-theory and heterotic theory
all of the SU(6) and SU(7) 6D models can be realized through
global constructions, and there
are matter transitions 
along paths in the space of theories
that change between different values of $r$
without changing the gauge group ({\it i.e.}, without involving
Higgsing processes).  We have identified explicit constructions only
for a subset of the possible SU(8) models, as discussed in further
detail below, and it is less clear whether there is a UV-consistent
description of the SU(8) transitions.

We briefly summarize the situation for SU($N$) blocks with $N > 8$.
For SU(9), the generic
$g = 0$ model has $16-n$ fundamentals ({\bf 9}'s) and
  $(n + 2)$ antisymmetrics ({\bf 36}'s).  The SU(9) model with a
  $\Lambda^3$ representation and the smallest number of matter fields
  has $n = 3$ and at least 327 matter fields (1 $\times {\bf 84}$ + 27
$  \times$ {\bf 9}).  Thus, in the absence of other gauge factors,
  $n_H-n_V\geq 327-80 = 247$.  SU(9) models with a $\Lambda^3$
    representation appear to be consistent in $n_T = 0$ supergravity (with $n_H-n_V=
    273$), though these models cannot have a heterotic description on
    a smooth K3.
It seems just barely possible to construct SU(9) models with $n_T = 1$,
where $n_H-n_V= 244$, which should in principle have an F-theory
description and heterotic duals.  For example, by adding an SU(3)
factor, such as can be done in a heterotic model with $12 \pm 3$
instantons in the two $E_8$ factors, corresponding to F-theory on
$\F_3$, there is just enough room to satisfy the gravitational anomaly
bound.  As we discuss in the following sections, however, it is not
clear whether or how these models may be realized in either the
F-theory or heterotic constructions.  Similar considerations of the
global gravitational anomaly condition show that SU(9) models with
$\Lambda^4$ or SU(10) models with $\Lambda^3$ representations are not
possible even with $n_T = 0$, assuming there are no further gauge
factors.  Note that this argument does not completely rule out 6D
supergravity models in which the fundamentals of {\it e.g.} SU(9) with
a $\Lambda^4$ are also charged under additional gauge factors,
effectively increasing $n_V$ without changing $n_H$, or models where there
is a second, non-Higgsable gauge factor, which contributes to $n_V$
without a corresponding contribution to $n_H$. But such models seem very
difficult to construct in a consistent fashion.  Note also that the
constraints just discussed rely on the purely gravitational anomaly
cancellation condition, and do not in principle constrain the
existence of {\it e.g.} SU(10) models with $\Lambda^3$ representations
in 6D field theory.

\subsection{Higgsing processes}

The different models with matter transitions that we consider are
connected to one another, and to other related models, by a network of
Higgsing transitions, many of which were also considered in
\cite{Bershadsky-all}.  In general, Higgsing a gauge group SU($N$) in
a supersymmetric theory requires turning on a vacuum expectation value
(VEV) for matter fields that transform in nontrivial representations.
In the language of 4D ${\cal N} = 1$ theories, such expectation values
must be turned on in such a way that the D-term constraints of the
form $\sum_{i} \bar{\phi}_iT_A \phi_i = 0$ are satisfied for each
generator $T_A$ of the Lie algebra.

\subsubsection{Higgsing fundamentals} \label{higsfund}

The simplest Higgsing of SU($N$) is done by giving vacuum expectation
values to two fundamental fields.  Note that two
fundamental fields must be given
VEVs in order to satisfy the D-term constraints in the supersymmetric
theory.  
Supersymmetric Higgsing on a single fundamental of SU($N$) is not
possible.  Recall that each full 6D ${\cal N} = 1$ hypermultiplet in a
given representation $R$ contains fields in both the representation
$R$ and its conjugate, so when Higgsing two fundamental fields we are
really giving expectation values to a field component in the
representation $R$ and  another field component in the conjugate
representation $\bar{R}$, allowing the D-term constraint to be
satisfied.  For example, for the fundamental representation any VEV
can be rotated into the canonical form $(0, 0, \ldots, 0, v)$.  This
can be described in the language of Young tableaux by a single box
containing the value $[N]$.  When two fundamental fields are given
VEVs in this way, the gauge group is broken down to SU($N -1$).  The
Goldstone bosons of the Higgsed matter fields are ``eaten'' by the
broken gauge generators in the usual fashion.  In 6D supergravity
theories this matches with the gravitational anomaly cancellation
condition $n_H-n_V = 273-29n_T$.

Explicitly,
we can match the number of Goldstone bosons $2 N -1$ with the number
of broken gauge generators $[N^2 -1] -[(N -1)^2 -1] = 2 N -1$.  After
breaking SU($N$) $\rightarrow$ SU($N -1$), the other representations
branch as follows:
\begin{eqnarray}
{\tiny\yng(1)}_N & \rightarrow & {\tiny\yng(1)}_{N -1}+ {\bf 1}\\
{\tiny\yng(1,1)}_N & \rightarrow & {\tiny\yng(1,1)}_{N -1}+
{\tiny\yng(1)}_{N -1}\\
{\tiny\yng(1,1,1)}_N & \rightarrow & {\tiny\yng(1,1,1)}_{N -1}+
{\tiny\yng(1,1)}_{N -1}\\
{\tiny\yng(1,1,1,1)}_8 & \rightarrow &
{\tiny\yng(1,1,1,1)}_{7} + {\tiny\yng(1,1,1)}_{7}
=
\overline{{\tiny\yng(1,1,1)}}_{7} + {\tiny\yng(1,1,1)}_{7}\,.
\end{eqnarray}
This can easily be understood by looking at the Young tableaux for
each representation; the $k$-antisymmetric representation with an $N$
entry goes to a $(k-1)$-antisymmetric representation, and the tableaux
without an $N$ entry go to a $k$-antisymmetric representation of SU($N
-1$). The dimensions of these sets of representations match as, for example in the case of triple antisymmetrics,
\begin{equation}
  \frac{N ( N -1) (N -2)}{6}  =
\frac{(N -1) (N -2) (N -3)}{6}  + \frac{(N -1) (N -2)}{2}  \,. 
\end{equation}

We can see that this Higgsing process takes an SU($N$) model with
a specific value of $n$ to an SU($N -1$) model with the same value of
$n$ between the models (\ref{eq:model-6}, \ref{eq:model-7},
\ref{eq:model-8}).  Higgsing the SU(7) models with a value $r_7 > 0$
  gives the corresponding SU(6) model with twice that value of $r_6
  = 2r_7$.
Higgsing an SU(8) model with $r_8 > 0, r_8' \geq 0$ gives the SU(7) model with
$r_7 = r_8 + r_8'$.

In short, in breaking SU($N$) to SU($N-1$) on a fundamental representation, we lose two full hypermultiplets in the $\mathbf{N-1}$ (fundamental) representation
of ${\rm SU}(N-1)$ and one singlet full hypermultiplet. This accounts for a reduction of $2N-1$ full hypermultiplet degrees of freedom, one for each
gauge boson that becomes massive, as required by the six-dimensional super-Higgs mechanism.

\subsubsection{Higgsing antisymmetric representations} \label{antihigs}

We can similarly Higgs two antisymmetric representations, giving
VEVs to states with Young tableau entries $[N -1, N]$.  This breaks
\begin{equation}
SU(N) \rightarrow SU(N-2) \times {\rm SU}(2)
\label{eq:Higgsing-anti}
\end{equation}
The two antisymmetrics that break the gauge group have $ 2(N -2)$
Goldstone bosons each, corresponding to states with Young tableaux having
entries $[i, j]$ with $i \leq N- 2, j \geq N -1$,
and the number of generators in the group is reduced by
\begin{equation}
[N^2 -1] -[(N -2)^2 -1]-[3] = 4 N-7=2 \times (2N -4)+1\,.
\end{equation}
Together with the loss of an additional singlet hypermultiplet (containing one goldstone boson and 3 degrees of freedom fixed by D-terms) the number of degrees of freedom match.  The antisymmetric
representations break down in the decomposition
(\ref{eq:Higgsing-anti}) as
\begin{eqnarray}
{\tiny\yng(1)}_N & = & 
  {\tiny\yng(1)}_{N -2} \times {\bf
  1}  + {\bf 1} \times   {\tiny\yng(1)}_2\\
{\tiny\yng(1,1)}_N & = & 
  {\tiny\yng(1,1)}_{N -2} \times {\bf
  1} + {\tiny\yng(1)}_{N -2} \times 
{\tiny\yng(1)}_2 + {\bf 1} \times {\bf 1}\\
{\tiny\yng(1,1,1)}_N & = & 
  {\tiny\yng(1,1,1)}_{N -2} \times {\bf
  1} + {\tiny\yng(1,1)}_{N -2} \times 
{\tiny\yng(1)}_2 +{\tiny\yng(1)}_{N-1}  \times {\bf 1} \label{eq:Higgs-a-3}
\end{eqnarray}
For the two $\Lambda^2$ fields that take VEVs, one singlet is the VEV
component, and
the  other singlet and the bifundamental degrees of freedom are the ones that are lost, so
the number of antisymmetric representations stays unchanged in the
breaking.

More generally, Higgsing two $k$-antisymmetric representations breaks
\begin{equation}
SU(N) \rightarrow SU(N -k) \times {\rm SU}(k) \,.
\end{equation}
There are $2k (N -k)+1$ degrees of freedom eaten by the lost gauge
bosons, and again the $k$-antisymmetrics are preserved under the
breaking and the bifundamentals and singlets are lost.

\subsection{Product groups and transitions}

6D models with multiple SU($N$) gauge factors can be constructed in
close parallel to the single-block models with one SU($N$) factor.  We
consider models where SU($N$) and SU($M$) are both realized as in
\S\ref{sec:blocks}, with the same value of $b$ in the anomaly lattice, so that
$b_1 \cdot b_1 = b_2 \cdot b_2 = b_1 \cdot b_2 = n$.  These are 
product group models with a smooth heterotic dual.
Generically, the intersection condition
(\ref{eq:intersection-condition}) is satisfied by including $n$
bifundamental $({\bf N}, {\bf M})$ fields in the spectrum.  For 
example, for
$n
\geq 0$, 
the spectrum for the generic SU($4$) $\times$ SU($2$) model
is
\begin{equation}
(n + 2) \times\left({\tiny\yng(1,1)}_4
\times {\bf 1} \right)
+ n \times \left({\tiny\yng(1)}_4\times {\tiny\yng(1)}_2 \right)
+ (16 + 2n) \times \left({\tiny\yng(1)}_4\times {\bf 1}\right)
+ (16 + 2n) \times \left({\bf 1}\times {\tiny\yng(1)}_2\right)\,.
\end{equation}

Just as Higgsing SU(8) or SU(7) models on a pair of fundamentals
relates models with a non-generic ({\it e.g.} $\Lambda^3$)
representation to other such models, Higgsing on antisymmetric
representations also gives rise to exotic matter structures for
product groups.  For example, consider breaking an SU(6) model with
$r$ half-hyper $\Lambda^3$ representations on a pair of antisymmetric
($\Lambda^2$) representations.  Then, from (\ref{eq:Higgs-a-3}), each
of the $r$ $\frac{1}{2}${\bf 20}'s breaks as
\begin{equation}
\frac{1}{2}{\tiny\yng(1,1,1)}_6
\rightarrow
({\tiny\yng(1)}_4 \times {\bf 1})
+ (\frac{1}{2}{\tiny\yng(1,1)}_4 \times
{\tiny\yng(1)}_2) \,.
\end{equation}
In general, this gives an SU(4) $\times$ SU(2) model with spectrum
\begin{eqnarray}
 &  & 
r \times (\frac{1}{2}{\tiny\yng(1,1)}_4 \times {\tiny\yng(1)}_2) 
+(n + 2-r) \times ({\tiny\yng(1,1)}_4 \times {\bf 1})
+(n-r) \times ({\tiny\yng(1)}_4 \times {\tiny\yng(1)}_2) 
\\
 &  & 
\hspace*{0.1in}
+(16 + 2n + 2r) \times ({\tiny\yng(1)}_4 \times  {\bf 1}) 
+(16 + 2n + r) \times ( {\bf 1} \times {\tiny\yng(1)}_2)  \,.
\end{eqnarray}
From this spectrum we see that there must be an anomaly equivalence
\begin{equation}
 (\frac{1}{2}{\tiny\yng(1,1)}_4 \times {\tiny\yng(1)}_2)
+ 2 \times  ({\tiny\yng(1)}_4 \times  {\bf 1}) 
+  ( {\bf 1} \times {\tiny\yng(1)}_2)
\leftrightarrow
({\tiny\yng(1,1)}_4 \times {\bf 1})+
 ({\tiny\yng(1)}_4 \times {\tiny\yng(1)}_2)
+ 2 \times ({\bf 1} \times {\bf 1})\,.
\label{eq:equivalence-42}
\end{equation}
Indeed, this relation also follows from a direct Higgsing of
(\ref{eq:equivalence-6}) under the breaking 
SU(6) $\rightarrow$ SU(4) $\times$ SU(2).

A similar transition can be found for SU(5) $\times$ SU(2) theories by
breaking SU(7):
\begin{equation}
 ({\tiny\yng(1,1)}_5 \times {\tiny\yng(1)}_2)
+ 6 \times  ({\tiny\yng(1)}_5 \times  {\bf 1}) 
+  5 \times( {\bf 1} \times {\tiny\yng(1)}_2)\,
\leftrightarrow
2 \times ({\tiny\yng(1,1)}_5 \times {\bf 1})+
3 \times ({\tiny\yng(1)}_5 \times {\tiny\yng(1)}_2)
+ 10 \times ({\bf 1} \times {\bf 1}) \,.
\label{eq:equivalence-52}
\end{equation}

And for SU(6) $\times$ SU(2) by breaking SU(8) on a pair of
antisymmetric representations:

\begin{equation}
 ({\tiny\yng(1,1)}_6 \times {\tiny\yng(1)}_2)
+ 8 \times  ({\tiny\yng(1)}_6 \times  {\bf 1}) 
+  9 \times( {\bf 1} \times {\tiny\yng(1)}_2)\,
\leftrightarrow
2 \times ({\tiny\yng(1,1)}_6 \times {\bf 1})+
4 \times ({\tiny\yng(1)}_6 \times {\tiny\yng(1)}_2)
+ 18 \times ({\bf 1} \times {\bf 1}) \,.
\label{eq:equivalence-62}
\end{equation}

These relations suggest that  transitions
between theories with these different matter contents should be
possible; in the later sections of this paper we explicitly identify
these transitions in F-theory and heterotic theories.
While we find that the SU(4) $\times$ SU(2) and SU(5) $\times$ SU(2)
transitions are described nicely in both F-theory and the heterotic
theory, the SU(6) $\times$ SU(2) transition is less clear.
The representation that transforms as an antisymmetric field under
SU($N$) and a fundamental under SU($M$) is an exotic representation,
analogous in many ways to the triple-antisymmetric representation of
SU($N$).
Generalizing these constructions, we can Higgs on a pair of
triple-antisymmetric representations; this gives interesting
transitions in SU($N$) $\times$ SU(3) theories, as we explore to some
extent in the following sections.

It would be interesting to explore further product representations and
associated transitions.  For example, Higgsing the relation
(\ref{eq:equivalence-82}) on a pair of antisymmetric representations
suggests that there should be a transition
in SU(6) $\times$ SU(2) theories
\begin{equation}
 (\frac{1}{2}{\tiny\yng(1,1,1)}_6 \times {\tiny\yng(1)}_2)
+ 8 \times  ({\tiny\yng(1)}_6 \times  {\bf 1}) 
+  8 \times( {\bf 1} \times {\tiny\yng(1)}_2)
\leftrightarrow
2 \times ({\tiny\yng(1,1)}_6 \times {\bf 1})+
3 \times ({\tiny\yng(1)}_6 \times {\tiny\yng(1)}_2)
+ 16 \times ({\bf 1} \times {\bf 1}) \,.
\end{equation}
Since we have not identified an explicit realization of the
$\Lambda^4$ representation of SO(8) in either the F-theory or
heterotic pictures, however, it is not clear whether this transition
can be realized in consistent 6D supergravity theories.  One could
also use this approach to explicitly construct representations such as
the trifundamental of SU(2) $\times$ SU(2) $\times$ SU(2) by breaking
the $({\bf 6}, {\bf 2})$ of SU(4) $\times$ SU(2) to $({\bf 2}, {\bf
  2}, {\bf 2})$ by breaking SU(4) $\rightarrow$ SU(2) $\times$ SU(2).
And the relations (\ref{eq:equivalence-42})-(\ref{eq:equivalence-62})
suggest that similar transitions may occur for higher groups such as
SU(7) $\times$ SU(2).  We leave further exploration of these
possibilities to future work.

\section{F-theory description of matter transitions}
\label{sec:F-theory}

F-theory provides powerful methods for realizing stringy constructions
of supergravity theories. In this section, we give F-theory
realizations of the models with matter transitions described above,
building on the work of \cite{mt-singularities}. These F-theory models
provide further insights into the mechanisms behind the matter
transitions.

\subsection{F-theory overview}

Here, we briefly describe those aspects of F-theory that will be
important in the upcoming analysis. More extensive reviews of F-theory
can be found in \cite{Morrison-TASI, Denef-F-theory,Taylor:2011wt}.

\subsubsection{Weierstrass models}

F-theory \cite{Vafa-F-theory, Morrison-Vafa-I, Morrison-Vafa-II}
is a method for compactifying type IIB string theory in
situations where the axiodilaton is allowed to vary over the
compactification space. The axiodilaton and the corresponding
SL(2, $\mathbb{Z}$) symmetry appear geometrically through an elliptic
fibration  over the compactification base. Specifically, an F-theory
compactification to $10-2d$ dimensions is given by an elliptic
fibration over a base $B$ of complex dimension $d$. These elliptic
fibrations can be described using the Weierstrass form
\begin{equation}
y^2 = x^3 + f x + g.
\end{equation}
Here, $y$ and $x$ are coordinates describing an elliptic curve, while
$f$ and $g$ vary over the base $B$. To preserve supersymmetry, the
total elliptic fibration must be a Calabi-Yau manifold. As a result,
$f$ and $g$ must respectively be sections of $\mathcal{O}(-4K_B)$ and
$\mathcal{O}(-6K_B)$, where $K_B$ is the canonical class of the base
$B$. Note that $B$ does not necessarily need to be a Calabi-Yau
manifold on its own.

The sections $f$ and $g$ can be described locally near a divisor
$\Sigma$ on $B$ using the formalism of
\cite{Anderson:2014gla,mt-4D-clusters}.  Let $\Sigma$ have an associated
coordinate $\sigma$, so that $\Sigma =\{\sigma = 0\}$. We can then
expand $f$ and $g$ in terms of 
$\sigma$ as
\begin{equation}
f = f_0 + f_1 \sigma + f_2 \sigma^2 + \ldots \label{eq:fexp}
\end{equation}
and 
\begin{equation}
g = g_0 + g_1 \sigma + g_2 \sigma^2 + \ldots . \label{eq:gexp}
\end{equation}
The coefficients $f_k$ and $g_k$ are respectively sections of
$\mathcal{O}(-4K_\Sigma +(4-k) N_\Sigma)$ and $\mathcal{O}(-6K_\Sigma
+(6-k) N_\Sigma)$, where $-K_\Sigma$ and $N_\Sigma$ are the
anti-canonical and normal line bundles for $\Sigma$.  
Note that in general the coordinate $\sigma$ is not globally defined,
and only the first nonvanishing $f_k, g_k$ is uniquely defined, as
discussed in more detail in \cite{mt-4D-clusters}.  These subtleties
are irrelevant for the discussion in this paper; for toric bases, in
particular, $\sigma$ can always be taken as a global toric coordinate.

While the analysis in this section is quite general, and applies to 4D
as well as 6D F-theory models,
in
this paper we focus particularly on F-theory compactifications to 6D with
heterotic duals, where the compactification base $B$ is  one of the
Hirzebruch surfaces $\mathbb{F}_n$. For a description of Hirzebruch
surfaces from an F-theory perspective, see
\cite{Taylor:2011wt,Morrison-Vafa-I}. Hirzebruch surfaces are complex
two-dimensional surfaces that can be described as $\mathbb{P}_1$
bundles over $\mathbb{P}_1$. The divisors of $\mathbb{F}_n$ have a basis consisting of divisors $S$ and $F$. 
$F$ and $S$ span the cone of effective curves on $\F_m$.
$S$ is a section
of the fibration, with a 1-1 map to the $\mathbb{P}_1$ base of $\mathbb{F}_n$, and has self-intersection
number $-n$ ({\it i.e.} $S\cdot S=-n$). $F$ meanwhile refers to the fiber
$\mathbb{P}_1$, and satisfies $S\cdot F = 1$ and $F\cdot F =
0$. Another important divisor class for $\mathbb{F}_n$ is
$\tilde{S}=S+nF$, which satisfies $S\cdot \tilde{S} = 0$
and $\tilde{S}\cdot \tilde{S}= n$.
The anti-canonical class of $\mathbb{F}_n$ is
\begin{equation}
-K_{\mathbb{F}_n} = 2S+(n+2)F,
\end{equation} 
so $f$ and $g$ are sections of $\mathcal{O}(8S + (4n+8)F)$ and
$\mathcal{O}(12S + (6n+12)F)$. 
In most of our analysis we will tune the relevant gauge groups on
$\tilde{S}$.  We thus associate the coordinate $\sigma$ with $\tilde{S}$ and $z$ with $F$,
$f$ and $g$ are therefore polynomials of order 8 and 12 in $\sigma$, at
maximum. Using Equations \eqref{eq:fexp} and \eqref{eq:gexp}, we can
expand $f$ and $g$ around $\tilde{S}$ as
\begin{align*}
f &= \sum_{k=0}^8 f_k \sigma^k,\\
g &= \sum_{k=0}^{12} g_k \sigma^k.
\end{align*}
$f_k$ and $g_k$ can be described in terms of a common line bundle on $\P^1$,
which we take to be $\mathcal{O}(1)$. $-K_{\tilde{S}}$ is then
equivalent to $2$, while $N_{\tilde{S}}$ is $n$. $f_k$ and $g_k$ are
thus polynomials in $z$ of order $8+n(4-k)$ and $12+n(6-k)$. For
$n\geq 3$, some of the $f_k$ or $g_k$ may be ineffective for $k\leq 8$
or $k\leq 12$, meaning that higher order terms in $\sigma$ must
vanish. This signals the presence of a non-Higgsable cluster
(described below) on $S$. The expressions for $f$ and $g$ given above
are in fact the generic expansions near a $+n$ curve, so long as the
limits of $k$ are adjusted for the appropriate situation.  

\subsubsection{Gauge groups in F-theory}

In F-theory, the total elliptically-fibered compactification space may
admit certain types of singularities.  
 Singularities of the fibration
occur at loci where the
discriminant $\Delta$, defined as
\begin{equation}
\Delta = 4 f^3+27 g^2,
\end{equation}
equals zero. 
Some of these singularities in the fibration give singularities in the
total space as well.  In general,
not all singularities in the total space can be resolved to give a smooth
Calabi-Yau manifold. Kodaira classified all of the resolvable
singularities for situations where the singularity occurs along a
codimension one locus on the base \cite{Kodaira}. As summarized in
Table \ref{tab:kodaira}, the resulting singularities can be described
by the orders of vanishing of $f$, $g$, and $\Delta$ on the
codimension one locus. If $f$ and $g$ vanish to orders 4 and 6 or
greater on a codimension one locus, the resulting singularity has no
Calabi-Yau resolution.
For F-theory compactifications,  singularities that arise over
codimension one loci on the base can always be resolved to give a
smooth Calabi-Yau manifold and obey the Kodaira classification.

\begin{table}[tbp]
\centering
\begin{tabular}[t]{|c|c|c|c|c|c|}\hline
Fiber Type & $\text{ord}(f)$ & $\text{org}(g)$ & $\text{ord}(\Delta)$ & Singularity Type & Gauge Algebra \\\hline
$I_0$ & 0 & 0 & 1 & none & none \\
$I_n$ & 0 & 0 & $n$ & $A_{n-1}$ & $\agsu(n)$ or $\agsp(\fl{\frac{n}{2}})$ \\
$II$ & $\geq 1$ & 1 & 2 & none & none \\
$III$ & 1 & $\geq 2$ & 3 & $A_1$ & $\agsu(2)$ \\
$IV$ & $\geq 2$ & 2 & 4 & $A_2$ & $\agsu(2)$ or $\agsu(3)$ \\
$I_0^*$ & $\geq 2$ & $\geq 3$ & 6 & $D_4$ & $\aggg_2$, $\agso(7)$ or $\agso(8)$ \\
$I^*_n$ & 2 & 3 & $n+6$ & $D_{n+4}$ & $\agso(2n+7)$ or $\agso(2n+8)$\\
$IV^*$ & $\geq 3$ & 4 & 8 & $E_6$ & $\agf_4$ or $\age_6$\\
$III^*$ & 3 & $\geq 5$ & 9 & $E_7$ & $\age_7$ \\
$II^*$ & $\geq 4$ & 5 & 10 & $E_8$ & $\age_8$ \\\hline
\end{tabular}
\caption{Kodaira classification of codimension one singularities in
  elliptic fibrations and corresponding gauge algebras. When multiple
  gauge algebras are given, the gauge algebra is determined by
  monodromy conditions.} 
\label{tab:kodaira}
\end{table}

When an F-theory compactification has such a codimension one
singularity, the physical theory has a corresponding nonabelian gauge
symmetry.  
Resolving the codimension one singularity will produce a set
of 2-cycles whose intersection pattern can be mapped to a Dynkin
diagram. The Dynkin diagram then identifies the algebra for the
physical model's gauge symmetry. 
In this paper we generally describe theories using
the Lie group, with the understanding that only the algebra is
actually fixed definitively by F-theory, and that the given Lie group
may be subject to a quotient by a finite discrete subgroup. 
Note that monodromy effects need to
be considered when compactifying to 6D or 4D, meaning that the same
Kodaira fiber type can give different gauge algebras. In such cases,
determining the resulting gauge group requires a more in-depth
analysis of $f$, $g$, and $\Delta$ (see, for example,
\cite{Bershadsky-all, Katz-etal-Tate}). An example is the split condition, where an
$I_n$ fiber can correspond to either an $\agsu$ algebra or an $\agsp$
algebra depending on the form of $f_0$ and $g_0$.

The coefficients in $f$ and $g$ can be tuned to special values that
satisfy the conditions in Table \ref{tab:kodaira}. In such cases, the
gauge symmetry can be Higgsed by deforming the coefficients away from
their special values. However, when constructing Weierstrass models on
rigid divisors, some coefficients in the expansion of $f$ and $g$
around that divisor may be forced to
vanish, giving a gauge symmetry that cannot be removed by altering
coefficients. These gauge symmetries are known as non-Higgsable
clusters and are discussed further in \cite{mt-clusters,
  mt-4D-clusters}. Most of the examples we focus on here will involve
tuned gauge symmetries rather than non-Higgsable clusters.

The gauge symmetry discussion has focused on local features of
the Weierstrass model near a particular divisor. Local considerations
are sufficient to determine the symmetry algebra of a nonabelian,
continuous gauge symmetry.
 Producing abelian or some discrete
symmetries involves creating an extra section in the elliptically
fibered compactification space, which, however,
requires an analysis of
global behavior. Our F-theory analysis will mostly be concerned with
nonabelian algebras,  and local analyses involving expansions such as
\eqref{eq:fexp} and \eqref{eq:gexp} will suffice.

\subsubsection{Matter}

If an F-theory compactification to $d\leq 6$ has a codimension one
singularity, there generally will be codimension two loci within the
codimension one locus where the singularity type is enhanced. These
codimension two loci indicate
that the model has matter charged under the
corresponding gauge algebra. In some cases, it is easy to determine
the representations of the charged matter. For example, if the
singularity undergoes a rank-one enhancement to a standard Kodaira
fiber, the resulting charged matter can be found using the Katz-Vafa
method \cite{Katz-Vafa}. The enhanced singularity type has a
corresponding gauge algebra; breaking the adjoint of this enhanced
gauge algebra to the original gauge algebra gives the matter
content. Importantly, the enhanced singularity does not represent an
actual enhancement of the gauge group. Breaking the adjoint of the
enhanced singularity's gauge algebra simply provides a convenient way
of determining the matter content. As an example, consider a situation
where a codimension one $A_{n-1}$ singularity enhances to an $A_{n}$
singularity on a codimension two locus. The adjoint of $A_{n}$ breaks
as
\begin{equation}
\textbf{Adj} \rightarrow \textbf{Adj} + \mathbf{n}+\mathbf{\bar{n}} + \mathbf{1}.
\end{equation}
The $\mathbf{n}+\mathbf{\bar{n}}$ term in the above breaking pattern represents the charged matter contributed by each $A_{n}$ locus. In particular, matter in 6D F-theory models must come in quaternionic representations, and the $\mathbf{n}$ and $\mathbf{\bar{n}}$ combine to form a full multiplet in the $\Lambda^1$ or ${\tiny\yng(1)}$ representation. The Katz-Vafa analysis can be used for $A_{n-1}\rightarrow D_{n}$ and $A_6 \rightarrow E_7$ enhancements as well.

However, the Kodaira classification is strictly valid only for
codimension one, and Kodaira codimension-one singularities can enhance
to non-standard codimension-two singularities. Moreover, there can be
codimension-two enhancements that do not enhance the rank by exactly
one. Determining the resulting matter in these cases requires a more
detailed analysis, as described in \cite{mt-singularities, Esole-Yau,
  Lawrie-sn, Hayashi-ls, hlms, Esole-sy, Braun-sn}. Examples that will
be of interest here are the $A_5 \rightarrow E_6$ and $A_7 \rightarrow
E_8$ enhancements described in
\cite{mt-singularities}. The work of \cite{Grassi-Halverson-Shaneson} presents an
alternative method of determining the matter content that does not
require an explicit resolution of singularities.

The number of singlet hypermultiplets corresponds to
(one more than) the $h^{2,1}$ of
the compactification space. Alternatively, the number of singlets can
be found by counting the number of complex degrees of freedom in the
Weierstrass model. There is not a direct
1-1 equivalence between the
Weierstrass degrees of freedom and the number of neutral
hypermultiplets, as automorphisms on the base and the effects of $-2$
curves must be taken into account. The complete expression for
relating the Weierstrass degrees of freedom to the number of neutral
hypermultiplets is given in \cite{Martini-WT}. 

\subsubsection{Superconformal points and tensionless string transitions}

In some situations, a codimension two locus can have an enhanced
singularity such that $f$ and $g$ vanish to orders 4 and 6. Resolving
these codimension two singularities requires a blow-up on the base, as
described in \cite{Morrison-Vafa-II,  Witten-phases}. In
contrast, the codimension-two enhanced singularities giving matter can
be resolved using blow-ups only on the elliptic fiber. From a field
theory perspective, blowing up the base introduces an additional
tensor multiplet, and the size of the new $\mathbb{P}^1$ corresponds
to the expectation value $\expec{S}$ of the tensor multiplet's
scalar. The original codimension-two locus can then be thought of as
describing the limit where $\expec{S}$ approaches zero. $\expec{S}$
governs the tension of strings that couple to the tensor. In the limit
where $\expec{S}$ goes to zero, these strings becomes tensionless
\cite{Bershadsky-Johansen}. The blow-up and blow-down processes
associated with these codimension two loci are thus the tensionless
string transitions described in the literature \cite{Ganor-Hanany,
  Seiberg-Witten}. $\expec{S}$ also controls the couplings of any
gauge groups that appear after the blowup, and shrinking the
exceptional curves to zero size takes any such gauge group to its
strongly coupled limit. These loci are associated with 6D
superconformal field theories \cite{Seiberg} and have been the focus
of much recent work
\cite{Heckman-Morrison-Vafa,DelZotto-Heckman-Tomasiello-Vafa,Haghighat-Klemm-Lockhart-Vafa,DelZotto-Heckman-Morrison-Park,Heckman-Morrison-Rudelius-Vafa}. For
this reason, we will refer to these codimension two loci as
superconformal points.

The tensionless string transitions provide a means of connecting
F-theory compactifications on different bases, playing a similar role
as the small instanton transitions of heterotic string theory
\cite{Seiberg-Witten, Morrison-Vafa-II}. For instance, one can blow up
$\mathbb{F}_1$ at a point and perform a subsequent blow-down to obtain
an F-theory compactification on $\mathbb{F}_2$. 
In fact, all 6D F-theory compactifications are connected into a single
moduli space by such transitions \cite{Seiberg-Witten,
KMT-II}.
In some cases, the
matter content can change during
such a tensionless string transition, as
in the chirality-changing 4D
phase transitions studied in
\cite{Kachru-Silverstein, Ovrut-Pantev-Park,Donagi:1999jp,Buchbinder:2002ji}. 
Such transitions involve a change in the number of  tensor multiplets.
The transitions we consider here differ in that they
also involve passing through a superconformal point in
the moduli space of vacua, but do not involve a change in the number
of tensor multiplets, and are also distinct from
Higgsing/unHiggsing transitions, which always involve a change in the
gauge field content of the theory and generally do not involve passing
through a superconformal point.

\subsection{F-theory tunings of $\gsu(6)-\gsu(8)$ models}
\label{subsec:ftheorytunings}
In this section, we derive the explicit local forms of the Weierstrass
models for $\gsu(6)-\gsu(8)$ that will be used to analyze the
transitions. We mostly use the conventions of \cite{mt-singularities},
where gauge groups were tuned on a divisor $D$ with associated
coordinate $\sigma$.  We assume in particular that the divisor $D$ is
smooth and that we are working over a unique factorization domain
(UFD).  When $D$ is singular there are more complicated ways of
realizing SU($N$) gauge groups, which generally involve
representations other than the adjoint and $k$-index antisymmetric
representations studied in this section; we describe one such example
in \S\ref{sec:SU(3)}.  This general analysis is valid for an arbitrary
choice of F-theory base $B$, which could be of complex dimension two
or three and an arbitrary smooth effective divisor $D$.  For the rest
of this F-theory section, $-K_B$ will
refer to the anti-canonical class of the base, and
$-K$ and $N$ will refer to the anti-canonical class and the normal line
bundle of $D$ unless stated otherwise. Additionally, we will often
refer to the situation where the gauge group is tuned on a $+n$ curve
in a 6D F-theory model (such as $\tilde{S}$ in $\F_n$);
$n$ will be this self-intersection number for the rest of the F-theory
section, and in this case $-K = 2$ and $N = m$.
\subsubsection{$\gsu(6)$}\label{sec:su6fth}

General tunings for $\gsu$ models up to $\gsu(6)$ were found in
\cite{mt-singularities}. Here, we repeat the derivation of the
$\gsu(6)$ model as a warmup for subsequent tunings. As described in
\cite{mt-singularities}, the $f$ and $g$ for an
$\gsu(5)$ Weierstrass
model are
\begin{align}
f =& \frac{-\phi_0^4}{48} - \frac{1}{6}\phi_0^2 \phi_1 \sigma + \Biggparen{\frac{\phi_0 \psi_2}{2}-\frac{\phi_1^2}{3}}\sigma^2 +\mathcal{O}(\sigma^3), \label{eq:fsu5}\\
g =& \frac{\phi_0^6}{864} + \frac{1}{72}\phi_0^4 \phi_1 \sigma + \Biggparen{\frac{\phi_0 ^2\phi_1^2}{18}-\frac{\phi_0^3 \psi_2}{24}} \sigma^2 + \frac{1}{108}\Biggparen{8\phi_1^3 -18 \phi_0 \phi_1 \psi_2 - 9 \phi_0^2 f_3} \sigma^3 \notag\\
&+ \frac{1}{12}\Biggparen{3 \psi_2^2 -4\phi_1 f_3 - \phi_0^2 f_4} \sigma^4 + \mathcal{O}(\sigma^5).\label{eq:gsu5} 
\end{align}
Here $\phi_0, \phi_1, \ldots$ are sections of certain line bundles
over $D$, which can be thought of locally as polynomials in a local
set of variables on $D$.
For example, since $f$ is a section of ${\cal O} (-4K_B)$, $\phi_0$ must
be a section of ${\cal O} (-K_B)$, which descends to ${\cal O} (-K +
N)$ on $D$.
The discriminant of this $\gsu(5)$ tuning vanishes to order $\sigma^5$, as expected:
\begin{equation}
\Delta = \frac{\phi_0^4}{192}\Biggsquare{12 \phi_1 \psi_2^2 -12 f_3 \psi_2 \phi_0 + \phi_0^2\Bigparen{12 g_5 + 4 f_4 \phi_1 + f_5\phi_0^2}} \sigma^5 + \mathcal{O}(\sigma^6). \label{eq:discsu5}
\end{equation}

For the $\gsu(6)$ tuning, the discriminant must vanish to
$\mathcal{O}(\sigma^6)$, while $f$ and $g$ cannot be proportional to
$\sigma$. These requirements demand that the term in square brackets
in Equation \eqref{eq:discsu5} vanishes. $\phi_1 \psi_2^2$ must
therefore be proportional to $\phi_0$, as every other term in the square
brackets is at least first order in $\phi_0$. However, the various
factors in $\phi_0$ can be distributed in any way between $\phi_1$ and
$\psi_2^2$. We can rewrite $\phi_0$, $\phi_1$ and $\psi_2$ in a way that
explicitly resolves this ambiguity.  Defining $\alpha$ to be the GCD
of $\phi_0$ and $\psi_2$, we have
\begin{align}
\phi_0 &= \alpha \beta, \label{eq:6-0}\\
\phi_1 &= \beta \nu, \\
\psi_2 &= -\frac{1}{3} \alpha \phi_2. \label{eq:6-2}
\end{align}

With these redefinitions, the discriminant now reads
\begin{equation}
\Delta = \frac{\alpha^6 \beta^5}{576}\Biggsquare{4\phi_2\Bigparen{3 f_3 + \nu \phi_2}+ 36 \beta g_5 + 3 \beta^2 \Bigparen{4 f_4 \nu + \alpha^2 \beta f_5}} \sigma^5 + \mathcal{O}(\sigma^6). \label{eq:discsu5a}
\end{equation}
Removing the lowest order term in $\beta$ requires that
there exists a $\lambda$ such that
\begin{equation}
f_3 = -\frac{1}{3}\nu \phi_2 - 3 \beta \lambda.
\end{equation}
Note that  $\phi_2$
does not share any factors with
$\beta$, under the assumption that $\alpha$
is the GCD of $\phi_0$ and $\psi_2$. After tuning $f_3$, all the terms under
consideration are sixth-order in $\beta$, and we can solve for the
$g_5$ that makes $\Delta$ vanish to $\mathcal{O}(\sigma^6)$.
\begin{equation}
g_5 = \lambda \phi_2 - \frac{1}{3} \beta \nu f_4 -\frac{1}{12}\alpha^2 \beta^2 f_5. 
\end{equation}
The final $f$ and $g$ for the SU(6) tuning are
\begin{equation}
f = -\frac{\alpha^4\beta^4}{48} - \frac{1}{6}\alpha^2 \beta^3 \nu
\sigma - \frac{\beta}{6}\Biggparen{\alpha^2 \phi_2+2\beta \nu^2}
\sigma^2 - \Biggparen{3 \beta \lambda + \frac{\nu \phi_2}{3}}
\sigma^3+ \mathcal{O}(\sigma^4)
\label{eq:tuning-f-6}
\end{equation}
and
\begin{equation}
\begin{split}
g = &\frac{\alpha^6 \beta^6}{864} + \frac{\alpha^4\beta^5}{72} \nu \sigma + \frac{\alpha^2\beta^3}{72}\Biggparen{4 \beta \nu^2 + \alpha^2 \phi_2} \sigma^2 + \frac{\beta^2}{108}\Biggparen{8\beta \nu^3 +9 \alpha^2 \nu \phi_2 + 27 \alpha^2 \beta \lambda} \sigma^3\\
 &+ \frac{1}{36}\Biggparen{4\beta\nu^2\phi_2  + \alpha^2 \phi_2^2 + 36 \beta^2 \nu \lambda -3\alpha^2\beta^2 f_4} \sigma^4\\
 &+ \frac{1}{12}\Biggparen{12\lambda \phi_2 - 4 \beta \nu f_4 - \alpha^2 \beta^2 f_5} \sigma^5+\mathcal{O}(\sigma^6)\label{eq:gsu6} 
\end{split}
\end{equation}

\begin{table}[tbp]
\centering
\begingroup
\renewcommand*\arraystretch{1.5}
\begin{tabular}[t]{|c|c|c|c|}\hline
Parameter & Divisor Class & Order on $+n$ curve & Associated Matter\\\hline
$\alpha$ & $-K + N - L$ & $n+2-r$ & ${\tiny \yng(1,1)}$ (\textbf{15})\\
$\beta$ & $L$ & $r$ & $\frac{1}{2}{\tiny \yng(1,1,1)}$ ($\frac{1}{2}$\textbf{20})\\ 
$\nu$ & $-2K + N - L$ & $n+4-r$ & \textemdash\\
$\phi_2$ & $-2K + L$ & $4+r$ & \textemdash\\
$\lambda$ & $-4K + N - L$ & $n+8-r$ & \textemdash \\\hline
\end{tabular}
\endgroup
\caption{Free parameters in
the general $\gsu(6)$ Weierstrass model. $r$ must be greater than
or equal to 0 and less than $n+2$}
\label{tab:su6dofs}
\end{table}

The $\gsu(6)$ model has five free  parameters (apart from the untuned
$f_k$ and $g_k$), which are summarized in Table \ref{tab:su6dofs}. 
Each parameter is a section of a line bundle over $D$.
One parameter, which we have chosen to be the divisor class
$L$ associated
with $\beta$, is independent.
Once this divisor class is fixed, all the other divisor classes can be
computed from the form of the expansion and the divisor classes of $f,
g$.
On
the zeroes of $\alpha$, $(f,g,\Delta)$ vanish to orders $(2,3,8)$, and
the singularity type is enhanced to $D_6$. Every zero of $\alpha$
therefore contributes a full multiplet in the ${\tiny\yng(1,1)}$
representation. A zero of $\beta$ meanwhile enhances the singularity
type to $E_6$, giving a half-multiplet in the ${\tiny\yng(1,1,1)}$
representation. Fundamental matter comes from codimension two loci
where $(f,g,\Delta)$ vanish to orders (0,0,7) and the singularity type
enhances to $A_6$. Finally, the number of neutral multiplets can be
found by counting the number of degrees of freedom, as described
previously. The resulting multiplicities agree
with the expected supergravity spectrum of Equation \eqref{eq:model-6}.

\subsubsection{$\gsu(7)$}\label{sec:su7fth}

Some aspects of $\gsu(7)$ tunings were discussed in \cite{mt-singularities}, but a general $\gsu(7)$ tuning was not given there. Instead, tunings were presented for two limiting cases of the matter spectrum; the models lacked either ${\tiny \yng(1,1)}$ (\textbf{21}) matter or ${\tiny \yng(1,1,1)}$ (\textbf{35}) matter. We present a more general $\gsu(7)$ tuning that can exhibit any of the antisymmetric matter spectra consistent with anomaly conditions. 

The discriminant of the $\gsu(6)$ model has the form
\begin{equation}
\Delta = \frac{\alpha^4
  \beta^3}{432}\Biggsquare{\alpha^2\Bigparen{\phi_2^3 +9\beta^2 \phi_2 f_4+27
    \beta^3 g_6 + 9 \beta^4 f_5 \nu + 9\frac{\alpha^2
      \beta^5}{4}f_6}-3\beta\Bigparen{\nu\phi_2-9 \beta \lambda}^2}
\sigma^6 + \mathcal{O}(\sigma^7).\label{eq:discsu6} 
\end{equation}
For the $\gsu(7)$ tuning, $\Delta$ must vanish to
$\mathcal{O}(\sigma^7)$. As noted in \cite{mt-singularities}, this demands that $\alpha^2$ is proportional to $\beta$. We can implement
this requirement by 
rewriting $\alpha$ and $\beta$ as 
\begin{align*}
\beta &= \gamma \delta^2, \\
\alpha &= \gamma \delta \xi.
\end{align*}
This decomposition is uniquely defined if we impose the condition that
$\gamma$ be square-free.
These redefinitions would in turn require $\nu$ to be proportional to $\gamma$, so we will temporarily write $\nu$  as $\gamma \zeta$. The term in square brackets in Equation \eqref{eq:discsu6} would then be equivalent to
\begin{equation}
\gamma^2 \delta^2 \Biggsquare{\xi^2\gamma^2\delta^4\Bigparen{9 f_4\phi_2+ 27 \gamma \delta^2 g_6 + 9\gamma^3 \delta^4 f_5 \zeta + \frac{9}{4}\xi^2\gamma^5\delta^8}-3\gamma\Bigparen{\zeta\phi_2-9\delta^2\lambda}^2 +\xi^2\phi_2^3}.
\end{equation}
$\phi_2$ cannot share any factor with $\gamma$, since 
it shares no factors with $\beta$. $\xi^2$ must therefore
be proportional to $\gamma$, and since $\gamma$ is square-free, $\xi$
must be proportional to $\gamma$. Then, the term lowest order in
$\gamma$ would be $3\gamma(\zeta\phi_2-9\delta^2\lambda)^2$, implying
that $\zeta\phi_2-9\delta^2\lambda$ must also be proportional to
$\gamma$. Then, $3\gamma(\zeta\phi_2-9\delta^2\lambda)^2$ would be
order $\gamma^3$, in turn demanding that $\xi$ must be proportional to
$\gamma^2$. The tuning is stuck in a never-ending cycle where the two
terms must be proportional to greater and greater powers of
$\gamma$. The only way out of this cycle is if $\gamma$ is a perfect
square, which would violate the earlier square-free
assumption. $\gamma$ (and $\zeta$)
should therefore be ignored in the $\alpha$ and
$\beta$ redefinitions, leaving
\begin{align}
\beta &= \delta^2 \label{eq:deltaxi:beta}\\
\alpha &= \delta \xi \label{eq:deltaxi:alpha}
\end{align}
We will continue to refer to $\nu$ and will not use
$\gamma$ or $\zeta$ from this point.

Equations \eqref{eq:deltaxi:beta} and \eqref{eq:deltaxi:alpha} could
have been anticipated from field theory considerations alone. While
each zero of $\beta$ gives $\frac{1}{2}{\tiny \yng(1,1,1)}$ of
$\gsu(6)$ matter, matter in the $\gsu(7)$ ${\tiny \yng(1,1,1)}$
(\textbf{35}) representation must come in full multiplets. When the
$\gsu(7)$ model is Higgsed to $\gsu(6)$, each \textbf{35} multiplet
will give two $\frac{1}{2}$\textbf{20}s of $\gsu(6)$. Any zeroes that produce
\textbf{35}s in the $\gsu(7)$ must therefore provide $\beta$ with two
identical factors when the Weierstrass model is deformed to an
$\gsu(6)$ model. In other words, $\beta$ must be a perfect square, in
agreement with Equation \eqref{eq:deltaxi:beta}. Each \textbf{35} also
gives a full \textbf{15} multiplet. From  \eqref{eq:deltaxi:beta}, we
expect 
that the zeroes of $\delta$ will give the \textbf{35}s of $\gsu(7)$, so $\alpha$ must be proportional $\delta$. The above redefinitions additionally imply that some $\gsu(6)$ models cannot be enhanced to $\gsu(7)$. \eqref{eq:deltaxi:beta} and \eqref{eq:deltaxi:alpha} require the divisors $L^\prime \sim \frac{1}{2}L$ and $-K + N - 3 L^\prime$ to be effective. For the $+n$ curve situation, only $\gsu(6)$ models where $\rsixexp$ is even and where $n+2 \geq \frac{3}{2}\rsixexp$ can be enhanced to $\gsu(7)$. These restrictions follow naturally from the $\gsu(7)\rightarrow\gsu(6)$ branching patterns.

With the redefinitions for $\alpha$ and $\beta$, the discriminant is
now given by 
\begin{equation}
\Delta =
\frac{\delta^{12}\xi^4}{432}\Biggsquare{\xi^2\delta^4\Bigparen{9 f_4
    \phi_2+27 \delta^2 g_6 + 9\delta^4 f_5 \nu +
    \frac{9}{4}\xi^2\delta^8}-3\Bigparen{\nu\phi_2-9\delta^2\lambda}^2
  +\xi^2\phi_2^3}.
\label{eq:discriminant-7}
\end{equation}
Considering the lowest order term in $\delta$ gives the constraint
\begin{equation}
\phi_2^2\Bigparen{\xi^2\phi_2 - 3\nu^2} \propto \delta^2.
\end{equation}
To satisfy this condition, one could argue that when $\xi=0$, 
from (\ref{eq:discriminant-7})
$\nu$
should be proportional to $\delta^2$. This would suggest that we
redefine $\nu$ as 
\begin{equation}
\nu = \zeta_1 \xi + \zeta_2 \delta^2,\label{eq:nuredef}
\end{equation}
where $\zeta_1$ and $\zeta_2$ are independent, untuned
polynomials. Note that we could have also considered the possibility
that $\xi$ and $\nu$ share some common factors with $\delta$. However,
this possibility will end up being a special point in the moduli space
of tunings found using \eqref{eq:nuredef}. More specifically, the
tuning involving common factors can be derived by taking the tuning
found using \eqref{eq:nuredef} and further demanding that $\xi$,
$\nu$, and $\delta$ share common factors. Therefore, Equation
\eqref{eq:nuredef} will be sufficient to find a general tuning, and we
do not need to consider the possibility of shared factors in this
step.

Ultimately, we will want to solve for $g_6$ to make the sixth order term of $\Delta$ vanish. There are still lower order terms to deal with before we can solve for $g_6$, but these terms can be removed using a set of standard polynomial redefinitions:
\begin{align}
\phi_2 &= 3 \zeta_1^2 + \delta^2 \omega,\label{eq:phi2def}\\
\lambda &= \frac{1}{3}\zeta_1^2 \zeta_2 -\frac{1}{18}\zeta_1 \xi \omega + \frac{1}{9}\zeta_2 \delta^2 \omega + \xi \delta^2 \lambda_1,\\
f_4 &= -6 \zeta_1 \lambda_1 -\frac{1}{12}\omega^2 + \psi_4 \delta^2.
\end{align}
$\omega$, $\lambda_1$, and $\psi_4$ are free parameters. We can then solve for $g_6$:
\begin{equation}
g_6 = \frac{-1}{108}\Biggsquare{\omega^3 +108 \zeta_1\Bigparen{\lambda_1 \omega + \zeta_1 \psi_4} + 36 \delta^2\Bigparen{\psi_4 \omega -27\lambda_1^2+f_5 (\zeta_1 \xi + \zeta_2 \delta^2) + \frac{\xi^2 \delta^4}{4}f_6} }. \label{eq:g6def}
\end{equation}
The final tunings are
\begin{equation}
\begin{split}
f =& -\frac{\delta^{12}\xi^4}{48} - \frac{\delta^8 \xi^2}{6}\Bigparen{\zeta_1 \xi + \zeta_2 \delta^2} \sigma - \frac{\delta^4}{6}\Bigparen{2\delta^4 \zeta_2^2 + 4 \delta^2\zeta_1 \zeta_2 \xi + \xi^2(5 \zeta_1^2 + \delta^2\omega)} \sigma^2\\
& - \frac{1}{6}\Bigparen{4\delta^2(3\zeta_1^2 \zeta_2+\delta^2\zeta_2\omega)+\xi(6\zeta_1^3+\delta^2 \zeta_1\omega+18\delta^4\lambda_1)} \sigma^3\\
&-\frac{1}{12}\Bigparen{\omega^2+ 72 \zeta_1 \lambda_1 - 12 \delta^2 \psi_4}\sigma^4 +\mathcal{O}(\sigma^5)\label{eq:fsu7}
\end{split} 
\end{equation}
and 
\begin{equation}
\begin{split}
g =& \frac{\delta^{18} \xi^6}{864} + \frac{\delta^{14}\xi^4}{72} \Bigparen{\zeta_1 \xi + \zeta_2\delta^2} \sigma + \frac{\delta^{10} \xi^2}{72}\Bigparen{4 \delta^4 \zeta_2^2 + 8 \delta^2 \xi \zeta_1 \zeta_2 + 7 \zeta_1^2 \xi^2 + \delta^2 \xi^2 \omega} \sigma^2 \\
&+ \frac{\delta^6}{216}\Bigparen{16 \delta^6 \zeta_2^3 +48 \delta^4 \xi \zeta_1 \zeta_2^2 +120 \delta^2 \zeta_1^2 \zeta_2 \xi^2 + 70 \zeta_1^3 \xi^3 \\  & \hspace{0.35\columnwidth}+54 \delta^4 \xi^3\lambda_1 + 24 \delta^4 \zeta_2\xi^2 \omega+15 \delta^2\xi^3 \zeta_1\omega} \sigma^3\\
&+ \frac{\delta^2}{144}\Bigparen{84 \zeta_1^4 \xi^2+\delta^4(96 \zeta_1^2 \zeta_2^2+5\xi^2\omega^2+8\zeta_1\xi(27 \lambda_1 \xi + 5 \zeta_2 \omega))\\& \hspace{0.18\columnwidth}+16\delta^2\zeta_1^2\xi(9 \zeta_1\zeta_2 + 2 \xi\omega)+4\delta^6(36 \zeta_2 \lambda_1 \xi -3 \xi^2 \psi_4 + 8 \zeta_2^2 \omega)} \sigma^4 \\
 &+ \frac{1}{36}\Bigparen{2\Bigparen{3 \zeta_1^2 + \delta^2\omega}\Bigparen{6\zeta_1^2\zeta_2 - \zeta_1\xi\omega + 2\delta^2(9\lambda_1 \xi + \zeta_2 \omega)}-3\delta^6 \xi^2 f_5\\&\hspace{0.28\columnwidth}-\delta^2\Bigparen{\delta^2\zeta_2+\zeta_1\xi}\Bigparen{-72\zeta_1\lambda_1+12 \delta^2 \psi_4-\omega^2}} \sigma^5\\
 &-\Biggparen{\frac{\omega^3}{108} +\zeta_1\Bigparen{\lambda_1 \omega + \zeta_1 \psi_4} + \frac{\delta^2}{3}\Bigparen{\psi_4 \omega -27\lambda_1^2+f_5 \nu + \frac{\xi^2 \delta^4}{4}f_6} }\sigma^6 + \mathcal{O}(\sigma^7).\label{eq:gsu7} 
\end{split}
\end{equation}

\begin{table}[tbp]
\centering
\begingroup
\renewcommand*\arraystretch{1.5}
\begin{tabular}[t]{|c|c|c|c|}\hline
Parameter & Divisor Class & Order on $+n$ curve & Associated Matter\\\hline
$\delta$ & $L^\prime$ & $\rseven$ & ${\tiny \yng(1,1,1) + \yng(1)} $\\
$\xi$ & $-K + N - 3L^\prime$ & $n+2-3\rseven$ & ${\tiny \yng(1,1)}$\\ 
$\zeta_1$ & $-K  + L^\prime$ & $\rseven+2$ & \textemdash\\
$\zeta_2$ & $-2K +N-4 L^\prime$ & $n+4-4\rseven$ & \textemdash\\
$\omega$ & $-2K$ & $4$ & \textemdash \\
$\lambda_1$ & $-3K - L^\prime$ & $6-\rseven$ & \textemdash \\
$\psi_4$ & $-4K -2L^\prime$ & $8-2\rseven$ & \textemdash \\\hline
\end{tabular}
\endgroup
\caption{Degrees of freedom in $\gsu(7)$ Weierstrass model. $\rseven$ and $n$ must satisfy $n+2\geq 3\rseven$.}
\label{tab:su7dofs}
\end{table}

There are seven free polynomials apart from the untuned $f_k$ and $g_k$, which are given in Table \ref{tab:su7dofs}. On the zeroes of $\xi$, the $A_6$ singularity is enhanced to an $D_7$ singularity, giving matter in the ${\tiny \yng(1,1)}$(\textbf{21}). The zeroes of $\delta$, meanwhile, enhance the singularity to $E_7$; decomposing the $E_7$ adjoint into $A_6$ representations shows that each zero of $\delta$ corresponds to one full ${\tiny \yng(1,1,1)}$ (\textbf{35}) multiplet and one full multiplet in the fundamental (\textbf{7}) representation. When the gauge group is tuned on a $+n$ curve, there are an additional $16+n+4\rseven$ fundamentals coming from codimension two loci where the discriminant vanishes to order 8. The charged matter content agrees with the results from gauge anomaly cancellation conditions, as given in Equation \eqref{eq:model-7}. 

As before, the number of neutral hypermultiplets corresponds to the total number of complex degrees of freedom. A naive counting gives more degrees of freedom than expected from the gravitational anomaly cancellation condition. However, the polynomials in Table \ref{tab:su7dofs} can be redefined in the following way without changing $f$ or $g$:
\begin{equation}
\begin{split}
\zeta_1 &\rightarrow \zeta_1^\prime = \zeta_1 + \delta^2 \epsilon, \\
\zeta_2 &\rightarrow \zeta_2^\prime = \zeta_2 - \xi \epsilon, \\
\omega &\rightarrow \omega^\prime = \omega - 6 \zeta_1 \epsilon - 3\delta^2 \epsilon^2,\\
\lambda_1 &\rightarrow \lambda_1^\prime = \lambda_1 + \frac{1}{6}\omega \epsilon - \frac{1}{2}\zeta_1\epsilon^2 - \frac{1}{6}\delta^2\epsilon^3,\\
\psi_4 &\rightarrow \psi_4^\prime = \psi_4 + 6 \lambda_1 \epsilon + \frac{1}{2}\omega \epsilon^2 - \zeta_1 \epsilon^3 -\frac{1}{4}\delta^2\epsilon^4.   
\end{split}
\end{equation}
$\epsilon$ is a section of $\mathcal{O}(-K - L^\prime)$, and on a $+n$
curve, $\epsilon$ is a polynomial of order $2-\rseven$. Therefore,
$2-\rseven+1$ of the complex degrees
of freedom can be thought of as redundant and should not be considered
when finding the number of neutral hypermultiplets. Subtracting these
redundant degrees of freedom in fact leads to a number of neutral
hypermultiplets consistent with the anomaly cancellation
conditions. It is unclear whether the redundancies can be avoided in
an alternative tuning or whether they are a necessary part of the
tuning with some physical interpretation.

\subsubsection{$\gsu(8)$} \label{fsu8}
The $\gsu(7)$ discriminant has the form
\begin{equation}
\Delta = \delta^8\xi^4\Bigparen{-\frac{1}{8}\zeta_1^7 \xi +
\frac{1}{4} \zeta_2 \zeta_1^6 \delta^2 +
 \mathcal{O}(\delta^4, \xi \delta^2)}\sigma^7 + \mathcal{O}(\sigma^8)
\end{equation}
In order for $\Delta$ vanish to $\mathcal{O}(\sigma^8)$, $\zeta_1^7 \xi$ must be proportional to $\delta^2$. If $\delta$ and $\zeta_1$ share a common factor, the resulting Weierstrass model will have a codimension two $(4,6)$ singularity. $\xi$ must therefore be proportional to $\delta^2$, and we can rewrite $\xi$ as
\begin{equation}
\xi = \delta^2 \tau. \label{eq:xiredef}
\end{equation}
$\Delta$ now reads
\begin{equation}
\Delta = \tau^4\delta^{18}\Biggsquare{\frac{\zeta_1^6}{4}\Bigparen{\zeta_2 - \frac{1}{2}\zeta_1\tau}+\mathcal{O}(\delta^2)}\sigma^7 + \mathcal{O}(\sigma^8),
\end{equation}
implying that
\begin{equation}
\zeta_2 = \frac{1}{2}\zeta_1\tau + \zeta_3 \delta^2. \label{eq:zeta2redef}
\end{equation}

At this point, the leading order term in the discriminant is given by
\begin{equation}
\Delta = \tau^4 \delta^{20}\Biggsquare{\frac{\zeta_1^5}{16}\Bigparen{4 \zeta_1 \zeta_3 - \tau\omega}+\mathcal{O}(\delta^2)}\sigma^7 + \mathcal{O}(\sigma^8). \label{eq:deltazeta1}
\end{equation}
$4 \zeta_1 \zeta_3 - \tau\omega$ must be therefore proportional to $\delta^2$. Following the same strategy as in the $\gsu(7)$ tuning, we could argue that, when $\delta=0$, $\zeta_1$ is proportional to $\tau$. $\zeta_1$ would then decompose as
\begin{equation}
\zeta_1  = \tau \zeta_4 + \delta^2 \zeta_5.
\end{equation}
However, Equation \eqref{eq:zeta2redef} and the discussion of redundant degrees of freedom in the $\gsu(7)$ tuning suggests that $\zeta_5$ in the above decomposition could be absorbed into $\zeta_3$ and other variables. Performing a full tuning with $\zeta_5$ indeed shows that $\zeta_5$ can be absorbed into other polynomials with removing any degrees of freedom. $\zeta_1$ can therefore be redefined as
\begin{equation}
\zeta_1 = \zeta_4 \tau
\end{equation}
without any loss of generality. Substituting this $\zeta_1$ expression into \eqref{eq:deltazeta1} gives a solution for $\omega$:
\begin{equation}
\omega = 4 \zeta_3 \zeta_4 + \delta^2 \omega_1.
\end{equation}

\eqref{eq:deltazeta1} now becomes
\begin{align*}
\Delta = \frac{1}{16}\delta^{22}\tau^4\Big[&-{\zeta_4\tau^5}\Bigparen{6\lambda_1 + \zeta_4 \tau \omega_1} -2{\zeta_4^3\tau^3}\Bigparen{\tau \psi_4 + 12 \zeta_3 \lambda_1+2\zeta_3\zeta_4\tau\omega_1}\delta^2\\
&-{\zeta_4 \tau}\Bigparen{4 \zeta_3\zeta_4\tau\psi_4 -18\lambda_1^2\tau +\zeta_4^2\tau^3\omega_1^2 +4\zeta_3^2\zeta_4(6\lambda_1 + \zeta_4 \tau \omega_1) }\delta^4 \\
&+\Bigparen{f_5 \zeta_4^2 \tau^4 - \zeta_4\tau^2\omega_1(\psi_4 + 2 \zeta_3 \zeta_4 \omega_1)+6\lambda_1(6\zeta_3\lambda_1 + \tau\psi_4)}\delta^6 \\
&+ \tau\Bigparen{g_7 \tau + \frac{1}{4}\omega_1^2(6\lambda_1 - \zeta_4 \tau \omega_1)}\delta^8 + \mathcal{O}(\tau^2\delta^8)  \Big]\sigma^7 + \mathcal{O}(\sigma^8).
\end{align*}
Only standard redefinitions are needed from this point forward to get $\Delta$ in a form to solve for $g_7$:
\begin{align}
\lambda_1 &= -\frac{1}{6}\zeta_4\omega_1\tau + \delta^2\tau\lambda_2,\\
\psi_4 &= -3\zeta_4\lambda_2 \tau^2 -\frac{1}{4}\delta^2\omega_1^2 -6 \zeta_3 \lambda_2 \delta^2+\phi_4\delta^4,\\
f_5 &= 2 \zeta_4\phi_4 + \psi_5 \delta^2.
\end{align}
Finally, we solve for $g_7$:
\begin{equation}
\begin{split}
g_7 = \frac{1}{12}\Big(&16\zeta_3\zeta_4^2\phi_4 -12\zeta_4^2\tau^2\psi_5-16\delta^2\zeta_4(\zeta_3\psi_5-\phi_4\omega_1) \\
&-4\delta^4(18 \lambda_2\phi_4 + \psi_5\omega_1)-2\delta^4f_6(2\zeta_3\delta^2 + 3\zeta_4\tau^2) - f_7\delta^{10} \tau^2\Big).
\end{split}
\label{eq:g7def}
\end{equation}
For the sake of brevity, we do not rewrite the full $f$ and $g$ here. 

\begin{table}[tbp]
\centering
\begingroup
\renewcommand*\arraystretch{1.5}
\begin{tabular}[t]{|c|c|c|c|}\hline
Degree of Freedom & Line Bundle & Order on $+n$ curve & Associated Matter\\\hline
$\delta$ & $L^\prime$ & $\reight$ & ${\tiny \yng(1,1,1)+\yng(1,1)+\yng(1)}$\\
$\tau$ & $-K + N - 5L^\prime$ & $n+2-5\reight$ & ${\tiny \yng(1,1)}$\\ 
$\zeta_3$ & $-2K +N-6 L^\prime$ & $n+4-6\reight$ & \textemdash\\
$\zeta_4$ & $-N+6L^\prime$ & $-n+6\reight$ & \textemdash\\
$\omega_1$ & $-2K-2L^\prime$ & $4-2\reight$ & \textemdash \\
$\lambda_2$ & $-2K -N +2L^\prime$ & $-n+4+2\reight$ & \textemdash \\
$\phi_4$ & $-4K -6L^\prime$ & $8-6\reight$ & \textemdash \\
$\psi_5$ & $-4K -N-2L^\prime$ & $-n+8-2\reight$ & \textemdash \\\hline
\end{tabular}
\endgroup
\caption{Degrees of freedom in $\gsu(8)$ Weierstrass model, excluding untuned $f_k$ and $g_k$. Note that $\reight$ and $n$ must satisfy $n+2 \geq 5\reight$}
\label{tab:su8dofs}
\end{table}

The free polynomials for this $\gsu(8)$ model are shown in Table \ref{tab:su8dofs}. When $\tau=0$, the singularity type enhances from $A_7$ to $D_8$, so each zero of $\tau$ contributes a full multiplet in the ${\tiny \yng(1,1)}$(\textbf{28}) representation. Meanwhile, the discriminant takes the form
\begin{equation}
\Delta = \delta^{12}\tau^4\Bigparen{\frac{3}{16}\zeta_4^8 \tau^8 + \mathcal{O}(\delta^2)}\sigma^8 + \mathcal{O}(\sigma^9).
\end{equation}
The factor in parentheses represents the discriminant locus where the
singularity is enhanced to $A_8$. This discriminant locus is a section
of $\mathcal{O}(-8K+8L)$ and is of order $16+8\reight$. There are therefore
$16+8\reight$ full multiplets in the fundamental (\textbf{8}) representation
coming from the divisor locus. On the zeroes of $\delta$, the
singularity type enhances to $E_8$. As mentioned in
\cite{mt-singularities}, if the $A_7$ singularity structure is
embedded in the standard $E_8$ Dynkin diagram, each zero of $\delta$
contributes one ${\tiny \yng(1,1,1)}$ multiplet, one ${\tiny
  \yng(1,1)}$ multiplet, and one ${\tiny \yng(1)}$
multiplet. 
In principle,
the $A_7$ singularity structure could also be
embedded in the extended $E_8$ Dynkin diagram in a different fashion,
to give a half-multiplet in the ${\tiny
  \yng(1,1,1,1)}$ representation and two full-multiplets in the
${\tiny \yng(1,1)}$ representation. 
In this non-standard embedding, one of the roots of the $A_7$ is
mapped to the ``extra'' root of the affine $\hat{E}_8$.
Either possibility
is consistent with gauge anomaly cancellation conditions.  Here,
in the specific tunings we have identified,
the $A_7$ is enhanced to an $E_8$ singularity, 
which we assume gives the most generic matter content, in the
\textbf{56}, \textbf{28}, and \textbf{8} representations. The
resulting charged matter content is summarized in Table
\ref{tab:su8matter}.

\begin{table}[tbp]
\centering
\begingroup
\renewcommand*\arraystretch{1.5}
\begin{tabular}[t]{|c|c|}\hline
Representation & Multiplicity (in full multiplets)\\\hline
${\tiny \yng(1,1,1)}$ (\textbf{56}) & $\reight$\\
${\tiny \yng(1,1)}$ (\textbf{28}) & $n+2-4\reight$\\ 
${\tiny \yng(1)} (\textbf{8})$ & $16+9\reight$\\\hline
\end{tabular}
\endgroup
\caption{Matter content of the $\gsu(8)$ F-theory model. Note that $n+2 \geq 5\reight$. The $\gsu(8)$ model we have constructed does not seem to give all of the possible spectra listed in Equation \eqref{eq:model-8}.}
\label{tab:su8matter}
\end{table}

To find the number of neutral multiplets, we again count the number of complex degrees of freedom in the Weierstrass model and subtract any redundant degrees of freedom. As with the $\gsu(7)$ model, the $f$ and $g$ expressions for the $\gsu(8)$ tuning are invariant under the following transformations in the polynomials:
\begin{equation}
\begin{split}
\zeta_3 &\rightarrow \zeta_3^\prime = \zeta_3 + \tau^2 \epsilon,\\
\zeta_4 &\rightarrow \zeta_4^\prime = \zeta_4 - \frac{2}{3}\delta^2\epsilon,\\
\omega_1&\rightarrow \omega_1^\prime = \omega_1 + \frac{8}{3}\zeta_3\epsilon + \frac{4}{3}\tau^2\epsilon^2,\\
\lambda_2&\rightarrow \lambda_2^\prime= \lambda_2 - \frac{18}{81} \omega_1\epsilon - \frac{24}{81}\zeta_3\epsilon^2 - \frac{8}{81}\tau^2\epsilon^3,\\
\psi_5 &\rightarrow \psi_5^\prime = \psi_5 + \frac{4}{3}\phi_4\epsilon.
\end{split}
\end{equation}
In these transformations $\epsilon$ is a section of
$\mathcal{O}(-N+4L^\prime)$, so there are $-n+4\reight+1$ redundant
degrees of freedom. After taking the transformations into account, the
number of neutral hypermultiplets agrees with anomaly conditions only
when $\delta$ gives ${\tiny \yng(1,1,1)}$,${\tiny \yng(1,1)}$, and
${\tiny \yng(1)}$ multiplets. This gives concrete evidence that the generic
charged matter content of Table \ref{tab:su8matter} is correct for
this form of $E_8$ singularity.

The tuning presented here does not seem to give matter in the ${\tiny
\yng(1,1,1,1)}$ (\textbf{70}) representation. 
As discussed earlier, the
supergravity models with (\textbf{70}) matter seem to be consistent
with the anomaly conditions, posing the question of whether these
models have valid F-theory realizations. At this point, it is unclear
if there is some F-theory constraint that forbids the $\Lambda^4$
models or if the tuning presented here can be extended to give 4-index
antisymmetric matter. We will return to this issue in later sections.

There are also certain supergravity $\gsu(8)$ models with ${\tiny
  \yng(1,1,1)}$ matter that are not allowed in our F-theory
  construction. Equation \eqref{eq:xiredef} requires that $-K + N - 5
  L^\prime$ be effective; for the $+n$ curve, $n+2 \geq 5\reight$. The
  F-theory $\gsu(8)$ models with $\reight > 0$
{\bf 56} multiplets
therefore have at least one ${\tiny
  \yng(1,1)}$ multiplet, even though models without ${\tiny
  \yng(1,1)}$ multiplets can still be consistent with anomaly
  cancellation conditions. For instance, a model on a $+2$ curve with
  1 \textbf{56} multiplet and 25 \textbf{8} multiplets is anomaly-free
  but cannot be realized in our F-theory constructions.  This
  represents another potentially interesting point of disagreement
  between the low-energy anomaly analysis and what can be realized in
  F-theory.  The restriction also implies that only some of the
  $\gsu(7)$ models can be enhanced to $\gsu(8)$. In particular, an
  $\gsu(7)$ model cannot be enhanced unless it has at least two
  $\mathbf{21}$ multiplets for every $\mathbf{35}$ multiplet. These restrictions on enhancement are also consistent with what can be observed for models from a heterotic perspective (see, for example, Table \ref{atranstab2} in Appendix \ref{su8app}).
\subsubsection{$\gsu(9)$}

The $\gsu(8)$ discriminant takes the form
\begin{equation}
\Delta = \frac{\delta^{12}\tau^4}{192}\left(36 \zeta_4^8 \tau^8 + \mathcal{O}(\delta^2)\right)\sigma^8 + \mathcal{O}(\sigma^9)
\end{equation}
To tune an $\gsu(9)$ singularity, $\Delta$ must vanish to order
$\sigma^9$, which would require $\zeta_4^8 \tau^8$ to be proportional
to $\delta^2$. But from the $\gsu(8)$ tuning, $\delta$ cannot share
factors with either $\tau$ or $\zeta_4$ without introducing
superconformal points where $f$ and $g$ vanish to order $4$ and
$6$. The only way to tune an $\gsu(9)$ gauge symmetry seems to be to have
$\delta$ be a constant. If one proceeds with constant $\delta$, the
resulting $\gsu(9)$ models have only $\tiny{\yng(1,1)}$, fundamental,
and singlet matter.

The $\gsu(9)$ tuning presents a similar challenge as the $\gsu(8)$
models. From the anomaly conditions, there appear to be consistent
$\gsu(9)$ supergravity models with $\tiny{\yng(1,1,1)}$ matter. Yet
our tunings seem to forbid F-theory constructions of these models if
one wishes to avoid superconformal points. Both situations also
require the gauge group Dynkin diagram to be embedded in an extended
Dynkin diagram. For instance, the $\Lambda^3$ representation of
$\gsu(9)$ comes from the enhancement $A_8 \rightarrow \hat{E}_8$. We
will further discuss these missing cases later.
Note that in \cite{Katz-etal-Tate}, it is argued that all SU($N$)
models except those with $N = 6, 7, 8, 9$ can be put in Tate form, but
that in these four cases there are non-Tate realizations.
This suggests that a more sophisticated F-theory construction may
indeed realize SU($9$) models with exotic matter content, despite the
analysis here.

\subsubsection{$\gsu(10)$}

There is no clear way in which a $\Lambda^3$ representation of SU(10)
could be realized in F-theory, since the Dynkin diagram for $A_9$ does
not embed in $\hat{E}_8$. So there is no way to enhance the Kodaira $A_9$
singularity to an exceptional singularity that might carry exotic
matter.  Put differently, {\rm SU}(10) is a rank 9 group and cannot
possibly be a subgroup of $E_8$.  The impossibility of realizing 
theories with SU(10) or higher gauge groups and $\Lambda^3$
representations in F-theory matches nicely with the low-energy theory
where such models are essentially ruled out by anomaly cancellation.
One puzzle here, however, is that the low-energy condition ruling out
these models seems to rely on the pure gravitational anomaly
cancellation condition, while the F-theory obstacle seems to arise
from purely local considerations.  
Note that the absence of SU($10$) models with $\Lambda^3$
representations is also consistent with the results of
\cite{Katz-etal-Tate}.

\subsection{Realization of the transitions} \label{ftrans}

The free polynomials for $\gsu(6)$, $\gsu(7)$, and $\gsu(8)$ models
have ambiguous divisor classes, as parametrized by the divisors $L$
and $L^\prime$. One can imagine a process where $L$ or $L^\prime$ is
allowed to change
while keeping the gauge group and codimension one singularity
structure fixed. The matter content will change as a result; in
SU(6), for instance, parameters such as $\alpha$ and $\beta$ that
control the number of antisymmetric multiplets will have different
 classes if $L$ changes. In fact, these
transformations are the F-theory realization of the transitions
described previously in the supergravity context.

As an example, consider the $\gsu(6)$ model on a $+n$ curve. Different
values of $r$ parametrize the space of models with different
$L$. There is a process by which we can transition between two models
with different $r$. First, let $\alpha$, $\nu$, and $\lambda$ develop
a common factor $a$: 
\begin{align}
\alpha &\rightarrow a \alpha', \nonumber\\
\nu &\rightarrow a \nu',\label{eq:alphafact}\\
\lambda &\rightarrow a \lambda'. \nonumber
\end{align}
This can be done in particular by following a continuous family of models
parameterized by a variable $\hat{\varepsilon} > 0$, with the factorization
occurring as $\hat{\varepsilon} \rightarrow 0$. 
At this point in the transition, $f$ and $g$ vanish to orders 4 and 6 wherever $a=\sigma=0$, indicating that $a$ is a superconformal point. Then, we can regroup $a$ into $\beta$ and $\phi_2$:
\begin{align}
a\beta &\rightarrow \beta',\label{eq:betafact}\\ 
a\phi_2 &\rightarrow \phi_2'. \nonumber
\end{align}
Note that regrouping this factor does not involve any change in the
Weierstrass model, it is simply a new labeling of the factors in the
Weierstrass model that leaves the individual terms in $f, g$ such as
$\alpha^4 \beta^4$ unchanged, since {\it e.g.}  $\alpha \beta = a
\alpha^\prime \beta = \alpha^\prime \beta^\prime$.  When $a$ is regrouped in this way,
the new $\beta^\prime$ and $\phi_2^\prime$ share a common factor, while the theory
is still at the superconformal point. But $\beta^\prime$ and $\phi_2^\prime$ are
free parameters that can now be varied to remove the common factor. A complex structure deformation that ``absorbs" $a$ into $\beta'$ (i.e. deforms this coefficient so that it no longer factors) can be realized for example by following a continuous family of models parameterized
by the variable $\hat{\varepsilon} < 0$, with the superconformal point at
  $\hat{\varepsilon} = 0$. 
  Once the common factor is removed, the model
  no longer has a superconformal point, and the transition is
  complete. If $a$ is a polynomial of degree $1$, then $r$ has increased by 1 during
  this transition. The process can be reversed as well so that $r$
  decreases by one; we simply let $\beta$ and $\phi_2$ develop a
  common factor and absorb the factor into $\alpha$, $\nu$ and
  $\lambda$.  We thus see that the transition process can be
  physically realized by a one-parameter family of Weierstrass models,
  with a superconformal field theory at $\hat{\varepsilon} = 0$ and
  theories with $\gsu(6)$ gauge group and two distinct matter contents for
  $\hat{\varepsilon} > 0$ and $\hat{\varepsilon} < 0$.
  
  The one-parameter nature of the transition can be seen more directly by writing expressions for the parameters in terms of $\hat{\varepsilon}$. To illustrate the first step in the transition, where $\alpha$, $\nu$ and $\lambda$ obtain common factors, we could write
\begin{equation}
\alpha = \left\{\begin{array}{ll}a \alpha^\prime + \alpha^{\prime\prime} \hat{\varepsilon} & \quad \hat{\varepsilon}>0\\ \alpha^\prime & \quad \hat{\varepsilon}\leq 0\end{array} \right.,\label{eq:alphaepsilon}
\end{equation}
with similar expressions for $\nu$ and $\lambda$. Such expressions
show that the factorization step of Equation \eqref{eq:alphafact}
involves moving along a path of models parameterized by
$\hat{\varepsilon}$. Likewise, we could describe the side as
$\hat{\varepsilon}$ approaches zero from a negative value by defining
$\beta^\prime$ as 
  \begin{equation}
  \beta^\prime = \left\{\begin{array}{ll} \beta & \quad \hat{\varepsilon}>0\\ a \beta + \beta^{\prime\prime}\hat{\varepsilon} & \quad \hat{\varepsilon}\leq 0\end{array} \right..\label{eq:betaepsilon}
  \end{equation}
  As expected, Equations \eqref{eq:alphaepsilon} and \eqref{eq:betaepsilon} both lead to a common superconformal point when $\hat{\varepsilon}$ is taken to 0. With these expressions, the change in multiplicities can be seen directly. Consider the term $\beta^2\nu^2\sigma^2$ in Equation \eqref{eq:tuning-f-6}; it vanishes on the ${\tiny \yng(1,1,1)}$ loci but not on the ${\tiny \yng(1,1)}$ loci and, along with $\sigma^3$ terms in $g$, distinguishes between the two matter possibilities. From the $\hat{\varepsilon}$ expressions, the $\beta^2\nu^2\sigma^2$ term could be written as
  \begin{equation}
  \left(a \beta \nu^\prime + \hat{\varepsilon}\beta \nu^{\prime\prime}\Theta(\hat{\varepsilon}) + \hat{\varepsilon}\beta^{\prime\prime}\nu^\prime\Theta(-\hat{\varepsilon}) \right)^2\sigma^2
  \end{equation}
  where
  \begin{equation}
  \Theta(\hat{\varepsilon}) = \left\{\begin{array}{ll}1 & \quad \hat{\varepsilon}>0\\ 0 & \quad \hat{\varepsilon}\leq 0\end{array}\right..
  \end{equation}
  Notice that this term vanishes on the locus $a\beta +
  \hat{\varepsilon}\beta^{\prime\prime}=\sigma=0$ locus for
  $\hat{\varepsilon} < 0$ but only on the smaller locus $\beta =
    \sigma = 0$ when $\hat{\varepsilon} > 0$, indicating that the
  number of ${\tiny \yng(1,1,1)}$ multiplets has changed during the
  transition. Other terms in $f$ and $g$ can be shown to have similar
  behavior, although for brevity we will not write out the full $f$
  and $g$ in terms of $\hat{\varepsilon}$. Nevertheless, the
  transition can be explicitly described as a single-parameter path
  through a family of models. It should be noted that in the resolved
  geometry associated to the two sides of the transition we would see
  that two smooth CY 3-folds with distinct topology could be tuned in
  their complex structure moduli spaces ({\it i.e.}, three-cycles collapsed)
  to share a common singular locus (involving the superconformal
  point). We will return to the notion of these matter transitions as
  topology changing transitions in later sections when they are
  realized in the dual heterotic theories. 
  
We can track the matter participating in the transition through each of these steps. When $\alpha$, $\nu$, and $\lambda$ lose a factor of $a$, one ${\tiny \yng(1,1)}$ multiplet is lost because of the change in the order of $\alpha$. Three of the complex degrees of freedom in $\alpha$, $\nu$ and $\lambda$ are traded for one degree of freedom in $a$, so two neutral multiplets are lost. Finally, the number of fundamentals is determined by the discriminant. Prior to the appearance of the shared factor, $\Delta$ takes the form
\begin{equation}
\Delta = \alpha^4 \beta^3 \Delta_6 \sigma^6 + \mathcal{O}(\sigma^7),
\end{equation}
where the order of $\Delta_6$ corresponds to the number of fundamental multiplets in the model. At the transition point, this expression becomes
\begin{equation}
\Delta = a^6 \alpha^4 \beta^3 \Delta_6^\prime \sigma^6 + \mathcal{O}(\sigma^7).
\end{equation}
Two of the factors of $a$ have come from $\Delta_6$, indicating that moving to the transition point causes two ${\tiny \yng(1)}$ multiplets to disappear.  
The first step in the transition can therefore be thought of as one
${\tiny \yng(1,1)}$ multiplet, two ${\tiny \yng(1)}$ multiplets, and
two singlets combining to form superconformal matter represented by
$a$. Importantly, there are total of $29$ multiplets participating in
the transition, reminiscent of the appearance of an extra tensor
multiplet in the tensionless string transitions.  
Thus, if we imagine going off onto the tensor branch of the theory by
blowing up at the superconformal $(4, 6)$ point, all the matter fields
at that point would be absorbed in the associated transition.

When $a$ is subsequently reabsorbed into $\beta$ and $\phi_2$, a new
$\frac{1}{2}{\tiny \yng(1,1,1)}$ multiplet comes from the now
enlarged $\beta$. The single degree of freedom in $a$ is traded for
two new degrees of freedom in $\beta$ and $\phi_2$, signaling the
appearance of a new singlet. And as the discriminant returns to its
previous form, three factors of $a$ are absorbed into $\beta$, while
the remaining three factors are absorbed into $\Delta_6^\prime$. Three
fundamentals have therefore appeared in the reabsorption step. The
transition can be summarized as
\begin{equation}
{\tiny \yng(1,1)}+2\times{\tiny \yng(1)}+2\times\mathbf{1}\rightarrow \textbf{Superconformal Matter} \rightarrow \frac{1}{2}{\tiny \yng(1,1,1)} + 3\times {\tiny \yng(1)} + \mathbf{1}.\label{eq:su6transitionftheory}
\end{equation}
The net change in matter content is
\begin{equation} \label{ftran1trim}
{\tiny \yng(1,1)} + \mathbf{1} \rightarrow \frac{1}{2}{\tiny \yng(1,1,1)} + { \tiny \yng(1)},
\end{equation}
in exact agreement with the expected transition from supergravity. Note that the total number of multiplets is the same before and after the transition. 

The $\gsu(7)$ transition happens in a similar fashion. First, $\xi$, $\zeta_2$, $\lambda_1$, and $\psi_4$ develop common factors:
\begin{align*}
\xi &\rightarrow a^3 \xi'\\
\zeta_2 &\rightarrow a^4 \zeta_2'\\
\lambda_1 &\rightarrow a \lambda_1'\\
\psi_4 &\rightarrow a^2 \psi_4'.
\end{align*}
$a$ is once again a superconformal point where $(f,g)$ vanish to order $(4,6)$. The common factor is then absorbed into $\delta$ and $\zeta_1$:
\begin{align*}
a \delta &\rightarrow \delta'\\
a\zeta_1 &\rightarrow \zeta_1'.
\end{align*}
We can once again track the matter in the transitions through a procedure similar to that of the $\gsu(6)$ transition, although we will only summarize the results here. To produce $a$ in the first step, 3 ${\tiny \yng(1,1)}$ multiplets, 3 fundamentals, and 8 singlets disappear. Note that the redundant degrees of freedom need to be considered when counting singlets. $a$ subsequently breaks into a ${\tiny \yng(1,1,1)}$ multiplet, 8 fundamentals, and one singlet. The $\gsu(7)$ transition is therefore
\begin{equation}
3\times{\tiny \yng(1,1)}+3\times{\tiny \yng(1)}+8\times\mathbf{1} \rightarrow \textbf{Superconformal Matter} \rightarrow {\tiny \yng(1,1,1)} + 8\times {\tiny \yng(1)} + \mathbf{1}, \label{eq:su7transitionftheory}
\end{equation}
with a net change in matter content of
\begin{equation} \label{su7ftrans}
3\times{\tiny \yng(1,1)}+7\times\mathbf{1} \rightarrow {\tiny \yng(1,1,1)} + 5\times {\tiny \yng(1)}.
\end{equation}
The net matter change agrees exactly with the supergravity expectations. Just as with $\gsu(6)$, the transition can occur in reverse as well. 

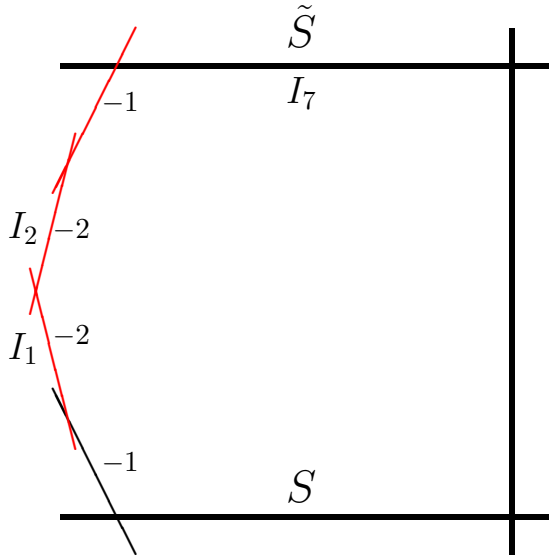
\begin{figure}[tbp]
\centering
\begingroup
\setlength{\unitlength}{1cm}
\begin{picture}(8,7)(-4,-3.5)
\thicklines
\linethickness{0.7mm}
\put(3,-3.5){\line(0,1){7}}
\put(3.5,3){\line(-1,0){6.5}}
\put(0,3.25){\LARGE $\tilde{S}$}
\put(0,2.5){\Large$I_7$}
\put(3.5,-3){\line(-1,0){6.5}}
\put(0,-2.8){\LARGE $S$}
\put(-2.0,3.5){\color{red}\line(-1,-2){1.1}}
\put(-2.45,2.4){$-1$}
\put(-2.8,2.1){\color{red}\line(-1,-4){0.6}}
\put(-3.1,0.7){$-2$}
\put(-3.7,0.7){\Large$I_2$}
\put(-2.0,-3.5){\line(-1,2){1.1}}
\put(-2.45,-2.4){$-1$}
\put(-2.8,-2.1){\color{red}\line(-1,4){0.6}}
\put(-3.1,-0.7){$-2$}
\put(-3.7,-0.9){\Large$I_1$}
\end{picture}
\endgroup

\caption{$\gsu(7)$ transition point when blown up. Here, the
  compactification base was taken to be $\mathbb{F}_n$, while the
  original $\gsu(7)$ gauge group was tuned on $\tilde{S}$. The blow-up
  procedure introduces three exceptional curves shown in red, one of
  which has an $I_2$ singularity. The $I_2$ singularity indicates
  there is a strongly coupled $\gsu(2)$ at the transition point.} 

\label{fig:su7blowup}
\end{figure}

A total of 92 multiplets participate in the transition, which is not a multiple of 29. While the transition does not explicitly require a blowup on the base, the general wisdom of superconformal points and tensionless string transitions would suggest that the multiplets in the transition should somehow fit into new tensor multiplets. To see the source of the mismatch, we can move to the transition point and resolve $a$ using blow-ups on the base. The blow-ups are performed using the procedure of \cite{Morrison-Vafa-II}, but we will not go through the details of the blow-up process here. In the end, a total of three blow-ups are required to resolve $a$, leading to a situation illustrated in Figure \ref{fig:su7blowup}. One of the three exceptional curves carries an $I_2$ singularity, signaling the presence of a new $\gsu(2)$ gauge algebra with 4 fundamentals. From anomaly considerations, the change in matter content due to the blowups should satisfy
\begin{equation}
\delta n_H - \delta n_V = -29 \delta n_T. 
\end{equation}
87 multiplets are traded for the three tensor multiplets, while a net of 5 multiplets are needed to create the 4 fundamentals and 3 vector multiplets of the $\gsu(2)$ gauge algebra. This adds up to a total of 92 multiplets, in exact agreement with \eqref{eq:su7transitionftheory}. In the limit where the new exceptional curves shrink to zero size, any gauge groups on the exceptional divisors become strongly coupled. Hence, the transition point $a$ for $\gsu(7)$ should involve a superconformal field theory with three tensor multiplets and a strongly coupled $\gsu(2)$ gauge symmetry.

Finally, we turn to the $\gsu(8)$ transition. To convert ${\tiny \yng(1,1)}$ matter to ${\tiny \yng(1,1,1)}$ matter, we first let the following parameters obtain common factors:
\begin{align*}
\tau &\rightarrow a^5 \tau,\\
\zeta_3 &\rightarrow a^6 \zeta_3,\\
\omega_1 &\rightarrow a^2 \omega_1,\\
\phi_4 &\rightarrow a^6 \phi_4.\\
\psi_5 &\rightarrow a^2 \psi_5.
\end{align*}
5 ${\tiny \yng(1,1)}$ multiplets, 10 fundamentals, and 24 singlets disappear to form $a$. $a$ is once again a superconformal point, but $(f,g)$ vanish to orders $(6,8)$. $a$ is then reabsorbed into $\delta$, $\zeta_4$, and $\lambda_2$:
\begin{align*}
a \delta &\rightarrow \delta\\
a^6 \zeta_4 &\rightarrow \zeta_4\\
a^2 \lambda_2 &\rightarrow \lambda_2.
\end{align*}
Once $\delta$, $\zeta_4$, and $\lambda_2$ are allowed to vary independently, $a$ breaks into a ${\tiny \yng(1,1,1)}$ multiplet, a ${\tiny \yng(1,1)}$ multiplet, 19 fundamentals, and 8 singlets. The complete transition is therefore
\begin{equation}
5\times{\tiny \yng(1,1)}+10\times{\tiny \yng(1)}+24\times\mathbf{1}\rightarrow \textbf{Superconformal Matter} \rightarrow {\tiny \yng(1,1,1)} + {\tiny \yng(1,1)} + 19\times {\tiny \yng(1)} + 8\times\mathbf{1}, \label{eq:su8transitionftheory}
\end{equation}
with a corresponding net matter change of
\begin{equation}
4\times {\tiny \yng(1,1)} + 16\times\mathbf{1} \rightarrow {\tiny \yng(1,1,1)} + 9\times {\tiny \yng(1)}.
\end{equation}
At the transition point, however, the codimension two singularity at
$a=\sigma=0$ does not seem to be resolvable even with blowups on the
base. If one tries to perform the resolution, there will be a
codimension one singularity along one of the exceptional curves where
$f$ and $g$ vanish to orders $4$ and $6$. This result suggests that
the $\gsu(8)$ transition may not be valid. However, it is unclear
whether our $\gsu(8)$ tuning is completely general, and a different
tuning may admit a resolvable transition point. We leave the question
of whether the $\gsu(8)$ transition is valid for future work.

But the $\gsu(6)$ and $\gsu(7)$ transitions do seem to be valid. The
analysis here does not give a full description of the mechanism behind
these transitions. Superconformal points seem to be key to the actual
transition.  Importantly, it does not seem necessary to have a
strongly coupled gauge group associated with the superconformal
point. While the $\gsu(7)$ transition does include a strongly coupled
$\gsu(2)$, the $\gsu(6)$ transition seems to not require any
additional gauge group. A better understanding of the superconformal
points could perhaps lead to a more complete picture of the transition
mechanism.

\subsection{Higgsing processes in F-theory}

Here, we examine how the $\gsu(6)$-$\gsu(8)$ F-theory models fit into
the Higgsing structure discussed earlier in the supergravity
context. In F-theory, Higgs transitions occur when the coefficients in
the Weierstrass model are deformed from particular values, breaking a
gauge symmetry in the process. 
Such deformations were recently
explored in
\cite{Grassi-Halverson-Shaneson,Grassi-Halverson-Shaneson-2014}. We
 discuss the F-theory deformations that represent Higgsing
processes where fundamental, ${\tiny \yng(1,1)}$, and ${\tiny
\yng(1,1,1)}$ multiplets obtain VEVs. We also examine how
Higgsing affects the $\gsu(6)-\gsu(8)$ matter
transitions. In particular,
some the Higgsed models will have transitions where the product-group
representations change, as discussed previously in the supergravity
section.

There will often be several parameters in the Weierstrass model that
can be deformed to give the same Higgsing process. We can therefore
associate each Higgsing process with a set of deformations, with each
deformation corresponding to a particular degree of freedom in the
Weierstrass model.  Suppose that the number of deformations for a
particular Higgsing process is $n_D$. 
From anomaly cancellation, we know that $n_H-n_V$ stays unchanged in
the Higgsing process.  This corresponds to the fact that $\Delta n_V$
of Goldstone bosons are eaten by the gauge field, so from a $\Delta
n_V+ 1$ dimensional space of deformations related by gauge symmetry,
only one remains as a deformation.  The remaining $n_D -1$ deformation
directions arise when originally charged fields become singlets after
the reduction in gauge group.  In the cases considered here, this
seems to occur in a similar way with a simple characterization.
From the field theory
perspective, Higgsing occurs when a certain number of
multiplets in a
particular representation $R$ obtain expectation values. Let us say
that the unHiggsed model has $n_R$ multiplets in this representation
and that $n_{vev}$ of them need to obtain expectation values. For the
Higgsing processes we consider in this paper, the number of
deformations $n_D$ seems to be related to $n_R$:
\begin{equation}
n_D = n_R - n_{vev} +1. \label{eq:numdeformations}
\end{equation}
Most of the Higgsing processes we examine will involve two multiplets
obtaining VEVs, so that $n_{vev}=2$. In these cases, we will find that
there is one fewer deformation than the number of multiplets that can
obtain VEVs.  For most of the Higgsing processes in this paper, this
holds since each singlet in the Higgsed model arises from a single
non-Higgsed field in the representation $R$. But it is unclear if
\eqref{eq:numdeformations} should hold more generally, and we do not
present a more general proof of this conjecture here.

\subsubsection{Higgsing on fundamental matter}
As described earlier, an $\gsu(N)$ symmetry can be broken to $\gsu(N-1)$ by giving VEVs to two fundamental multiplets. Because the tuning process of section \ref{subsec:ftheorytunings} proceeds through each $\gsu(N)$ algebra sequentially, the Higgsing deformations can be read off directly from the individual tuning steps. 

The $\gsu(6)$ model can be Higgsed to $\gsu(5)$ through a set of
possible deformations contained in
$\deform{\phi}_1$,$\deform{\psi}_2$,$\deform{f}_3$, and
$\deform{g}_5$: 
\begin{align}
\phi_1 &= (\beta + \deform{\phi}_1) \nu \label{eq:su6def1}\\
\psi_2 &= -\frac{1}{3}(\alpha + \deform{\psi}_2) \phi_2\\
f_3 &= -\frac{1}{3}\nu\phi_2 - 3(\beta+\deform{f}_3)\lambda\\
g_5 &= -\frac{1}{12}\alpha^2\beta^2 - \frac{1}{3}f_4\beta\nu + \lambda\phi_2 + \deform{g}_5.\label{eq:su6def4}
\end{align}
Each of the parameters in the four polynomials is an independent
deformation that can be adjusted separately; as long as at least one
parameter is non-zero, the $\gsu(6)$ gauge symmetry will be Higgsed to
$\gsu(5)$. $\deform{\phi}_1$ and $\deform{f}_3$ are sections of
$\mathcal{O}(L)$, $\deform{\psi}_2$ is a section of
$\mathcal{O}(-K+N-L)$, and $\deform{g}_5$ is a section of
$\mathcal{O}(-6K + N)$. When the $\gsu(6)$ is tuned on a $+n$ curve,
there are a combined $2n+15+r$ complex degrees of freedom, one fewer
than the number of fundamental multiplets in the $\gsu(6)$ model. This
fits with the expectation that \eqref{eq:numdeformations} is satisfied
for the $\gsu(6)\rightarrow\gsu(5)$ Higgsing process.  Note that in
4D, this set of deformations is not in general complete since, for example, quantities such as $\phi_1$ and $\psi_2$ need not factorize.

From the branching patterns, all of the possible $\gsu(6)$ models on a $+n$ curve should give the same $\gsu(5)$ charged matter content when Higgsed. To see whether this is true in F-theory, we can plug \eqref{eq:su6def1}-\eqref{eq:su6def4} into the $\gsu(5)$ model of \eqref{eq:fsu5} and \eqref{eq:gsu5} (while setting $\phi_0 = \alpha\beta$) and examine the resulting matter. We can consider a maximally-deformed model in which all of the deformations are turned on. In this case, $f$ and $g$  vanish to orders 2 and 3 when either $\alpha=\sigma=0$ or when $\beta=\sigma=0$, so both $\alpha$ and $\beta$ will give $\mathbf{10}$ matter in the resulting $\gsu(5)$ model. The discriminant meanwhile takes the form
\begin{equation}
\Delta = \alpha^4\beta^4\Delta_5\sigma^5 + \mathcal{O}(\sigma^6).
\end{equation}
The $\gsu(5)$ model thus has a total of $n+2$ $\mathbf{10}$ multiplets and $3n+16$ fundamental multiplets regardless of the initial $\gsu(6)$ model, indicating all of the $\gsu(6)$ models have Higgsed to the same $\gsu(5)$ model. If some of the deformations are turned off, the resulting $\gsu(5)$ model may have codimension-two loci with rank-two singularity enhancements. These loci may contribute more than one multiplet of charged matter, making the analysis more involved. While we do not go through all of the possible situations, all of these specialized situations should give the same charged matter content. 

The $\gsu(7)\rightarrow \gsu(6)$ Higgsing pattern has a more interesting structure, as a given $\gsu(7)$ model can Higgs only to a particular $\gsu(6)$ model. Moreover, only a subset of the $\gsu(6)$ models can be reached by Higgsing an $\gsu(7)$ model, as discussed previously. In F-theory, the corresponding deformations are
\begin{align}
\beta &= (\delta+\deform{\beta})\delta,\\
\alpha &= (\delta+\deform{\alpha})\xi,\\
\phi_2 &= 3 \zeta_1^2+(\delta^2+\deform{\phi}_2)\omega,\\
\lambda &= \frac{1}{3}\zeta_1^2\zeta_2-\frac{1}{18}\zeta_1\xi+\frac{1}{9}\delta^2\zeta_2\omega+(\xi\delta^2+\deform{\lambda})\lambda_1,\\
f_4 &= -6 \zeta_1\lambda_1 - \frac{1}{12}\omega^2+(\delta^2+\deform{f}_4)\psi_4,
\end{align}
along with a deformation of $g_6$ by adding $\deform{g}_6$ to  \eqref{eq:g6def}. The resulting set of deformations has $n+16+5\rseven$ complex degrees of freedom, one fewer than the number of $\mathbf{7}$ multiplets for $\gsu(7)$. Note that there is no deformation associated with the $\nu$ redefinition \eqref{eq:nuredef}. The $\nu$ redefinition does not seem to remove degrees of freedom like the other tuning steps; instead, the step simply reorganizes $\nu$ to have a particular structure.  A $\nu$ deformation could be included, but we would then need to account for the redundant degrees of freedom described earlier in the counting. Once the redundant degrees of freedom are subtracted off, the total number of deformations is the same as before. We therefore do not include a $\nu$ deformation in the above set of deformations.

 $\xi$ will give $\mathbf{15}$ matter in the $\gsu(6)$ model, as $(f,g)$ vanish to orders $(2,3)$ on $\xi=\sigma=0$ regardless of which deformations are turned on.  When considering the maximally deformed situation, the loci where $\delta+\deform{\alpha}=\sigma=0$ also contribute $\mathbf{15}$ matter. Finally, the singularity type enhances to an incompletely resolved $E_6$ when $\delta=\sigma=0$ or when  $\delta+\deform{\beta}=\sigma=0$, giving $\frac{1}{2}\mathbf{20}$ matter. In total, the resulting $\gsu(6)$ model has $2\rseven$ $\frac{1}{2}\mathbf{20}$ multiplets, $n+2-2\rseven$ $\mathbf{15}$ multiplets, and $16+2n+2\rseven$ fundamentals, so each $\gsu(7)$ model is Higgsed to the $\gsu(6)$ model with $\rsixexp = 2 \rsevenexp$. 

Higgsing relates the $\gsu(6)$ and $\gsu(7)$ transitions in a non-trivial way. Directly applying the $\gsu(7)\rightarrow \gsu(6)$ branching patterns to the $\gsu(7)$ transition \eqref{eq:equivalence-7} would seem to imply that the $\gsu(6)$ theory undergoes minimal transitions of the form
\begin{equation}
\mathbf{20}\left({\tiny \yng(1,1,1)}\right) + 2 \times \mathbf{6} \left({\tiny \yng(1)}\right) \leftrightarrow 2\times \mathbf{15}\left({\tiny \yng(1,1)}\right) + 2\times \mathbf{1}.\label{eq:su6transfromsu7}
\end{equation}
However, Equation \eqref{eq:equivalence-6} and the F-theory $\gsu(6)$ model both suggest the minimal $\gsu(6)$ transition should involve half this amount of matter. This discrepancy reflects the fact mentioned in Sections \ref{higsfund} and \ref{sec:su7fth} that only $\gsu(6)$ models with an even number of $\frac{1}{2}\mathbf{20}$ multiplets can be unHiggsed to $\gsu(7)$. Applying the transition \eqref{eq:equivalence-6} only once could produce an $\gsu(6)$ spectrum with an odd number $\frac{1}{2}\mathbf{20}$ multiplets that cannot be unHiggsed to $\gsu(7)$. Thus, \eqref{eq:equivalence-6} should not be directly visible from the $\gsu(7)$ model. Instead, \eqref{eq:equivalence-6} must be applied twice to move between two $\gsu(6)$ vacua that can be enhanced to $\gsu(7)$, as reflected in  \ref{eq:su6transfromsu7}. Said differently, the smallest $\gsu(7)$ transition changes $\rsevenexp$ by 1, which becomes a $\Delta \rsixexp = 2$ change in the resulting $\gsu(6)$ model. The heterotic analogue of this phenomenon will be discussed in Section \ref{sec:smallinstHiggs}.

For the $\gsu(8)\rightarrow\gsu(7)$ Higgsing process, the corresponding deformations are
\begin{subequations}
\begin{align}
\xi &= (\delta^2+\deform{\xi})\tau,\\
\zeta_2 &= \frac{1}{2}\zeta_4\tau^2+(\delta^2+\deform{\zeta}_2)\zeta_3,\\
\omega &= 4\zeta_3\zeta_4+(\delta^2+\deform{\omega})\omega_1,\\
\lambda_1 &= -\frac{1}{6}\zeta_4\tau\omega_1+(\tau\delta^2+\deform{\lambda}_1)\lambda_2,\\
\psi_4 &= -3 \zeta_4\lambda_2 - \frac{1}{4}\omega_1^2\delta^2 - 6\zeta_3\lambda_2\delta^2+(\delta^4+\deform{\psi}_4)\phi_4,\\
f_5 &=2\zeta_4 \phi_4 +(\delta^2+\deform{f}_5)\psi_5,
\end{align}
\end{subequations}
along with a deformation of $g_7$ by adding $\deform{g}_7$ to Equation \eqref{eq:g7def}. This leads to a total of $15+9\reight$ possible deformations, which is one fewer than the number of $\mathbf{8}$ multiplets. Modifications to the $\zeta_1$ redefinition were not considered due to the redundant degrees of freedom in the tuning. In the Higgsed model, there are two codimension two loci, $\tau=\sigma=0$ and $\delta^2+\deform{\xi}=\sigma=0$, where the singularity type enhances to $D_7$; these loci give $\mathbf{21}$ matter. Loci where $\delta=\sigma=0$ have an enhanced $E_7$ singularity and contribute one $\mathbf{35}$ multiplet and one $\mathbf{7}$ multiplet. The Higgsed model therefore has $\reight$ $\mathbf{35}$ multiplets, $n+2-3\reight$ $\mathbf{21}$ multiplets, and $n+2+5\reight$ fundamental multiplets, indicating that each $\gsu(8)$ model is Higgsed to the $\gsu(7)$ model with $\rsevenexp=\reightexp$. Once again, the restriction that $5\reight\leq n+2$ for $\gsu(8)$ means that only some $\gsu(7)$ models can be reached by Higgsing an $\gsu(8)$ model.

\subsubsection{Higgsing on two-index antisymmetric matter}
From field theory, an $\gsu(N)$ gauge symmetry can be broken to
$\gsu(N-2)\times\gsu(2)$ by giving expectation values to two ${\tiny
  \yng(1,1)}$ multiplets. For $N=4$ and $N=5$, such Higgsing processes
can be realized in F-theory by tuning an $\gsu(N-2)$ gauge symmetry on
$\sigma=0$ and tuning an $\gsu(2)$ gauge symmetry on
$\sigma-\epsilon=0$. Here, $\epsilon$, like $\sigma$, is a section of
the line bundle
$\mathcal{O}(D)$ on $B$. When considering a compactification base
$\mathbb{F}_n$ with the coordinate $\sigma$ associated with
$\tilde{S}$, $\epsilon$ will be a polynomial in the coordinate
associated with $F$ of order $n$; the rest of the discussion in this
F-theory Higgsing section will focus mostly on this particular setup.
Note that in this discussion we treat all the coefficients in the
expansion of $f, g$ as sections of line bundles over $B$ and do not
pull them back to a given divisor.  This may be subtle in general
circumstances of is clear in the toric context at least where all
these coefficients can be expanded in a local coordinate system.  In
the situation just described,
the discriminant will then be proportional to
$\sigma^{N-2}(\sigma-\epsilon)^2$. In the limit where $\epsilon$ goes
to zero, the discriminant becomes proportional to $\sigma^N$,
unHiggsing the gauge symmetry to $\gsu(N)$. $\epsilon$ therefore
parametrizes the set of possible deformations corresponding to this
Higgsing process. Physically, the Higgsing process occurs by
separating the stack of coincident branes forming the $\gsu(N)$
singularity into two distinct sets, with $\epsilon$ representing the
separation between the two new sets.

Identifying the deformations requires tuning singularities on $\sigma=0$ and $\sigma=\epsilon$ simultaneously, which is difficult when $f$ and $g$ are expanded in $\sigma$ alone. The tuning process is easier when $f$ and $g$ are expanded in $\sigma (\sigma-\epsilon)$. Specifically, we write
\begin{align}
f & = F_0 + F_1 \sigma(\sigma-\epsilon) +  F_2 \sigma^2(\sigma-\epsilon)^2 + \ldots \\
g &= G_0 + G_1 \sigma(\sigma-\epsilon) +  G_1 \sigma^2(\sigma-\epsilon)^2 + \ldots , 
\end{align}
where
\begin{align}
F_i &= f_{2i} + f_{2i+1}\sigma\\
G_i &= g_{2i} + g_{2i+1}\sigma.
\end{align}
With this expansion, the $\gsu(N-2)$ and $\gsu(2)$ symmetries can be tuned through a process similar to that of tuning $\gsu(N)$ on $\sigma$. Some terms of the $\gsu(N)$ tuning process will obtain modifications proportional to $\epsilon$ when tuning $\gsu(N-2)\times\gsu(2)$. These additional terms represent the Higgsing deformations.

$\epsilon$ is the only free parameter in the $\gsu(N-2)\times\gsu(2)$ not present in the unHiggsed $\gsu(N)$ model. As an order $n$ polynomial, $\epsilon$ has $n+1$ degrees of freedom, whereas the unHiggsed $\gsu(4)$ and $\gsu(5)$ models both have $n+2$ ${\tiny \yng(1,1)}$ multiplets. Once again, the number of deformation parameters is one fewer than the number of matter multiplets that can obtain VEVs for this Higgsing pattern.  Since $\epsilon$ will be ineffective unless $n\geq 0$, the unHiggsed $\gsu(4)$ or $\gsu(5)$ model must have at least two ${\tiny \yng(1,1)}$ multiplets. Moreover, the $n$ zeroes of $\epsilon$ are the only source of bifundamental matter in the $\gsu(N-2)\times\gsu(2)$ model, reflecting the fact that two bifundamental multiplets are eaten during Higgsing. 

A similar story holds when Higgsing $\gsu(6)$ to $\gsu(4)\times\gsu(2)$. As noted earlier, the $\gsu(4)\times\gsu(2)$ model itself has a transition where the product-group representations are allowed to change. Such transitions can be achieved in F-theory by having $\epsilon$ depend on other parameters in the Weierstrass model. We tune $\gsu(4)$ on $\sigma=0$ and tune $\gsu(2)$ on $\sigma - \beta \epsilon_1=0$. The overall tuning process is similar to that of $\gsu(6)$, except that some of the tuned parameters may obtain additional terms proportional to $\epsilon_1$. When $\epsilon_1$ is taken to zero, we recover the general $\gsu(6)$ model; $\epsilon_1$ therefore represents the set of deformations corresponding to the Higgsing process.  Note that $\epsilon_1$ is a polynomial of order $n-r$ in our standard $\mathbb{F}_n$. There are $n+2-r$ \textbf{15} multiplets in the $\gsu(6)$ model, so we once again have a number of deformations that is one fewer than the number of Higgsable multiplets. As before, the original $\gsu(6)$ model must have at least $2$ \textbf{15} multiplets for $\epsilon_1$ to be effective and for the $\gsu(6)$ to be Higgsable. 

The zeroes of $\epsilon_1$ contribute bifundamental $(\mathbf{4},\mathbf{2})$ matter, while the zeroes of $\alpha$ give $(\mathbf{6},\mathbf{1})$ matter. When $\beta=0$, the singularity type enhances to $D_5$, indicating every zero of $\beta$ contributes a half multiplet of $(\mathbf{6},\mathbf{2})$ matter. The total matter content from the F-theory model agrees exactly with the spectra derived from the $\gsu(6)$ branching patterns. Once again, the fact that $\epsilon_1$ is of order $n-r$ indicates that two bifundamental multiplets are eaten in the Higgsing process. The transition between product-group representations occurs in a similar fashion as the $\gsu(6)$ transition. First, $\alpha$, $\epsilon_1$, $\nu$ and $\lambda$ obtain common factors:
\begin{align*}
\alpha &\rightarrow a \alpha', \\
\epsilon_1 &\rightarrow a \epsilon_1', \\
\nu &\rightarrow a \nu',\\
\lambda &\rightarrow a \lambda'.
\end{align*}
The common factor $a$ is once again a superconformal point; $(f,g)$ vanish to order $(4,6)$ on the locus $a=\sigma=0$. $a$ is then absorbed into $\beta$ and $\phi_2$ as before:
\begin{align*}
a \beta &\rightarrow \beta',\\
a \phi_2 &\rightarrow \phi_2' .
\end{align*}
The transition can be summarized as
\begin{equation}
(\mathbf{6},\mathbf{1}) + (\mathbf{4},\mathbf{2}) + 2\times (\mathbf{4},\mathbf{1})+ 2\times(\mathbf{1},\mathbf{2}) + 3\times(\mathbf{1},\mathbf{1}) \rightarrow \frac{1}{2}\times(\mathbf{6},\mathbf{2})+4\times(\mathbf{4},\mathbf{1})+3\times(\mathbf{1},\mathbf{2}) + (\mathbf{1},\mathbf{1}),
\end{equation}
which is the Higgsed version of \eqref{eq:su6transitionftheory}. Just as in the $\gsu(6)$ transition, there are a total of 29 multiplets participating in the transition. The net change in matter content is
\begin{equation}
(\mathbf{6},\mathbf{1}) + (\mathbf{4},\mathbf{2}) + 2\times(\mathbf{1},\mathbf{1}) \rightarrow \frac{1}{2}\times(\mathbf{6},\mathbf{2})+2\times(\mathbf{4},\mathbf{1})+(\mathbf{1},\mathbf{2}),
\end{equation}
as expected from supergravity. The transition can be reversed as well by inverting the steps. 

To realize the $\gsu(7)\rightarrow\gsu(5)\times\gsu(2)$ Higgsing process, we tune the $\gsu(5)$ symmetry on $\sigma=0$ and the $\gsu(2)$ symmetry on $\sigma - \delta^3\epsilon_2=0$. While $\beta$ is still defined to be $\delta^2$, $\alpha$ is redefined as
\begin{equation}
\alpha = \xi \delta - \zeta_1 \epsilon_2.
\end{equation}
The rest of the tuning process is similar to that of $\gsu(7)$, but the parameter redefinitions may have additional terms proportional to $\epsilon_2$. $\epsilon_2$ is of order $n-3 \rseven$; again, there is one fewer deformation parameter than the number of $\mathbf{21}$'s of $\gsu(7)$. Taking the $\epsilon_2 \rightarrow 0$ limit gives the $\gsu(7)$ tuning, and the zeroes $\epsilon_2$ therefore contribute bifundamental $(\mathbf{5},\mathbf{2})$ matter. $(\mathbf{10},\mathbf{1})$ matter comes from the codimension two locus where $\xi\delta-\zeta_1\epsilon_2=\sigma=0$ with $\delta\neq0$, so there are a total of $n+2-3\rseven$ $(\mathbf{10},\mathbf{1})$ multiplets. Finally, the singularity type enhances to $E_6$ on the $\delta=\sigma=0$ loci; each zero of $\delta$ therefore contributes a $(\mathbf{10},\mathbf{2})$ multiplet and a $(\mathbf{5},\mathbf{1})$ multiplet. The resulting charged matter content agrees exactly with that from field theory considerations.

The $\gsu(5)\times\gsu(2)$ model inherits the $\gsu(7)$ transition. The steps in the transition are the same as those for the $\gsu(7)$ transition, only $\epsilon_2$ develops the common factor $a$ along with $\xi$, $\zeta_2$, $\lambda_1$, and $\psi_4$:
\begin{equation}
\epsilon_2 \rightarrow a^3 \epsilon_2'.
\end{equation}
The complete transition is therefore
\begin{multline}
3\times (\mathbf{10},\mathbf{1}) + 3\times(\mathbf{5},\mathbf{2}) + 3\times (\mathbf{5},\mathbf{1}) + 3\times(\mathbf{2},\mathbf{1}) + 11\times(\mathbf{1},\mathbf{1}) \\
\rightarrow (\mathbf{10},\mathbf{2}) + (\mathbf{10},\mathbf{1})+ 9\times(\mathbf{5},\mathbf{1}) + 8\times(\mathbf{1},\mathbf{2}) + (\mathbf{1},\mathbf{1}),
\end{multline}
which is the Higgsed version of the $\gsu(7)$ transition. The corresponding net change in matter content is
\begin{equation}
2\times (\mathbf{10},\mathbf{1}) + 3\times(\mathbf{5},\mathbf{2}) + 10\times(\mathbf{1},\mathbf{1}) \rightarrow (\mathbf{10},\mathbf{2}) + 6\times(\mathbf{5},\mathbf{1}) + 5\times(\mathbf{1},\mathbf{2}).
\end{equation}

For $\gsu(8)\rightarrow\gsu(6) \times\gsu(2)$, we enhance the symmetry tuned on $\sigma=0$ to $\gsu(6)$ while having the $\gsu(2)$ occur on $\sigma-\delta^4\epsilon_3=0$. $\xi$ is redefined to be
\begin{equation}
\xi = \delta^2-\zeta_1 \epsilon_3,
\end{equation}
while some of the other $\gsu(8)$ parameter redefinitions get additional terms proportional to $\epsilon_3$. $\epsilon_3$ contains the set of deformations corresponding to the Higgsing process, and taking $\epsilon_3$ to zero recovers the $\gsu(8)$ model.

$\epsilon_3$ contributes bifundamental $(\mathbf{6},\mathbf{2})$ matter, and $\tau$ gives $(\mathbf{15},\mathbf{1})$ matter. When $\delta=\sigma=0$, the singularity type enhances to $E_7$, so each zero of $\delta$ gives a $(\mathbf{15},\mathbf{2})$ multiplet, a $(\mathbf{15},\mathbf{1})$ multiplet, and a $(\mathbf{1},\mathbf{2})$ multiplet. Finally, the singularity type enhances to $E_6$ on the locus $\delta^2-2\epsilon_3 \zeta_4$, contributing $\reight$ $(\mathbf{20},\mathbf{1})$ multiplets. There does not seem to be a locus that gives $(\mathbf{20},\mathbf{2})$ matter, which can only come from ${\tiny \yng(1,1,1,1)}$ matter in the $\gsu(8)$ model. This is further evidence that our $\gsu(8)$ tuning does not have 4-index antisymmetric matter. 

The $\gsu(6)\times\gsu(2)$ model has a transition similar to the $\gsu(8)$ transition. The only difference is that $\epsilon_3$ participates in the first step along with $\tau$, shedding four factors of $a$:
\begin{equation}
\epsilon_3 \rightarrow a^4 \epsilon_3.
\end{equation}
In terms of the matter content, the transition is
\begin{multline}
5\times(\mathbf{15},\mathbf{1})+4\times(\mathbf{6},\mathbf{2})+10\times(\mathbf{6},\mathbf{1})+10\times(\mathbf{1},\mathbf{2}) + 28\times(\mathbf{1},\mathbf{1}) \\
\rightarrow (\mathbf{20},\mathbf{1})+(\mathbf{15},\mathbf{2}) + (\mathbf{15},\mathbf{1}) + 20\times(\mathbf{6},\mathbf{1}) + 19 \times(\mathbf{1},\mathbf{2}) + 8\times(\mathbf{1},\mathbf{1}).
\end{multline}
This is not the Higgsed version of the $\gsu(8)$ transition, as there is a missing $(\mathbf{6},\mathbf{2})$ multiplet in the transition. However, just as with the $\gsu(8)$ transition, the transition point does not seem to be resolvable even with blow-ups on the base. 

\subsubsection{Higgsing on three-index antisymmetric matter}
\label{subsubsec:ftheory3asHiggs}

We can find deformations for the $\gsu(N)\rightarrow\gsu(N-3)\times\gsu(3)$ Higgsing process using a similar strategy; we tune an $\gsu(N-3)$ symmetry on $\sigma=0$ and an $\gsu(3)$ symmetry on $\sigma-\epsilon=0$. As in the ${\tiny \yng(1,1)}$ Higgsing process, $\epsilon$ will take particular forms when breaking $\gsu(6)$ through $\gsu(8)$ in order to accommodate the matter transitions. Because the ${\tiny \yng(1,1,1)}$ representation appears only at $\gsu(6)$ and above, the ${\tiny \yng(1,1,1)}$ Higgsing processes will exclusively involve situations where $\epsilon$ has a specialized form. 

To find the deformations that break $\gsu(6)$ to $\gsu(3)\times\gsu(3)$, we tune one $\gsu(3)$ algebra on $\sigma=0$ and the other on $\sigma - \alpha \epsilon_1=0$. $\epsilon_1$ is a polynomial of order $r-2$. Performing this tuning requires a modified redefinition of $\phi_0$:
\begin{displaymath}
\phi_0 = \alpha \beta - \epsilon_1 \nu.
\end{displaymath}
With this redefinition, the split condition is satisfied for both of the codimension-one singularities. The other steps of this tuning are similar to those of the $\gsu(6)$ tuning, but some parameter redefinitions may involve additional terms proportional to $\epsilon_1$. The new terms dependent on $\epsilon_1$ are the Higgsing deformations, and $\epsilon_1$ thus parametrizes the possible deformations. As expected, there is one fewer deformation than the number of $\frac{1}{2}\mathbf{20}$ multiplets in the $\gsu(6)$ model. Moreover, there must be at least two $\frac{1}{2}\mathbf{20}$ multiplets for $\epsilon_1$ to be effective and for the deformations to be possible. 

$\alpha$ and $\epsilon_1$ both contribute bifundamental matter in the $\gsu(3)\times\gsu(3)$ model, while $\beta$ does not contribute any matter. However, there are two ways to form bifundamental matter in the $\gsu(3)\times\gsu(3)$ model, as described in \cite{mt-singularities}; in half-hypermultiplets, bifundamental matter can be in the form $(\mathbf{3},\bar{\mathbf{3}})+(\bar{\mathbf{3}},\mathbf{3})$ or in the form $(\mathbf{3},\mathbf{3})+(\bar{\mathbf{3}},\bar{\mathbf{3}})$ . From the field theory perspective, $(\mathbf{3},\bar{\mathbf{3}})+(\bar{\mathbf{3}},\mathbf{3})$ matter should come from the $\frac{1}{2}\mathbf{20}$ multiplets of the $\gsu(6)$ model, whereas the $(\mathbf{3},\mathbf{3})+(\bar{\mathbf{3}},\bar{\mathbf{3}})$ matter should come from the $\mathbf{15}$ multiplets. For this reason, $\epsilon_1$, which represents bifundamentals originating from  $\mathbf{10}$ matter, should give $(\mathbf{3},\bar{\mathbf{3}})+(\mathbf{3},\bar{\mathbf{3}})$ bifundamental matter (which we will refer to as a $(\mathbf{3},\bar{\mathbf{3}})$ full multiplet). $\alpha$ meanwhile should contribute $(\mathbf{3},\mathbf{3})+(\bar{\mathbf{3}},\bar{\mathbf{3}})$ matter (or a full $(\mathbf{3},\mathbf{3})$ multiplet). These two realizations are physically indistinguishable in the $\gsu(3)\times\gsu(3)$ model, so the distinction is somewhat arbitrary. However, similar types of distinctions will be important in tunings considered later. 

The $\gsu(3)\times\gsu(3)$ model has a transition, although the transition does not have as interesting of a change in the representations. $\alpha$, $\nu$, and $\lambda$ obtain a common factor $a$ just as in the $\gsu(6)$ transition, while $\beta$, $\phi_2$ and $\epsilon_1$ each absorb one factor of $a$. $a$ is once again a superconformal point.  The transition is therefore
\begin{equation}
(\mathbf{3},\mathbf{3})+3\times(\mathbf{3},\mathbf{1})+3\times(\mathbf{1},\mathbf{3}) + 2\times(\mathbf{1},\mathbf{1}) \rightarrow (\mathbf{3},\bar{\mathbf{3}})+3\times(\mathbf{3},\mathbf{1})+3\times(\mathbf{1},\mathbf{3}) + 2\times(\mathbf{1},\mathbf{1}),
\end{equation}
which is the Higgsed version of the $\gsu(6)$ transition. The only net effect of this transition is to exchange the $(\mathbf{3},\mathbf{3})$
and $(\mathbf{3},\bar{\mathbf{3}})$ representations. Since the two
bifundamental representations are essentially equivalent, the
transition may not seem to be as interesting as the other
transitions. Nevertheless, the structure of the transition is the same
as that of the other transitions, and the $\gsu(3)\times\gsu(3)$
transition will be important when analyzing transitions in other
models.

Enhancing the $\gsu(3)$ algebra on $\sigma$ to $\gsu(4)$ allows us to find the $\gsu(7)\rightarrow \gsu(4)\times\gsu(3)$ Higgsing deformations. The $\gsu(3)$ singularity, which was previously on $\sigma-\alpha\epsilon_1=0$, now occurs on the locus $\sigma-\xi\delta^2\epsilon_2 = 0$. $\beta$ is no longer forced to be a perfect square, as it is redefined as
\begin{equation}
\beta = \delta^2 + \epsilon_2 \zeta_1.
\end{equation}
The rest of the $\gsu(4)\times\gsu(3)$ tuning process is similar to that of $\gsu(7)$, except that the $\phi_2$ and $g_6$ redefinitions of Equations \eqref{eq:phi2def} and \eqref{eq:g6def} obtain additional terms proportional to $\epsilon_2$. $\epsilon_2$, a polynomial of order $\rseven-2$, parametrizes the set of deformations.  

There is a matter-changing transition in this $\gsu(4)\times\gsu(3)$ model. The steps in the transition are nearly identical to the $\gsu(7)$ transition, only $\epsilon_2$ absorbs a single factor of $a$ along with $\delta$ and $\zeta_1$. Both $\epsilon_2$ and $\xi$ contribute bifundamentals, while each zero of $\delta$ gives a $(\mathbf{6},\mathbf{3})$ multiplet and a $(\mathbf{1},\mathbf{3})$. On the locus $\xi-\epsilon_2 \zeta_2 = \sigma=0$, the singularity type enhances to $D_4\times A_2$, giving $(\mathbf{6},\mathbf{1})$ matter. Including fundamentals and singlets, the matter transition is
\begin{multline}
3\times(\mathbf{6},\mathbf{1}) + 3\times(\mathbf{4},\mathbf{3}) + 3\times(\mathbf{4},\mathbf{1}) + 6\times(\mathbf{1},\mathbf{3}) + 8\times(\mathbf{1},\mathbf{1}) \\ \rightarrow (\mathbf{6},\mathbf{3}) + (\mathbf{4},\mathbf{3}) + 9\times(\mathbf{4},\mathbf{1}) +  8\times(\mathbf{1},\mathbf{3}) + 2\times(\mathbf{1},\mathbf{1}),
\end{multline}
leading to a net matter change of
\begin{equation}
3\times(\mathbf{6},\mathbf{1}) + 2\times(\mathbf{4},\mathbf{3}) + 6\times(\mathbf{1},\mathbf{1}) \rightarrow (\mathbf{6},\mathbf{3}) + 6\times(\mathbf{4},\mathbf{1}) + 2\times (\mathbf{1},\mathbf{3}).
\end{equation}
This transition is the Higgsed version of the $\gsu(7)$ transition. 

For the $\gsu(8)\rightarrow\gsu(5)\times\gsu(3)$ Higgsing process, we tune the $\gsu(5)$ symmetry on $\sigma$ and the $\gsu(3)$ symmetry on $\sigma -\delta^4\tau\epsilon_3$. While we are able to find an explicit F-theory realization of this Higgsing process, there are tight constraints on the possible $\gsu(8)$ models that can be Higgsed on $\mathbf{56}$ matter. For $\epsilon_3$ to be effective and for this Higgsing process to be possible, the original $\gsu(8)$ model must have at least 2 $\mathbf{56}$ multiplets. From anomaly cancellations alone, this requires $n\geq 6$; our F-theory tunings support two $\mathbf{56}$ multiplets only when $n\geq 8$. All of these models oversaturate the gravitational anomaly bound on their own, and it is necessary to include a second gauge symmetry. For instance, when the singularity is tuned on $\tilde{S}$ with compactification base $\mathbb{F}_n$, a non-Higgsable cluster on $S$ can allow the global model to satisfy the gravitational anomaly bound. In F-theory, we are able to find the deformations for  $\gsu(8)$ models on $\tilde{S}$ with explicit tunings given earlier. These models have $n\geq 8$ and only two $\mathbf{56}$ multiplets. The resulting $\gsu(5)\times\gsu(3)$ matter spectrum agrees exactly with that expected from the branching patterns. However, we cannot realize $\gsu(8)$ models with more than three $\mathbf{56}$ multiplets on $\tilde{S}$ using our constructions, as our F-theory tunings will not support three $\mathbf{56}$ multiplets for $n\leq12$. We therefore cannot see $\gsu(5)\times\gsu(3)$ transitions in the $\tilde{S}$ tunings. 

\section{Heterotic description of matter transitions}
\label{sec:heterotic}

In this section we describe the 
SU($N$) theories with exotic matter, as well as the
Higgsing and small instanton
transitions of interest in this paper in terms of the bundle geometry
on the heterotic side of the duality.  We begin in the next subsection
by considering the construction of SU($N$) theories where
$N=6,7,8$. In Subsection \ref{hettransitions} we describe the small
instanton transitions that are possible in such theories. Finally, in
Subsection \ref{Higgsingproc}, we describe how the Higgsing processes
are realized at the level of bundle geometry. Some of the results
presented here are, of course, not new and can be found in
\cite{Bershadsky-all}, for example. We find it useful to present them
again here, however, in the same language that we  use to describe
previously unstudied cases.

\subsection{SU($N$) theories in heterotic compactifications}

\subsubsection{SU(6)} \label{su6het}

The relevant group theory for studying a compactification of the $E_8 \times E_8$ heterotic string with an SU(6) gauge group in six dimensions is as follows.
\begin{eqnarray}  \label{expldecomp}
E_8 &\supset& {\rm SU}(6) \times {\rm SU}(3) \times {\rm SU}(2) \\ \nonumber
{\bf 248} &=& ({\bf 1},{\bf 1},{\bf 3})+({\bf 1},{\bf 8},{\bf 1}) + ({\bf 35},{\bf 1},{\bf 1}) + (\overline{{\bf 15}},{\bf 3},{\bf 1}) + ({\bf 15},\overline{{\bf 3}},{\bf 1})+ ({\bf 6},{\bf 3},{\bf 2}) +(\overline{{\bf 6}},\overline{{\bf 3}},{\bf 2})+({\bf 20},{\bf 1},{\bf 2})
\end{eqnarray}
From this we see that we need a bundle with structure group ${\rm SU}(3)\times {\rm SU}(2)$ to obtain an unbroken gauge group of SU(6). We take the gauge bundle to be
\begin{eqnarray} \label{explbundle}
{\cal V} = {\cal V}_3 \oplus {\cal V}_2
\end{eqnarray}
where ${\cal V}_3$ has structure group SU(3) and ${\cal V}_2$ has structure group ${\rm SU}(2)$. 

The matter content resulting from a sum of vector bundles such as (\ref{explbundle}) can be computed in terms of the first cohomology groups of combinations of those objects and their wedge powers. This can either be seen in terms of dimensional reduction of gaugino degrees of freedom in ten dimensions to give fermionic matter \cite{Green:1987mn}, or in terms of dimensional reduction of bosonic degrees of freedom. The bosonic components of the low-energy theory all descend from adjoint valued gauge fields in ten dimensions. We consider first cohomologies because these are associated with one forms - which can be used to account for the space-time index of the gauge field leading to scalar degrees of freedom in six dimensions. The particular combination of bundles that one considers is determined by the decomposition  (\ref{expldecomp}). For example, we can see from the final term in the second line of (\ref{expldecomp}) that if we want to obtain matter transforming in the ${\bf 20}$ representation of SU(6), then the relevant one forms in the dimensional reduction must carry an index in the fundamental of SU(2), leading us to consider $H^1({\cal V}_2)$.

The different first cohomology groups of interest can be computed, in the case of K3 compactifications, by using the Hirzebruch-Riemann-Roch theorem \cite{Hirz}. Using the fact that bundles in a heterotic compactification are required to be slope stable, the statement of this theorem can be reduced to the following formula for the dimension of the first cohomology of a bundle ${\cal F}$ in such a situation.
\begin{eqnarray}
-h^1({\cal F}) = \int_{K3} \textnormal{ch}({\cal F}) \wedge \left( 1+ \frac{c_2(K3)}{12} \right)
\end{eqnarray}
Applying this formula to the bundles that are relevant given the
decomposition (\ref{expldecomp}) gives rise to the results given in
Table \ref{tab9}.

\begin{table}[!h]
\begin{center}
\begin{tabular*}{14.1cm}{|c|c|c|}
\hline
Representation & Cohomology & Multiplicity \\ \hline
${\bf 15}$ & $H^1({\cal V}^{\vee}_3)$  & $c_2({\cal V}_3) -6$ \\
$\overline{{\bf 15}}$ & $H^1({\cal V}_3)$ & $c_2({\cal V}_3) -6$\\ 
${\bf 6}$ & $H^1({\cal V}_3 \otimes {\cal V}_2)$ & $2 c_2({\cal V}_3) + 3 c_2({\cal V}_2) -12$\\
$\overline{{\bf 6}}$ & $H^1({\cal V}_3^{\vee}\otimes {\cal V}_2)$ &  $2 c_2({\cal V}_3) + 3 c_2({\cal V}_2) -12$\\ 
${\bf 20}$ &$H^1({\cal V}_2)$ & $c_2({\cal V}_2) -4$\\ 
${\bf 1}$ &$ H^1(\textnormal{End}_0({\cal V}_3)) \oplus H^1(\textnormal{End}_0({\cal V}_2)) $ &$(4 c_2({\cal V}_2) -6) +(6 c_2({\cal V}_3) -16)$\\\hline
\end{tabular*}
\caption{{\it The cohomology 
and multiplicity
associated to each representation of the low-energy gauge group SU(6).}}
\label{tab9}
\end{center}
\end{table}
In the SU(6) case we are considering, the numbers of vector and tensor
multiplets are $n_V=35$ and $n_T=1$ respectively. The number of
half-hypermultiplets associated to the other (``hidden sector'') $E_8$ bundle is
$h^1(\textnormal{End}_0 {\cal V}_{E_8}) = c_2(\textnormal{End}_0 {\cal
V}_{E_8}) -496$. We also have an additional $20$ hypermultiplets from the metric moduli of K3. This information, together with the matter content
given in Table \ref{tab9} can be substituted into the six-dimensional
anomaly cancelation condition to give the following.
\begin{eqnarray}
n_H + 29 n_T -n_V =273 \\ \label{10danom2}
\Rightarrow  c_2({\cal V}_3) + c_2({\cal V}_2) + \frac{1}{60} c_2(\textnormal{End}_0({\cal V}_{E_8})) = 24
\end{eqnarray}
This is precisely the ten-dimensional anomaly cancelation condition as expected.

The matter content outlined in Table \ref{tab9} takes the form described in Equation (\ref{eq:model-6}), where we have
\begin{eqnarray} \label{su6nr}
n=c_2({\cal V}_3) + c_2({\cal V}_2) -12 \;\; \textnormal{and} \;\; r = c_2({\cal V}_2) -4\;.
\end{eqnarray}


\subsubsection{SU(7)} \label{su7het}

The relevant group theory in this case is
\begin{eqnarray}
E_8 &\supset& {\rm SU}(7) \times {\rm SU}(2) \times {\rm U}(1) \\  
{\bf 248} &=& ({\bf 1},{\bf 1})_0 + ({\bf 1},{\bf 3})_0 + ({\bf
  7},{\bf 2})_9 + (\overline{{\bf 7}},{\bf 2})_{-9} + ({\bf 48},{\bf
  1})_0 + ({\bf 7},{\bf 1})_{-12} + ({\bf 21},{\bf 2})_{-3} + ({\bf
  35},{\bf 1})_6  
\nonumber
\\  &&+ (\overline{{\bf 7}},{\bf 1})_{12} + (\overline{{\bf 21}},{\bf
  2})_3 + (\overline{{\bf 35}},{\bf 1})_{-6} \;.
\label{u1charges}
\end{eqnarray}
To obtain a bundle with structure group ${\rm SU}(2) \times {\rm U}(1) \cong {\rm S}({\rm U}(2) \times {\rm U}(1))$ embedded inside $E_8$, we take the gauge bundle to be
\begin{eqnarray} \label{su7bundle}
{\cal V} = {\cal V}_2  \oplus {\cal L} \;.
\end{eqnarray}
Here ${\cal V}_2$ is a ${\rm U}(2)$ bundle, ${\cal L}$ is a line bundle and
$c_1({\cal V}_2) = - c_1({\cal L})$.  The ${\rm U}(1)$ factor above appears both in the structure group and also in its commutant inside $E_8$. Naively, therefore, one might expect this Abelian group to be unbroken. However, as is well documented in this context \cite{Dine:1987xk,Sharpe:1998zu,Lukas:1999nh,Blumenhagen:2005ga,Anderson:2009sw,Anderson:2009nt}, this ${\rm U}(1)$ gains a mass through the Green-Schwarz mechanism. In this process the ${\rm U}(1)$ gauge boson is made massive and one entire hypermultiplet from the $K3$ metric moduli is removed from the low energy spectrum. One of the degrees of freedom in the hypermultiplet is eaten by the gauge boson and the remaining three are made massive by the triplet of D-terms of the six-dimensional ${\cal N}=1$ theory.

The matter content that results
from the bundle (\ref{su7bundle}) is given in Table \ref{tab10}.
\begin{table}[!h]
\begin{center}
\begin{tabular*}{14.6cm}{|c|c|c|}
\hline
Representation & Cohomology & Multiplicity \\ \hline
${\bf 1}$ & $H^1(\textnormal{End}_0({\cal V}_2))$  & $4c_2({\cal V}_2)- c_1({\cal L})^2 -6$ \\
${\bf 7}$ & $H^1({\cal V}_2^{\vee} \otimes {\cal L}) \oplus H^1 ({\cal L}^{\vee 2})$& $(c_2({\cal V}_2) -\frac{5}{2} c_1({\cal L})^2-4)+(-2c_1({\cal L})^2 -2)$\\
$\overline{{\bf 7}}$ & $H^1({\cal V}_2 \otimes {\cal L}^{\vee }) \oplus H^1 ({\cal L}^{ 2})$ &  $(c_2({\cal V}_2) -\frac{5}{2} c_1({\cal L})^2-4)+(-2c_1({\cal L})^2 -2)$\\ 
${\bf 35}$ & $H^1({\cal L})$ & $-\frac{1}{2} c_1({\cal L})^2 -2$\\
$\overline{{\bf 35}}$ & $H^1({\cal L}^{\vee})$&  $-\frac{1}{2} c_1({\cal L})^2 -2$\\ 
${\bf 21}$ &$ H^1 ({\cal V}_2)$ & $c_2({\cal V}_2) - \frac{1}{2}c_1({\cal L})^2 -4$\\ 
$\overline{{\bf 21}}$ &$H^1({\cal V}^{\vee}_2)$ & $c_2({\cal V}_2) - \frac{1}{2} c_1({\cal L})^2 -4$\\ 
\hline
\end{tabular*}
\caption{{\it The cohomology associated to each representation of the low-energy gauge group SU(7).}}
\label{tab10}
\end{center}
\end{table}

Following a similar procedure to that discussed in the SU(6) case,
 Table (\ref{tab10}) leads to the
following anomaly cancelation constraint
\begin{eqnarray} \label{oops2}
n_H+29 n_T-n_V &=&273 \\ \nonumber
\Rightarrow c_2({\cal V}_2)-2 c_1({\cal L})^2  + \frac{1}{60} c_2(\textnormal{End}_0({\cal V}_{E_8}))&=&24
\end{eqnarray}
Naively there is a mismatch  in the second equation, which is
corrected by including an additional vector multiplet, beyond those associated to ${\rm SU}(7)$, to account for the Green-Schwarz massive ${\rm U}(1)$. Alternatively, considering the theory below the mass scale associated to the Abelian factor, we can drop both the number of vectors and the number of metric moduli hypermultiplets by one to account for the effects of the Higgs process described above.

The matter content outlined in Table \ref{tab10} takes the form described in Equation (\ref{eq:model-7}) where we have
\begin{eqnarray} \label{su7nr}
n = c_2({\cal V}_2) - 2 c_1({\cal L})^2 -12\;\; \textnormal{and} \;\; r = -\frac{1}{2} c_1({\cal L})^2 -2 \;.
\end{eqnarray}

\subsubsection{SU(8) and beyond} \label{su8het}
The relevant group theory in this case is,
\begin{eqnarray} 
E_8 &\supset&  {\rm SU}(8) \times {\rm U}(1) \\ 
{\bf 248} &=&{\bf 1}_0 + {\bf 8}_9 + \overline{{\bf 8}}_{-9} +{\bf 28}_{-6} + \overline{{\bf 28}}_6 + {\bf 56}_3 + \overline{{\bf 56}}_{-3} + {\bf 63}_0 \;.
\end{eqnarray}
To embed a bundle with ${\rm U}(1)={\rm S}({\rm U}(1) \times {\rm U}(1))$ structure group inside $E_8$ we write the following.
\begin{eqnarray}
{\cal V} = {\cal L} \oplus {\cal L}^{\vee}
\end{eqnarray}
Here ${\cal L}$ is a simple line bundle. Once again computing the spectrum we obtain the result given in Table \ref{tab11}.
\begin{table}[!h]
\begin{center}
\begin{tabular*}{8.2cm}{|c|c|c|}
\hline
Representation & Cohomology & Multiplicity \\ \hline
${\bf 8}$ & $H^1({\cal L}^3)$  & $-\frac{9}{2} c_1({\cal L})^2 -2$ \\
$\overline{{\bf 8}}$ & $H^1({\cal L}^{\vee \,3})$& $-\frac{9}{2} c_1({\cal L})^2 -2$\\
${\bf 28}$ & $H^1({\cal L}^{\vee \,2})$ & $-2 c_1({\cal L})^2 -2$\\ 
$\overline{{\bf 28}}$ & $H^1({\cal L}^2)$ & $-2 c_1({\cal L})^2 -2$\\
${\bf 56}$ & $H^1({\cal L})$&  $-\frac{1}{2} c_1({\cal L})^2 -2$\\ 
$\overline{{\bf 56}}$ &$H^1 ( {\cal L}^{\vee})$ & $-\frac{1}{2} c_1({\cal L})^2 -2$\\ 
\hline
\end{tabular*}
\caption{{\it The cohomology associated to each representation of the low-energy gauge group SU(8).}}
\label{tab11}
\end{center}
\end{table}

Anomaly cancelation in this case is as follows.
\begin{eqnarray} \label{oops}
n_H+29 n_T-n_V =273 \\ \nonumber
\Rightarrow -4c_1({\cal L})^2+ \frac{1}{60} c_2(\textnormal{End}_0({\cal V}_{E_8}))=24
\end{eqnarray}
Here, again, there is a massive U(1), which can be thought of as
reducing the number of massless metric moduli on the K3 by 1, so only
19 K3 moduli are included in the anomaly matching condition.

The matter content given in Table \ref{tab11} takes the form described in Equation (\ref{eq:model-8}) where we have
\begin{eqnarray}\label{nexpp}
n = -4 c_1({\cal L})^2 -12 = 8r + 4 \;\;,\;\; r = -\frac{1}{2} c_1({\cal L})^2 -2 \;\; \textnormal{and} \;\; r' =0 
\end{eqnarray}

Note that we get an extremely non-generic spectrum in this case from a
six-dimensional field theory point of view. The spectrum of the SU(8)
charged matter in the heterotic compactification is controlled by a
single integer rather than by 3 as in (\ref{eq:model-8}). In fact, we
see from the expression for $n$ in (\ref{nexpp}), together with the
topological fact that $c_1({\cal L})^2$ is even on K3 for any ${\cal
  L}$, that $c_1({\cal L})^2 \leq -4$. Studying Equation (\ref{oops}),
we see that the smallest $c_1({\cal L})^2$ can be is $-6$. Thus the
two possibilities are $n=4$ and $n=12$. Neither of these two
possibilities leaves a large enough second Chern class available for
the hidden sector bundle to completely break the $E_8$ in the other
sector. We will have at minimum an SO(8) and $E_8$ ``hidden'' sector
gauge group respectively in these two cases
\cite{Anderson:2014gla}. These match the non-Higgsable clusters that
one would expect in a dual compactification of F-theory on an elliptic
fibration over $F_4$ and $F_{12}$.  As in the F-theory analysis of
Section \ref{fsu8}, we obtain no matter in the $\mathbf{70}$
representation of SU(8) in these perturbative heterotic
compactifications.

The gauge group SU(8) is on the edge of what can be achieved perturbatively in compactifications of the $E_8 \times E_8$ heterotic string. ${\rm SU}(9)$ is a subgroup of $E_8$ but its commutant inside $E_8$ is empty, meaning that it can not be achieved as the unbroken commutant of some continuous bundle structure group (although it can be achieved on singular K3 manifolds by using a $\mathbb{Z}_3$ structure group \cite{Aspinwall:1998xj}). The groups ${\rm SU}(10)$ and higher are simply not subgroups of $E_8$ and thus can't be achieved as gauge groups in the case of perturbative compactifications. It is interesting that the non-genericity of spectrum described in the proceeding paragraph arises in the boundary ${\rm SU}(N)$ case of the largest possible $N$.

\subsection{Realization of the transitions}\label{hettransitions}

The matter transitions described from a field theory perspective in Section \ref{sec:SUGRA}, and from an F-theory perspective in Section \ref{ftrans} are also realized concretely in these heterotic compactifications.

In the SU(6) case of Section \ref{su6het} we can utilize small
instanton transitions to swap second Chern class contributions between
${\cal V}_3$ and ${\cal V}_2$, subject to the overall constraint
(\ref{10danom2}). Note that we could also swap second Chern class with
the ``hidden sector'' bundle but this would lead to an intermediate
stage involving an increase in the number of tensor multiplets.

Studying Table \ref{tab9}, we see that lowering $c_2({\cal V}_2)$ by $1$ and raising $c_2({\cal V}_3)$ by $1$ causes the number of ${\bf 6}$ and $\overline{{\bf 6}}$'s to go down by a single unit and the number of ${\bf 20}$'s also to lower by $1$. Conversely we gain one ${\bf 15}$, one $\overline{{\bf 15}}$ and two singlets. This is precisely a transition of the form described in Equation (\ref{eq:equivalence-6}) from a field theory perspective and in Equation (\ref{ftran1trim}) in an F-theory context (it is important in making this comparison to realize that the multiplicities given in tables such as Table \ref{tab9} count half-hypermultiplets).

In the SU(7) case of Section \ref{su7het}, small instanton transitions within one $E_8$ factor swap  contributions to $c_2({\cal V}_2)$ with  $c_1({\cal L})^2$ in such a manner as to preserve Equation (\ref{oops2}). Note that not only the second, but also the first Chern class of ${\cal V}_2$ changes under such a transition. 

We see from Equation (\ref{oops2}) that if we increase $c_1({\cal L})^2$ by $2$\footnote{This is the minimum possible change given the expression for the number of $35$'s. In fact, that $c_1({\cal L})^2$ is even for an arbitrary line bundle ${\cal L}$ is enforced by the topology of $K3$.} in Table \ref{tab10} we must also increase $c_2({\cal V}_1)$ by $4$. Studying Table \ref{tab10}, we see that, under such a transition, the number of ${\bf 35}$ and $\overline{{\bf 35}}$ half hypermultiplets lowers by 1, the number of ${\bf 7}$ and $\overline{{\bf 7}}$ half hypermultiplets lowers by 5, the number of ${\bf 21}$ and $\overline{{\bf 21}}$ half hypermultiplets increases by 3 and the number of singlet half hypermultiplets increases by 14. This is precisely a transition of the form described in Equation (\ref{eq:equivalence-7}) from a field theory perspective and in Equation (\ref{su7ftrans}) in an F-theory context.

The SU(8) case of Section \ref{su8het} mirrors what was found in an
F-theory context in Section \ref{ftrans}. No small instanton
transitions purely in one $E_8$ factor, giving rise to either of the
forms (\ref{eq:equivalence-81}) or (\ref{eq:equivalence-82}) seen in
our field theory discussion, is possible, however, due to the constraint given in
Equation (\ref{oops}). Small instanton transitions are of course
possible if one allows a modification of the second $E_8$ bundle and
it is also possible that more generic results from a field theoretic
point of view could be obtained on a singular $K3$, where
non-perturbative contributions to the gauge charged sector are
possible.

\subsection{Higgsing processes} \label{Higgsingproc}

In heterotic compactification, Higgsing processes in the low-energy
field theory have a clear interpretation in terms of deformations of
the gauge bundle. Here, we describe one example each of the
deformations associated to Higgsing on fundamental, double
antisymmetric and triple antisymmetric matter, together with a table
detailing some of the key information in the other cases. The full
analysis of the remaining examples can be found in Appendix
\ref{hetHiggsapp}.

\subsubsection{Higgsing on fundamental matter} \label{hetfund}

Let us start by considering what happens to the bundle as we Higgs from SU(6) to ${\rm SU}(5)$. The relevant group theory in this case is
\begin{eqnarray}\label{groupbreaksu61}
{\rm SU}(6) &\to& {\rm SU}(5) \times {\rm U}(1) \\
{\bf 6} &=& {\bf 1}_{-5} + {\bf 5}_{1} \\ 
{\bf 15} &=& {\bf 5}_{-4} + {\bf 10}_2 \\
{\bf 20} &=& {\bf 10}_{-3} + \overline{{\bf 10}}_3 \;.
\end{eqnarray}
Clearly, we wish to turn on the singlet of ${\rm SU}(5)$ inside the fundamental of SU(6) to achieve the Higgsing. We see from Table \ref{tab9} that the ${\bf 6}$ and $\overline{{\bf 6}}$ half hypermultiplets, which combine to form a single hypermultiplet, are given by the cohomologies $H^1({\cal V}_3 \otimes {\cal V}_2)$ and $H^1({\cal V}_3^{\vee} \otimes {\cal V}_2)$ respectively.

In terms of bundle geometry, turning on fields descending from these cohomology groups corresponds to forming the following extension.
\begin{eqnarray} \label{su6v}
0 \to {\cal V}_2 \to {\cal V} \to {\cal V}_3 \to 0
\end{eqnarray}
In fact, the bundle that we form is a deformation of this extension
and its dual, as described
in the work of Li and Yau \cite{Li:2004hx}. 

The bundle ${\cal V}$ in (\ref{su6v}) has structure group ${\rm SU}(5)$, which is the relevant case to arrive at an ${\rm SU}(5)$ low-energy symmetry. One can check how the matter that one obtains from ${\cal V}$ compares to that which follows from a simple decomposition of the multiplets in the original SU(6) theory using the branching rules in (\ref{groupbreaksu61}).

We start by computing the matter content associated to an ${\rm SU}(5)$ bundle, with the general result being given in Table \ref{tab2bew}. In computing this table we have used the usual decomposition of the adjoint representation of $E_8$,
\begin{eqnarray}
E_8 &\supset& {\rm SU}(5) \times {\rm SU}(5) \\
{\bf 248} &=& ({\bf 5},{\bf 10})+ (\overline{{\bf 10}},{\bf 5})+({\bf 10},\overline{{\bf 5}}) +(\overline{{\bf 5}},\overline{{\bf 10}})+({\bf 24},{\bf 1})+({\bf 1},{\bf 24})
\end{eqnarray}
together with Hirzebruch-Riemann-Roch.
\begin{table}[!h]
\begin{center}
\begin{tabular*}{8.23cm}{|c|c|c|}
\hline
Representation & Cohomology & Multiplicity \\ \hline
${\bf 5}$ & $H^1(\wedge^2 {\cal V})$  & $3c_2({\cal V}) -20$ \\
$\overline{{\bf 5}}$ & $H^1(\wedge^2 {\cal V}^{\vee})$ & $3c_2({\cal V}) -20$\\ 
${\bf 10}$ & $H^1({\cal V}^{\vee})$ & $c_2({\cal V}) -10$\\
$\overline{{\bf 10}}$ & $H^1({\cal V})$ &  $c_2({\cal V}) -10$\\ 
${\bf 1}$ &$ H^1(\textnormal{End}_0({\cal V})) $ &$10 c_2({\cal V})-48$\\\hline
\end{tabular*}
\caption{{\it The cohomology associated to each representation of the low-energy gauge group SU(5).}}
\label{tab2bew}
\end{center}
\end{table}

Next we note that the second Chern class of the ${\cal V}$ resulting from a Higgsing transition such as that being considered in Equation (\ref{su6v}) is given by,
\begin{eqnarray} \label{su6c2v}
c_2({\cal V})= c_2({\cal V}_3) + c_2({\cal V}_2)\;.
\end{eqnarray}

Using these results we can compile Table \ref{transtab11dddd}, which compares the matter content associated to the bundle ${\cal V}$ in (\ref{su6v}) with a naive decomposition of the original SU(6) matter, as given in Table \ref{tab9}, using the branching rules (\ref{groupbreaksu61}).
\begin{table}[!ht]
\begin{center}
\begin{tabular*}{15.6cm}{|c|c|c|}
\hline
SU(5) Representation & \# from SU(6) multiplet decomposition & \# found after transition \\ \hline
${\bf 1}$ &  $10 c_2({\cal V}_3) +10 c_2({\cal V}_2) -46$  & $10 c_2({\cal V}_3) +10 c_2({\cal V}_2) -48$ \\
${\bf 5}$ & $3 c_2({\cal V}_3) +3 c_2({\cal V}_2) - 18$  & $3 c_2({\cal V}_3) +3 c_2({\cal V}_2) - 20$\\
$\overline{{\bf 5}}$ & $3 c_2({\cal V}_3) +3 c_2({\cal V}_2) - 18$ & $3 c_2({\cal V}_3) +3 c_2({\cal V}_2) - 20$\\
${\bf 10}$ & $c_2({\cal V}_3)+ c_2({\cal V}_2) - 10$ &  $c_2({\cal V}_3)+ c_2({\cal V}_2) - 10$\\ 
$\overline{{\bf 10}}$ & $c_2({\cal V}_3)+ c_2({\cal V}_2) - 10$ &  $c_2({\cal V}_3)+ c_2({\cal V}_2) - 10$\\
\hline
\end{tabular*}
\caption{{\it Matter content after Higgsing an SU(6) to an SU(5) theory, both via a naive decomposition of the initial SU(6) multiplets and via a direct computation from the resulting SU(5) bundle.}}
\label{transtab11dddd}
\end{center}
\end{table}

The differences in the last two columns of Table \ref{transtab11dddd}
consist of two full fundamental hypermultiplets and one scalar
hypermultiplet, and arise naturally due to degrees of freedom being absorbed by massive
gauge bosons, or being given a mass by D-terms, in the Higgsing
process. We can now confirm that this result matches the field theory
analysis given in section \ref{higsfund}.

\vspace{0.2cm}

Similar results are found by Higgsing SU(7) and SU(8) on their fundamental representations. The details of these computations can be found in Appendix \ref{hetHiggsapp}. The SU(7) case in particular has some different structure in that there are two different sources of $7$ representations in terms of cohomology, corresponding to two different bundles that are being deformed during the transition. Here we content ourselves with a presentation of the relevant bundle deformations in Table \ref{tab2trans}.

\begin{table}[!ht]
\begin{center}
\begin{tabular*}{15.0cm}{|c|c|c|}
\hline
Group transition & Bundle transition & Fields Gaining Vev \\ \hline \hline
${\rm SU}(6) \to SU(5)$ &$\begin{array}{c} {\cal V}_{SU(2)} \oplus {\cal V}_{SU(3)} \to \tilde{{\cal V}}_{SU(5)} \\ \textnormal{where}\;\; 0\to {\cal V}_{SU(2)} \to {\cal V}_{SU(5)} \to {\cal V}_{SU(3)} \to 0 \end{array}$& $\begin{array}{c} H^1({\cal V}_{SU(2)} \otimes {\cal V}_{SU(3)}) \end{array}$\\ \hline
${\rm SU}(7) \to SU(6) $ &$\begin{array}{c} {\cal V}_{U(2)} \oplus {\cal L} \to \tilde{{\cal V}}_{SU(3)} \oplus \tilde{{\cal V}}_{SU(2)} \\ \textnormal{where} \;\; 0\to {\cal L}^{\vee} \to {\cal V}_{SU(2)} \to {\cal L} \to 0 \\ \textnormal{and} \;\; 0\to {\cal L} \to V_{SU(3)} \to {\cal V}_{U(2)} \to 0\end{array}$& $\begin{array}{c} H^1({\cal V}_{U(2)}^{\vee} \otimes {\cal L} ) \\ H^1({\cal L}^{\vee 2}) \end{array}$ \\ \hline
${\rm SU}(8) \to SU(7) $ &$\begin{array}{c} {\cal L} \oplus {\cal L}^{\vee} \to \tilde{{\cal V}}_{U(2)} \oplus {\cal L} \\ \textnormal{where} \;\; 0 \to {\cal L}^{\vee 2} \to {\cal V}_{U(2)} \to {\cal L} \to 0  \end{array}$& $H^1({\cal L}^3)$\\\hline
\end{tabular*}
\caption{{\it Higgsing on the fundamental in various heterotic theories, and the resulting deformation of the gauge bundle. The tildes over some bundles in the second column indicate a Li-Yau type deformation of the untilded object and its dual \cite{Li:2004hx}.}}
\label{tab2trans}
\end{center}
\end{table}

\subsubsection{Higgsing on two-index antisymmetric matter} \label{hetanti2}

In this case we will give the example of Higgsing an SU(7) gauge group
on the ${\bf 21}$ dimensional representation.  The relevant group
theory in this case is the following.

\begin{eqnarray} \label{SU7asHiggsing1}
{\rm SU}(7) &\supset& {\rm SU}(5) \times {\rm SU}(2) \times {\rm U}(1) \\ \nonumber
{\bf 21} &=& ({\bf 1},{\bf 1})_{-10} +({\bf 5},{\bf 2})_{-3}+({\bf 10},{\bf 1})_4 \\\nonumber
{\bf 7} &=& ({\bf 1},{\bf 2})_{-5} + ({\bf 5},{\bf 1})_2 \\\nonumber
{\bf 35} &=& ({\bf 5},{\bf 1})_{-8} + (\overline{{\bf 10}},{\bf 1})_6 +({\bf 10},{\bf 2})_{-1} 
\end{eqnarray}
We see that giving a VEV to a {\bf 21} $\overline{{\bf 21}}$ pair will break SU(7) down to ${\rm SU}(5) \times {\rm SU}(2)$ with a Green-Schwarz massive ${\rm U}(1)$ also being present (this Green-Schwarz ${\rm U}(1)$ is the one originally present in the heterotic SU(7) model and does not correspond to the ${\rm U}(1)$ in (\ref{SU7asHiggsing1})). The ${\bf 21}$'s, according to Table \ref{tab10}, lie in the cohomology $H^1({\cal V}_2)$. In terms of bundle topology, giving an expectation value to such a field corresponds to forming the following bundle.
\begin{eqnarray} \label{higtrans1}
{\cal V} &=& {\cal Q}  \oplus {\cal L} \\ \nonumber
\textnormal{where}  &&0 \to {\cal V}_2 \to {\cal Q} \to {\cal O} \to 0
\end{eqnarray}
As in previous cases, one should really think of $Q$ as being a
deformation of this extension and its dual, a la Li-Yau
\cite{Li:2004hx}.  Here ${\cal Q}$ is a U(3) bundle and the line
bundle ${\cal L}$ is unaffected by the transition. The overall
structure group is ${\rm S}({\rm U}(3) \times {\rm U}(1))$ which does indeed break $E_8$
to ${\rm SU}(5) \times {\rm SU}(2) \times {\rm U}(1)$ (where the last factor is a common
Green-Schwarz anomalous factor between the structure group and the
visible gauge group). 

In order to compare the matter content before and after such a Higgsing transition, we must first compute the matter content in the ${\rm SU}(5) \times {\rm SU}(2)$ theory. The group theory for a general heterotic ${\rm SU}(5) \times {\rm SU}(2) $ case is as follows.
\begin{eqnarray} \label{su5group}
E_8 &\supset& {\rm SU}(5) \times {\rm SU}(2) \times {\rm SU}(3) \times {\rm U}(1) \\
{\bf 248} &=& ({\bf 5},{\bf 1},{\bf 1})_{-6} + ({\bf 5},{\bf 1},\overline{{\bf 3}})_4 + ({\bf 5},{\bf 2},{\bf 3})_{-1} + (\overline{{\bf 10}},{\bf 2},{\bf 1})_{-3} +(\overline{{\bf 10}},{\bf 1},{\bf 3})_2 \\ \nonumber
&&+ ({\bf 10},{\bf 2},{\bf 1})_3 + ({\bf 10},{\bf 1},\overline{{\bf 3}})_{-2} + (\overline{{\bf 5}},{\bf 1},{\bf 1})_{6}+(\overline{{\bf 5}},{\bf 1},{\bf 3})_{-4} + (\overline{{\bf 5}},{\bf 2},\overline{{\bf 3}})_1 \\ \nonumber
&& +({\bf 24},{\bf 1},{\bf 1})_0 + ({\bf 1},{\bf 1},{\bf 1})_0+({\bf 1},{\bf 3},{\bf 1})_0+({\bf 1},{\bf 2},{\bf 3})_5+({\bf 1},{\bf 2},\overline{{\bf 3}})_{-5} + ({\bf 1},{\bf 1},{\bf 8})_0
\end{eqnarray}
This leads us to the matter content given in Table \ref{tabastrans11} for such a theory.
\begin{table}[!h]
\begin{center}
\begin{tabular*}{11.57cm}{|c|c|c|}
\hline
Representation & Cohomology & Multiplicity \\ \hline
$({\bf 5},{\bf 1})$ & $H^1({\cal L}^{\vee 2}) \oplus H^1( {\cal Q}^{\vee} \otimes {\cal L})$  & $c_2({\cal Q}) -5 c_1({\cal L})^2 -8$ \\
$({\bf 5},{\bf 2})$ & $H^1({\cal Q} )$& $c_2({\cal Q}) - \frac{1}{2} c_1({\cal L})^2 -6$\\
$(\overline{{\bf 10}},{\bf 2})$ & $H^1 ({\cal L}^{\vee})$ &  $-\frac{1}{2} c_1({\cal L})^2 -2$\\ 
$(\overline{{\bf 10}},{\bf 1})$ & $H^1({\cal Q} \otimes {\cal L})$ & $c_2({\cal Q}) -  c_1({\cal L})^2-6$\\
$({\bf 1},{\bf 1})$ & $H^1(\textnormal{End}_0 ({\cal Q}))$&  $6 c_2({\cal Q})-2 c_1({\cal L})^2-16$\\ 
$({\bf 1},{\bf 2})$ &$ H^1({\cal Q} \otimes {\cal L}^{2})$ & $c_2({\cal Q}) - \frac{9}{2} c_1({\cal L})^2 -6$\\ 
\hline
\end{tabular*}
\caption{{\it The cohomology associated to each representation of the low-energy gauge group ${\rm SU}(5)\times {\rm SU}(2) \times U(1)$.}}
\label{tabastrans11}
\end{center}
\end{table}

For the particular case of an ${\rm S}({\rm U}(3) \times {\rm U}(1))$ bundle formed by a transition of the form given in equation (\ref{higtrans1}) we have:
\begin{eqnarray}
c_2({\cal Q})&=&c_2({\cal V}_2) \\
c_1({\cal L})&=&c_1({\cal L}) \;.
\end{eqnarray}
Given this we can form the same table as we did in the case of Higgsing on the fundamental: comparing a direct decomposition of the SU(7) multiplets under the symmetry breaking with the spectrum of the bundle after the transition. This is given in Table \ref{transtab1}.
\begin{table}[!h]
\begin{center}
\begin{tabular*}{17cm}{|c|c|c|}
\hline
${\rm SU}(5)\times {\rm SU}(2)$ Representation & \# from SU(7) multiplet decomposition & \# found after transition \\ \hline
$({\bf 1},{\bf 1})$ & $6 c_2({\cal V}_2) -2 c_1({\cal L})^2 -14$  & $ 6c_2({\cal V}_2) -2 c_1({\cal L})^2 -16$ \\
$({\bf 1},{\bf 2})$ & $ c_2({\cal V}_2) - \frac{9}{2} c_1({\cal L})^2 - 6$  & $ c_2({\cal V}_2) - \frac{9}{2} c_1({\cal L})^2 - 6$ \\
$({\bf 5},{\bf 1})$ & $ c_2({\cal V}_2) - 5 c_1({\cal L})^2 - 8$& $ c_2({\cal V}_2) - 5 c_1({\cal L})^2 - 8$\\
$({\bf 5},{\bf 2})$ & $c_2({\cal V}_2)- \frac{1}{2}c_1({\cal L})^2 - 4$ &  $c_2({\cal V}_2)- \frac{1}{2}c_1({\cal L})^2 - 6$\\ 
$(\overline{{\bf 10}},{\bf 1})$ & $c_2({\cal V}_2)- c_1({\cal L})^2 - 6$ &  $c_2({\cal V}_2)- c_1({\cal L})^2 - 6$\\
$(\overline{{\bf 10}},{\bf 2})$ & $- \frac{1}{2}c_1({\cal L})^2 -2$&  $- \frac{1}{2} c_1({\cal L})^2 -2$\\ 
\hline
\end{tabular*}
\caption{{\it Matter content after Higgsing an SU(7) to an $S(U(5) \times U(2))$ theory, both via a naive decomposition of the initial SU(7) multiplets and via a direct computation from the resulting $S(U(5)\times U(2))$ bundle.}}
\label{transtab1}
\end{center}
\end{table}
Once more, the differences between the second and third columns in Table \ref{transtab1} precisely match what we would expect from an analysis of the Higgs mechanism in such a situation. This Higgsing is precisely of the form described in a field theory context in Section \ref{antihigs}.

Similar results are found by Higgsing SU(6) and SU(8) on their two-index antisymmetric representations. The details of these computations can be found in Appendix \ref{hetHiggsapp}. Here we content ourselves with a presentation of the relevant bundle deformations in Table \ref{tab3trans}.

\begin{table}[!h]
\begin{center}
\begin{tabular*}{15.4cm}{|c|c|c|}
\hline
Group transition & Bundle transition & Fields Gaining VEV \\ \hline \hline
${\rm SU}(6) \to SU(4) \times {\rm SU}(2)$ &$\begin{array}{c} {\cal V}_{SU(2)} \oplus {\cal V}_{SU(3)} \to {\cal V}_{SU(2)} \oplus \tilde{{\cal V}}_{SU(4)} \\ \textnormal{where}\;\; 0\to {\cal V}_{SU(3)} \to {\cal V}_{SU(4)} \to {\cal O} \to 0 \end{array}$& $\begin{array}{c} H^1({\cal V}_{SU(3)}) \end{array}$\\ \hline
${\rm SU}(7) \to SU(5) \times {\rm SU}(2) $ &$\begin{array}{c} {\cal V}_{U(2)} \oplus {\cal L} \to \tilde{{\cal V}}_{U(3)} \oplus {\cal L} \\ \textnormal{where} \;\; 0\to {\cal V}_{U(2)} \to {\cal V}_{U(3)} \to {\cal O} \to 0 \end{array}$& $\begin{array}{c} H^1({\cal V}_{U(2)} ) \end{array}$ \\ \hline
${\rm SU}(8) \to SU(6) \times {\rm SU}(2) $ &$\begin{array}{c} {\cal L} \oplus {\cal L}^{\vee} \to \tilde{{\cal V}}_{SU(2)} \oplus {\cal L} \oplus{\cal L}^{\vee}\\ \textnormal{where} \;\; 0 \to {\cal L}^{\vee } \to {\cal V}_{SU(2)} \to {\cal L} \to 0  \end{array}$& $H^1({\cal L}^{\vee 2})$\\\hline
\end{tabular*}
\caption{{\it Higgsing on the two-index antisymmetric representation in various heterotic theories, and the resulting deformation of the gauge bundle. The tildes over some bundles in the second column indicate a Li-Yau type deformation of the untilded object and its dual \cite{Li:2004hx}.}}
\label{tab3trans}
\end{center}
\end{table}

\subsubsection{Higgsing on three-index antisymmetric matter} \label{hetanti3}

To illustrating Higgsing on three-index antisymmetric matter we will consider the example of SU(8). 
The relevant group theory in this case is as follows.
\begin{eqnarray}
\nonumber
{\rm SU}(8) &\supset& {\rm SU}(5) \times {\rm SU}(3) \times {\rm U}(1) \\\nonumber
{\bf 56} &=& ({\bf 1},{\bf 1})_{-15} +({\bf 5},\overline{{\bf 3}})_{-7} + (\overline{{\bf 10}},{\bf 1})_{9}+({\bf 10},{\bf 3})_1 \\\nonumber
{\bf 8} &=& ({\bf 1},{\bf 3})_{-5} + ({\bf 5},{\bf 1})_3 \\\nonumber
{\bf 28} &=& ({\bf 1},\overline{{\bf 3}})_{-10} + ({\bf 5},{\bf 3})_{-2} + ({\bf 10},{\bf 1})_6
\end{eqnarray}

Giving a VEV to the SU(5) singlets in a ${\bf 56}$, $\overline{{\bf 56}}$ pair will therefore break ${\rm SU}(8)\to {\rm SU}(5)\times {\rm SU}(3)$ with a Green-Schwarz massive ${\rm U}(1)$ also being present. The ${\bf 56}$'s, according to Table \ref{tab11}, lie in the cohomology $H^1({\cal L})$. In terms of bundle topology, giving an expectation value to such a field corresponds to forming the following bundle.
\begin{eqnarray} \label{lastcase}
&&{\cal V} = {\cal Q} \oplus {\cal L}^{\vee} \\
 \textnormal{where} && 0 \to {\cal L} \to {\cal Q} \to {\cal O} \to 0
\end{eqnarray}
Here ${\cal V}$ is an ${\rm S}({\rm U}(2)\times {\rm U}(1))$ bundle. The correct embedding of ${\rm S}({\rm U}(2)\times {\rm U}(1))$ does indeed break $E_8$ to ${\rm SU}(5) \times {\rm SU}(3) \times {\rm SU}(2) \times {\rm U}(1) $.

The group theory for the ${\rm SU}(5) \times {\rm SU}(3) $ case is as follows.
\begin{eqnarray} \label{su6group}
E_8 &\supset& {\rm SU}(5) \times {\rm SU}(3) \times {\rm SU}(2) \times {\rm U}(1) \\ \nonumber
{\bf 248} &=& ({\bf 5},{\bf 1},{\bf 1})_{-6}+ ({\bf 5},{\bf 3},{\bf 1})_4+({\bf 5},\overline{{\bf 3}},{\bf 2})_{-1}+(\overline{{\bf 10}},{\bf 1},{\bf 2})_{-3}+(\overline{{\bf 10}},\overline{{\bf 3}},{\bf 1})_2 \\ \nonumber
&&+({\bf 10},{\bf 1},{\bf 2})_3+({\bf 10},{\bf 3},{\bf 1})_{-2}+(\overline{{\bf 5}},{\bf 1},{\bf 1})_6+(\overline{{\bf 5}},\overline{{\bf 3}},{\bf 1})_{-4}+(\overline{{\bf 5}},{\bf 3},{\bf 2})_{1} \\ \nonumber
&&+({\bf 24},{\bf 1},{\bf 1})_{0}+({\bf 1},{\bf 1},{\bf 1})_0+({\bf 1},{\bf 1},{\bf 3})_0+({\bf 1},\overline{{\bf 3}},{\bf 2})_{5} + ({\bf 1},{\bf 3},{\bf 2})_{-5} + ({\bf 1},{\bf 8},{\bf 1})_0
\end{eqnarray}
This leads to a spectrum for such a heterotic compactification as given in Table \ref{tabastrans101}.
\begin{table}[!h]
\begin{center}
\begin{tabular*}{9.65cm}{|c|c|c|}
\hline
Representation & Cohomology & Multiplicity \\ \hline
$({\bf 5},{\bf 1})$ & $H^1 ({\cal L}^{\vee 3})  $ &  $-\frac{9}{2}c_1({\cal L})^2-2$\\ 
$({\bf 5},{\bf 3})$ & $H^1 ({\cal L}^2) $ &  $-2c_1({\cal L})^2-2$\\ 
$({\bf 5},\overline{{\bf 3}})$ & $H^1({\cal Q}^{\vee}) $&$c_2({\cal Q})-\frac{1}{2} c_1({\cal L})^2 -4 $\\
$({\bf 10},{\bf 1})$ & $ H^1({\cal Q}^{\vee}\otimes{\cal L}^{ 2} )$&$c_2({\cal Q})-\frac{5}{2} c_1({\cal L})^2 -4 $\\
$(\overline{{\bf 10}},\overline{{\bf 3}})$&$ H^1({\cal L})$&$ - \frac{1}{2} c_1({\cal L})^2-2$\\
$({\bf 1},{\bf 1})$&$H^1 (\textnormal{End}_0({\cal Q}))$&$4c_2({\cal Q}) - c_1({\cal L})^2-6$\\
$({\bf 1},\overline{{\bf 3}})$&$H^1 ({\cal Q} \otimes {\cal L}^2)$&$c_2({\cal Q}) - \frac{13}{2} c_1({\cal L})^2-4$\\
\hline
\end{tabular*}
\caption{{\it The cohomology associated to each representation of the low-energy gauge group ${\rm SU}(5)\times {\rm SU}(3) $. }}
\label{tabastrans101}
\end{center}
\end{table}

For the particular ${\rm S}({\rm U}(2) \times {\rm U}(1))$ bundle that we achieve after transition, as given in equation \ref{lastcase}, we have that,
\begin{eqnarray}
c_2({\cal Q})&=&0 \;. 
\end{eqnarray} 
With this information we can finally construct the table comparing the
break up of the SU(8) multiplets with the directly computed matter
spectrum after the bundle transition, as we did in the previous
cases. This is found in Table \ref{transtab1111}.

\begin{table}[!h]
\begin{center}
\begin{tabular*}{17.1cm}{|c|c|c|}
\hline
${\rm SU}(5)\times {\rm SU}(3)$ Representation & \# from SU(8) multiplet decomposition & \# found after transition \\ \hline
$({\bf 5},{\bf 1})$ & $ -\frac{9}{2} c_1({\cal L})^2 -2$  & $ -\frac{9}{2} c_1({\cal L})^2 -2$ \\
$({\bf 5},{\bf 3})$ & $ - 2 c_1({\cal L})^2 - 2$  & $- 2 c_1({\cal L})^2 - 2$ \\
$({\bf 5},\overline{{\bf 3}})$ & $   - \frac{1}{2} c_1({\cal L})^2 - 2$& $ - \frac{1}{2} c_1({\cal L})^2 - 4$\\
$({\bf 10},{\bf 1})$ & $- \frac{5}{2}c_1({\cal L})^2 - 4$ &  $- \frac{5}{2}c_1({\cal L})^2 - 4$\\ 
$(\overline{{\bf 10}},\overline{{\bf 3}})$ & $- \frac{1}{2} c_1({\cal L})^2 - 2$ &  $- \frac{1}{2} c_1({\cal L})^2 - 2$\\
$({\bf 1},{\bf 1})$ & $- c_1({\cal L})^2 -4$&  $ - c_1({\cal L})^2-6$\\ 
$({\bf 1},\overline{{\bf 3}})$ & $- \frac{13}{2}c_1({\cal L})^2 -4$&  $- \frac{13}{2} c_1({\cal L})^2-4$\\ 
\hline
\end{tabular*}
\caption{{\it Matter content after Higgsing an SU(8) to an ${\rm SU}(5)\times {\rm SU}(3)$ theory, both via a naive decomposition of the initial SU(8) multiplets and via a direct computation from the resulting $S(U(2) \times U(1))$ bundle.}}
\label{transtab1111}
\end{center}
\end{table}

As in all of the previous cases, this result is in perfect agreement with the field theory expectations given in Section \ref{antihigs}.

\vspace{0.1cm}

Similar results are found by Higgsing SU(6) and SU(7) on their three-index antisymmetric representations. The details of these computations can be found in Appendix \ref{hetHiggsapp}. Here we content ourselves with a presentation of the relevant bundle deformations in Table \ref{tab4trans}.

\begin{table}[!h]
\begin{center}
\begin{tabular*}{15.3cm}{|c|c|c|}
\hline
Group transition & Bundle transition & Fields Gaining VEV \\ \hline \hline
${\rm SU}(6) \to SU(3) \times {\rm SU}(3)$ &$\begin{array}{c} {\cal V}_{SU(2)} \oplus {\cal V}_{SU(3)} \to \tilde{{\cal V}}'_{SU(3)} \oplus \tilde{{\cal V}}_{SU(3)} \\ \textnormal{where}\;\; 0\to {\cal V}_{SU(2)} \to {\cal V}'_{SU(3)} \to {\cal O} \to 0 \end{array}$& $\begin{array}{c} H^1({\cal V}_{SU(2)}) \end{array}$\\ \hline
${\rm SU}(7) \to SU(4) \times {\rm SU}(3) $ &$\begin{array}{c} {\cal V}_{U(2)} \oplus {\cal L} \to \tilde{{\cal V}}'_{U(2)} \oplus {\cal V}_{U(2)} \\ \textnormal{where} \;\; 0\to {\cal L} \to {\cal V}'_{U(2)} \to {\cal O} \to 0 \end{array}$& $\begin{array}{c} H^1({\cal L} ) \end{array}$ \\ \hline
${\rm SU}(8) \to SU(5) \times {\rm SU}(3) $ &$\begin{array}{c} {\cal L} \oplus {\cal L}^{\vee} \to \tilde{{\cal V}}_{U(2)} \oplus {\cal L}^{\vee}\\ \textnormal{where} \;\; 0 \to {\cal L}\to {\cal V}_{U(2)} \to {\cal O} \to 0  \end{array}$& $H^1({\cal L})$\\\hline
\end{tabular*}
\caption{{\it Higgsing on the three-index antisymmetric representation in various heterotic theories, and the resulting deformation of the gauge bundle. The tildes over some bundles in the second column indicate a Li-Yau type deformation of the untilded object and its dual \cite{Li:2004hx}.}}
\label{tab4trans}
\end{center}
\end{table}

\vspace{0.2cm}

It should be noted that, due to the lack of ${\bf 70}$ (quadruple
antisymmetric) representations in the perturbative heterotic spectrum,
one can not break to ${\rm SU}(4)\times {\rm SU}(4) \times {\rm U}(1)$ in
this context and so our analysis of the possible Higgsing transitions
terminates here.  Once again, it is this boundary case of what is
possible in perturbative heterotic theory that fails to reproduce the
most general possibility from a field theory perspective. 

\subsection{Small Instanton transitions and Higgsing} \label{sec:smallinstHiggs}

It is interesting to note that the processes of Higgsing and undergoing small instanton transitions need not commute. We illustrate this here with the case of small instanton transitions before and after Higgsing SU(7) to SU(6).

As described in Section \ref{hettransitions}, before the Higgsing a minimal small instanton transition results in the following change in spectrum.
\begin{eqnarray} \label{mrtrans1}
1\times ({\bf 35}+ \overline{{\bf 35}}) + 5\times ({\bf 7}+ \overline{{\bf 7}}) \leftrightarrow 3 \times ({\bf 21}+\overline{{\bf 21}}) + 14 \times ({\bf 1})
\end{eqnarray}
The group theory governing a Higgsing of SU(7) on the fundamental representation is as follows.
\begin{eqnarray} \label{groupbreak1}
{\rm SU}(7) &\to& {\rm SU}(6) \times {\rm U}(1) \\ \nonumber
{\bf 7} &=&{\bf 1}_{-6} + {\bf 6}_1 \\\nonumber
{\bf 21} &=& {\bf 6}_{-5} + {\bf 15}_2 \\\nonumber
{\bf 35} &=& {\bf 15}_{-4} + {\bf 20}_{3}
\end{eqnarray}
Application of these branching rules to the transition
(\ref{mrtrans1})  results in the following SU(6)
transition
\begin{eqnarray} \label{double}
2 \times ({\bf 20}) +2 \times ({\bf 6} + \overline{{\bf 6}}) \leftrightarrow 4 \times ({\bf 1}) + 2 \times ({\bf 15} + \overline{{\bf 15}})
\end{eqnarray}

Purely from the point of view of an SU(6) bundle as in Section
\ref{su6het}, after Higgsing, we know that one can have the
following small instanton transition
\begin{eqnarray} \label{single}
  ({\bf 20}) +  ({\bf 6} + \overline{{\bf 6}}) \leftrightarrow 2 \times ({\bf 1}) +  ({\bf 15} + \overline{{\bf 15}})
\end{eqnarray}
which is more minimal than the one in (\ref{double}) above. 

If an SU(6) theory is obtained by Higgsing SU(7), however, then the bundle of structure group ${\rm SU}(2)$ will have the special form described in Table \ref{tab2trans}:
\begin{eqnarray} \label{mrtype}
0 \to {\cal L}^{\vee} \to {\cal V}_{SU(2)} \to {\cal L} \to 0 \;.
\end{eqnarray}
Such a bundle has $c_1({\cal V}_{SU(2)}) = -c_1({\cal L})^2$. The quantity $c_1({\cal L})^2$ for line bundles on $K3$ is always even and thus small instanton transitions involving an exchange of second Chern class between ${\rm SU}(3)$ and ${\rm SU}(2)$ bundles of the form (\ref{mrtype}) always results in a non-minimal transition of the type given in (\ref{double}) in the SU(6) theory.

After an SU(7) Higgsing, a subsequent small instanton transition in
the SU($6$) theory can change the bundle associated to the SU(6)
theory such that its ${\rm SU}(2)$ valued component is not of the form
(\ref{mrtype}). Such a transition could be of the more minimal type
(\ref{single}) and the resulting SU(6) theory then could not be
obtained by Higgsing any SU(7) model.


\section{Heterotic/F-theory Duality}\label{sec:matching}
The solutions presented in previous sections provide an interesting playground for heterotic/F-theory duality since they correspond to generically \emph{reducible} vector bundles in the heterotic theory. Such reducible vector bundles give rise to many interesting features not yet fully explored in even the $6$-dimensional duality, including small instanton transitions on a single $E_8$ fixed plane, the intricate intersection structure of reducible spectral covers and the presence of generically massive Green-Schwarz U(1) symmetries.

In this section we  consider the geometry of heterotic/F-theory dual pairs to be as follows \cite{Vafa-F-theory,Morrison-Vafa-I,Morrison-Vafa-II}:
\beq\label{theduals}
\text{Heterotic on}~~\pi_{h}: X_n \stackrel{\mathbb{E}}{\longrightarrow} B_{n-1}~~\Leftrightarrow ~~\text{F-theory on}~~ \pi_{f}: Y_{n+1} \stackrel{K3}{\longrightarrow} B_{n-1}
\eeq
where $X_n$ is elliptically fibered over $B_{n-1}$ and the K3-fibered manifold $Y_{n+1}$ admits a more
detailed description as an elliptically-fibered Calabi-Yau $(n+1)$-fold with section over a base ${\cal B}_n$ which is itself $\mathbb{P}^1$ fibered over $B_{n-1}$.

\subsection{The stable degeneration limit}\label{stab_degen}
To begin, we review briefly the standard arguments of heterotic/F-theory duality \cite{Morrison-Vafa-I,Morrison-Vafa-II,Bershadsky-all,Friedman:1997yq}. As discussed in the Introduction, in the case that the heterotic geometry is elliptically fibered and the F-theory geometry is $K3$ fibered, there exists a weakly coupled limit of both theories.

As is well-known in the literature (see \cite{Friedman:1997yq,Curio:1998bva,Anderson:2014gla} for reviews), this limit in parameter space corresponds to the large volume and weak coupling regime in the heterotic theory and is realized geometrically in the F-theory geometry via the following log semi-stable degeneration of the Calabi-Yau manifold, $Y$:
\beq\label{fib_prod}
Y_{n+1} \longrightarrow Y_{n+1}^{(1)} \cup_{D} Y_{n+1}^{(2)}
\eeq
where $Y^{(1)}$ and $Y^{(2)}$ are non-CY, $dP_9$ fibered $(n+1)$-folds, glued along a common divisor $D$ \cite{Aspinwall:1998bw, Curio:1998bva,Donagi:2012ts}. In the case of heterotic duality, $D=X_{n}$ is a CY variety of one lower dimension than $Y_{n+1}$ and forms the background of the heterotic geometry.

Now, the heterotic/F-theory dictionary says that if the $E_8 \times E_8$ heterotic theory is compactified on $D=X_{n}$ with vector bundles ${\cal V}_1, {\cal V}_2$ over $X_{n}$, the spacetime symmetries and matter spectrum should should match that of F-theory compactified on $Y^{(1)}\cup_D Y^{(2)}$. That is, for singularities leading to symmetries $G_i$ on $Y^{(i)}$ ($i=1,2$), we expect structure groups $H_i$ for ${\cal V}_i$ where $G_i \subset E_8$ is the commutant of $H_i$. Moreover, the full degrees of freedom of the theory can be counted and found to match (see Table \ref{dualtable} for a schematic review of the matching of the geometric degrees of freedom in the $6$-dimensional effective theories).
\begin{table}[ht]
\begin{align}
&\text{Het/Bundle} & &\text{Het/Spec. Cov.}& & \text{F-theory}&  \nonumber \\ \hline
&H^1(End_0({\cal V}_i))& &  Def(\mathcal{S})& &H^{2,1}(Y^{(i)}) & \nonumber \\
& & & Jac(\mathcal{S})& &H^{3}(Y^{(i)}, \mathbb{R}/\mathbb{Z}) & \nonumber \\
& && \text{Discrete data of}~\mathcal{L}_{\mathcal{S}} & & H^{2,2}(Y^{(i)},\mathbb{Z}) & \nonumber
\end{align}
\caption{A schematic matching of the duality in $6$-dimensions:
  heterotic vector bundle moduli, encoded as spectral data
  $(\mathcal{S}, \mathcal{L}_{\mathcal S})$ are matched to geometric
  moduli of the (resolved) F-theory $dP_9$-fibered geometry in the
  stable degeneration limit \cite{Friedman:1997yq, Aspinwall:1998bw,
    Curio:1998bva}.\label{dualtable}} 
\end{table}

Practically, to take the stable degeneration limit of the F-theory
geometry we must consider scaling the coefficients of $f,g$ such that
the $(n + 1)$-dimensional F-theory $K3$ fibration degenerates into a
fiber product of $dP_9$s. In terms of the Weierstrass model itself, if
$\sigma=0$ and $\sigma= \infty$ are chosen to define the loci where
the symmetries arise in each $E_8$ factor of the heterotic dual ({\it
  i.e.}, the two sections defining the base
$B_n$ of the $\mathbb{P}^1$ fibration), then we have
\beq
f\sim \sum_{i=0}^{8} f_i \sigma^i~~~~~,~~~~g \sim  \sum_{j=0}^{12} g_j \sigma^j
\eeq
and in the stable degeneration limit, it is possible to choose a scaling in which
\begin{align}\label{epsilons}
& f_i ~~\text{scales as}~~ \epsilon^{(i-4)} \\
& g_j ~~\text{scales as} ~~\epsilon^{(j-6)}~~. \nonumber
\end{align}
In the limit that $\epsilon \to 0$, it is clear that the zero locus of the discriminant $\Delta=4f^3+27g^2$ can be divided in "half" with nonabelian symmetry potentially present on each pole ($\sigma=0$ and $\sigma= \infty$) of the $\mathbb{P}^1$ fiber. In particular, the limit $\epsilon \to 0$ divides the $K3$ fiber into two rationally elliptically fibered surfaces ({\it i.e.} $dP_9$ surfaces) \cite{Friedman:1997yq, Aspinwall:1998bw}. This well-known limit is straightforward to apply for all Weierstrass models with ${\rm SU}(N)$ symmetry with $N\leq 6$. In these cases, if the complex structure is tuned to induce an ${\rm SU}(5)$ symmetry on the $\sigma=0$ locus for example, the stable degeneration limit isolates the ${\rm SU}(5) \subset E_8$ inside a single $E_8$ factor and effectively ``separates'' all $f_i$ with $i>4$ and $g_j$ with $j>6$ which correspond to the second $E_8$ factor (fully broken in this case). However, for ${\rm SU}(7), {\rm SU}(8)$ as we will see below, this limit can be somewhat subtle to take, since some $f_i, g_j$ with $i \leq 4, j \leq 6$ are determined in terms of $f_{i+n},g_{j+m}$ for some $n, m$ such that $i+n \geq 4, j+m \geq 6$ (See for example the Weierstrass coefficients in \eqref{eq:fsu7} and \eqref{eq:gsu7} in Section \ref{subsec:ftheorytunings}). In these cases, care must be taken with the powers of $\epsilon$ in each term. We will return to this issue in the following sections.

For now, we will begin our investigation of heterotic/F-theory dual solutions, as well as the ``Higgsing chains'' linking them, by reviewing a particular representation of the geometry of principal bundles in heterotic compactifications --  the so-called spectral cover construction \cite{fm_book,Friedman:1997ih}.

\subsection{A very brief review of the spectral cover construction}
In order to match the degrees of freedom in the heterotic geometry
$(X_n, \pi: {\cal V} \to X_n)$ with that in the F-theory
$(n+1)$-fold ({\it i.e.} $Y_{n+1}$ in \eqref{theduals}), it is
necessary to present the data of the heterotic bundle as a spectral
cover \cite{fm_book,Friedman:1997ih,Donagi:1997dp} or more generally,
a cameral cover \cite{Donagi:1997dp,donagi_cam,Donagi:1998vw}. For a
review of this bundle construction see \cite{Friedman:1997ih}. For
now, we will focus on the simple case of ${\rm SU}(N)$ structure
groups and spectral covers. It will suffice to recall that under the
Fourier-Mukai transform \cite{fm_book}, a vector bundle with structure
group ${\rm SU}(N)$ is equivalent to a pair $({\cal S}, {\cal L}_{\cal
  S})$ where ${\cal S}$ is a divisor in the elliptically fibered
heterotic CY geometry and ${\cal L}_{\cal S}$ is a rank $1$ sheaf over
${\cal S}$. The heterotic fibration can be presented in Weierstrass
form as \beq\label{het_wei} Y^2=X^3 +f_4 X +g_6~~.  \eeq Then, the
divisor ${\cal S}$ is called the ``spectral cover'' and can be
represented as the zero locus of a polynomial constraint of the form
\beq\label{spec_co} a_0 Z^N +a_2 XZ^{N-2} + a_3 YZ^{N-3} + \ldots =0
\eeq ending in $a_NX^{\frac{N}{2}}$ for $N$ even and $a_N
X^{\frac{N-3}{2}}Y$ for N odd \cite{Friedman:1997yq}. The coefficients
$a_j$ are sections of line bundles over the base $B$, of $\pi: X_n
\stackrel{\mathbb{E}}{\longrightarrow} B$, given by \beq a_j \in
H^0(B, K_{B}^{\otimes j} \otimes {\cal O}(\eta)) \eeq where in
$6$-dimensional compactifications, $\eta=c_2(V)$ and in
$4$-dimensional compactifications $c_2(V)=S_0 \wedge \eta +
\zeta^{2,2}$ with $S_0$ the form dual the zero section and
$\zeta^{2,2}$ the pullback of a $\{2,2\}$ form on $B$. The class of
${\cal S}$ is determined to be $[{\cal S}]=N[S_0]+ \pi^{*}(\eta)$. The
class of the rank one sheaf on the spectral cover ${\cal L}_{\cal S}$
is also determined entirely by the topology of ${\cal V}$ (see, for
example, \cite{Anderson:2014gla} for a review). With this in hand we
turn now to the matter transitions explored in Sections
\ref{sec:heterotic}, viewed now through the lens of the spectral cover
construction.

\subsection{Matter transitions in spectral covers} \label{sec:spectransition} 
To begin, let us consider briefly the form that the matter transitions
described in Sections \ref{sec:F-theory}-\ref{sec:heterotic} will take
in the context of heterotic/F-theory duality. Examples of other
related transitions have been explored in the heterotic literature
(see \cite{Ovrut:2000qi,Curio:1998vu} for representative examples). As
described in those works and in Section \ref{sec:heterotic}, small
instanton transitions on a single $E_8$ fixed plane can only arise
when the vector bundle geometry -- an $H$-principal bundle on $X_n$
with $H \subset E_8$ -- is reducible. That is, when $H=H_1 \otimes
H_2$ and the associated vector bundle decomposes as a direct sum
${\cal V}={\cal V}_1 \oplus {\cal V}_2$. In this case, the small
instanton transition takes the form \beq\label{spec_transi} {\cal V}_1
\oplus {\cal V}_2 \longrightarrow {\cal V'}_1 \oplus {\cal V}_2 \oplus
       {\cal I} \longrightarrow {\cal V'}_1 \oplus {\cal V'}_2 \eeq
       where ${\cal I}$ is a sky-scraper sheaf supported on the
       codimension 2 locus wrapped by $M5/N5$-branes in $X_n$. In the
       language of spectral covers then, we consider the simplest case
       where ${\cal V}_1, {\cal V}_2$ are both ${\rm SU}(N)$ bundles
       for some $N$. In particular, the situation in which instanton
       number is removed from ${\cal V}_1$ and then added to ${\cal
         V}_2$ can be summarized as \beq\label{spec_inst} ({\cal
         S}_1)({\cal S}_2)=0 \longrightarrow a(w) ({\cal S}'_1)({\cal
         S}_2) = 0 \longrightarrow ({\cal S}'_1)({\cal S}'_2)=0 \eeq where
       $w$ represents coordinates on the base $B$ of the heterotic
       elliptic fibration. Here $a(w)$ is a so-called ``vertical
       component" of the spectral cover -- corresponding to a small
       instanton in the heterotic theory \cite{Witten:1995gx}. The
       function $a(w)$ corresponds exactly to the function (also
       called "$a$") describing the F-theory matter transitions given
       in Section \ref{ftrans}.

As a concrete example, consider an $SO(12)$ theory. The commutant of this symmetry within $E_8$ is ${\rm SU}(2) \times {\rm SU}(2)$ which leads to a heterotic bundle geometry consisting of a sum of two ${\rm SU}(2)$ vector bundles ${\cal V}_1 \oplus {\cal V}_2$. For such reducible vector bundle, the spectral cover takes the form of a product
\beq\label{su2su2_eg}
{\cal S}_{{\cal V}}=(a_0 Z^2 +a_2 X)(b_0 Z^2+b_2X)=0
\eeq
In the $SO(12)$ theory, the matter consists of localized ${\bf 32}$ and ${\bf 32'}$ multiplets, located at the zeros of $a_2$ and $b_2$, respectively, and fundamental matter (${\bf 12}$s) associated with the intersection ${\cal S}_1 \cap {\cal S}_2$ of the two ${\rm SU}(2)$ components of the spectral cover. The simplest transitions then take the form of separating out a common factor of (for example) $a_2,a_0$ and then ``absorbing'' it into $b_0,b_2$. 

To illustrate this more concretely, consider $6$-dimensional heterotic/F-theory geometry with the spectral covers given above corresponding to curves inside a $K3$ surface. Since the minimal second Chern class for an ${\rm SU}(2)$ bundle on $K3$ is $c_2({\cal V})=4$, we will parameterize the topology of the pair ${\cal V}_1,{\cal V}_2$ as $c_2({\cal V}_1)=4+r$, $c_2({\cal V}_2)=8+n-r$ \cite{Bershadsky-all}. It follows that the degrees of the functions in \eref{su2su2_eg} are
\begin{align}
&a_0 \sim 4+r & b_0 \sim 8+n-r  & \\
&a_2 \sim r& b_2 \sim 4+n-r \\
\end{align}
With this in mind, we can denote an arbitrary function of degree $k$
on $\mathbb{P}^1$ via the subscript $f_k$. Then, as in
\eref{spec_inst}, a transition removing a small instanton from ${\cal
  V}_1$ and merging it into the bundle ${\cal V}_2$ takes the form
\begin{align}\label{spec_12_trans}
&(e_{4+r} Z^2 +f_{r} X)(g_{8+n-r} Z^2+h_{4+n-r}X) \nonumber \\
& \longrightarrow (a_{p})(e'_{4+r-p} Z^2 +f'_{r-p} X)(g_{8+n-r} Z^2+h_{4+n-r}X) \nonumber \\
&\longrightarrow (e'_{4+r-p} Z^2 +f'_{r-p} X)(g'_{8+n-r+p} Z^2+h'_{4+n-r+p}X)
\end{align}
where $p \leq r$. Note that in the process of realizing the transition in the spectral cover ((tuning a common factor in ${\cal S}_1$) $\to$ (identifying it as an overall factor of ${\cal S}_2$ )$\to$ (deforming ${\cal S}_2$ away from factorized form)), the topology of the bundles ${\cal V}_1$ and ${\cal V}_2$ has changed since $c_2({\cal V'}_1)=4+r-p$ and $c_2({\cal V'}_2)=8+n-r+p$. In this case the spectrum changes only between the ${\bf 32}$ and ${\bf 32}'$ fields\footnote{Note that the number of ${\bf 12}$'s is unchanged in this correspondence since it is determined solely by the geometric intersection number $[{\cal S}_1] \cdot [{\cal S}_2] \subset K3$. For the geometry above this gives $2(8+n)$ points which is independent of the transition given in \eref{spec_12_trans}.}:
\beq
(r)~\frac{1}{2} {\bf 32}'s + (4+n-r)~\frac{1}{2} {\bf 32's} \to (r-p)~\frac{1}{2} {\bf 32}'s + (4+n-r+p)~\frac{1}{2} {\bf 32's}
\eeq

As a concluding observation, it should be noted that while we
illustrated this example in $6$ dimensions for simplicity, entirely
analogous structure exists in $4$ dimensions. There the spectral covers
are complex surfaces inside Calabi-Yau $3$ folds and once again,
co-dimension two components (here curves) can be separated off one
component and the deformed smoothly into the other. One remarkable
difference in the $4$-dimensional theory however is that in the ${\cal
  N}=1$ theory, such transitions can also be chirality changing
\cite{Ovrut:2000qi,Curio:1998vu}. We will return to this point
later. For now, we continue towards understanding the basic structure of the
transitions described in Sections
\ref{sec:F-theory}-\ref{sec:heterotic}, and the Higgsing chains
linking them, beginning with the SU(6) theory.

\subsection{SU(6) heterotic/F-theory dictionary}
As described in Section \ref{sec:su6fth}, and equations \eqref{eq:tuning-f-6} and \eqref{eq:gsu6}, the SU(6) F-theory geometry is parameterized by 5 functions
\beq
\phi_2, ~\alpha,~\beta,~\nu~,\lambda \label{ffunc}
\eeq
associated, respectively with the divisors
\beq
-2K+L, -K+N-L, L, -2K+N-L, -4K+N-L
\eeq
The matching of the heterotic and F-theory effective physics can be made readily in both $6$- and $4$-dimensions. For concreteness, in the following paragraphs we will make the explicit correspondence between the degrees of freedom in the $6$-dimensional theory, but it should be kept in mind that the general structure is equally applicable to $4$-dimensions.

From Section \ref{su6het}, it is clear that the heterotic dual SU(6)
theory on $K3$ has commutant ${\rm SU}(2)\times {\rm SU}(3)$ inside
$E_8$. The bundle ${\cal V}$, with $c_2(V)=12+n$ can be denoted as
\beq\label{Redbun}
{\cal V}={\cal V}_2 \oplus {\cal V}_3
\eeq
As above, let ${\cal V}_2$ have structure group ${\rm SU}(2)$ and $c_2({\cal V}_2)=4+r$, while ${\cal V}_3$ has structure group ${\rm SU}(3)$ and $c_2({\cal V}_3)=8+n-r$. 

Using the spectral cover construction as described in the previous
section, the vector bundle ${\cal V}$ in \eqref{Redbun} can be
described as a reducible spectral cover, ${\cal S}_{V}={\cal S}_{1}
\cup {\cal S}_{2}$ of the form given in \eref{spec_co}, inside the
elliptically fibered $K3$ surface. Explicitly, \beq\label{su6specs}
{\cal S}_{V}=(a_0 Z^2 +a_2 X)(b_0 Z^3+b_2XZ+b_3 Y)=0 \eeq where the
degrees of the functions $a_i$ and $b_j$ (over the $\mathbb{P}^1$
base) are
\begin{align}
&a_0 \sim 4+r & b_0 \sim 8+n-r  & \\
&a_2 \sim r& b_2 \sim 4+n-r \\
&& b_3 \sim  2+n-r 
\end{align}
A key to matching these 5 functions with those in (\ref{ffunc}) is to note that the ${\bf 20}$'s of ${\rm SU}(6$) are located at the zeros of $a_2$ and the ${\bf 15}$'s at the zeros of $b_3$. It follows by inspection of \eqref{eq:tuning-f-6} and \eqref{eq:gsu6} then, that the parameter matching takes the form
\begin{equation}
\begin{array}{|c|c|c|c|c|c|}
\hline
& a_0 & a_2 &b_0 & b_2 & b_3\\ \hline
\sim &\phi_2 & \beta & \lambda & \nu & \alpha \\ \hline
\end{array}\label{dofmatch} 
\end{equation}
This matches with the association made in Table 
\ref{tab:su6dofs}.

Re-writing the spectral cover in this notation we have:
\beq\label{su6spec} S_{V}=(\phi_2Z^2 +\beta X)(\lambda Z^3+\nu
XZ+\alpha Y)=0 \eeq It is important to recall here that the matching
in (\ref{dofmatch}) is only
up to proportionality. There may be
constants/normalization in this matching that could be significant in
understanding the dual pair; we return to this topic below.

Finally, as observed above, it can be verified that the charged matter spectrum of the SU(6) theory is readily understood in terms of the spectral cover as the loci where $\alpha=0$ (${\bf 15}$ representations), $\beta=0$ (${\bf 20}$s)and the points where $S_1 \cap S_2$ (${\bf 6}$s). The matter transitions corresponding to \eref{spec_transi} in this case exactly match those found in the F-theory geometry given in Section \ref{sec:F-theory}. Recall, from Section \ref{ftrans} the transition was realized via
\begin{align}
\alpha &\rightarrow a \alpha', \nonumber\\
\nu &\rightarrow a \nu',\\
\lambda &\rightarrow a \lambda'. \nonumber
\end{align}
which corresponds to the separation of a small instanton from the ${\rm SU}(3)$ bundle ${\cal V}_3$ in \eqref{su6spec}. This was followed by a deformation which deformed the instanton into a smooth ${\rm SU}(2)$ bundle ${\cal V}'_2$:
\begin{align}
a\beta &\rightarrow \beta',\\
a\phi_2 &\rightarrow \phi_2'. \nonumber
\end{align}
exactly as expected from \eqref{su6spec}.
Given this exact matching between the parameter spaces of the
heterotic and F-theory descriptions, we can clearly follow the
heterotic transition just as the F-theory transition along a
one-parameter family of theories with the parameter $\hat{\varepsilon} =
0$ at the superconformal point and taking positive and negative signs
for the theories with two different matter contents.

\subsection{SU(7) tuning}
In this section we  explore heterotic/F-theory dual pairs in the SU(7) theory. Unlike the case of SU(6) given above, here the spectral cover is not a simple factorization of the form \eqref{su2su2_eg} and \eref{su6specs}, but a more interesting and complex object. To begin, we will compare the tunings/symmetry enhancements described in Sections \ref{sec:F-theory} and \ref{sec:heterotic} for SU(7) gauge symmetry. Recall that from \eqref{eq:fsu7} and \eqref{eq:gsu7}, the tuning taking ${\rm SU}(6) \rightarrow SU(7)$ is given by
\begin{align}
&\phi_2 = 3 {\zeta_1}^2 + \omega \delta^2 \label{su7}\\
& \beta=\delta^2 \nonumber \\
& \lambda=\frac{1}{3}{\zeta_1}^2\zeta_2 - \frac{1}{18}\zeta_1 \xi \omega+ \frac{1}{9}\delta^2 \zeta_2 \omega + \lambda_1\delta^2 \xi \nonumber \\
& \nu= \zeta_2 \delta^2+ \zeta_1 \xi \nonumber \\
&  \alpha=\delta \xi \nonumber
\end{align}
and the ``middle'' coefficients that will be related (in the stable degeneration limit) to the heterotic Weierstrass model, \eqref{het_wei} take the form:
\begin{align}\label{ffourgsix}
&f_4=-6 \zeta_1 \lambda_1 -\frac{1}{12} \omega^2 + \psi_4 \delta^2 \\
&g_6=\frac{1}{108} \left(-36 f_5 \delta^4 \zeta_2+972 \delta^2 {\lambda_1}^2-36 f_5 \delta^2 \zeta_1 \xi-9 f_6 \delta^6 \xi^2-108 \zeta_1^2 \psi_4-108 \zeta_1 \lambda_1 \omega-36 \delta^2 \psi_4 \omega-\omega^3 \right) \nonumber 
\end{align}

The key question that must be addressed is whether or not this choice of complex structure corresponds in the heterotic theory to the deformation ({\it i.e.} splitting) of vector bundles required by this symmetry enhancement described in Section \ref{su7het}?

First, it is useful to briefly review the enhancement of symmetry from the point of view of vector bundle geometry in the heterotic theory. As in \eqref{Redbun}, the reducible bundle with structure group ${\rm SU}(2) \times {\rm SU}(3)$ leading to commutant ${\rm SU}(6) \subset E_8$  is
\beq
{\cal V}={\cal V}_2 \oplus {\cal V}_3
\eeq
From Section \ref{sec:heterotic}, we recall that in the transition from ${\rm SU}(6) \to SU(7)$ the heterotic bundle structure group must reduce from ${\rm SU}(2) \times {\rm SU}(3)$ to $S[U(2) \times U(1)]$ via the ``splitting'' of the bundles ${\cal V}_2, {\cal V}_3$:
\begin{align}
& {\cal V}_2 \to {\cal L} \oplus {\cal L}^{\vee} \label{su7bun1}\\
& {\cal V}_3 \to {{\cal L}^{\vee}} \oplus {\cal U}_2 \label{su7bun2}
\end{align}
where ${\cal L}$ is a line bundle and $c_1({\cal U}_2)=c_1({\cal L})$ ({\it i.e.} ${\cal U}_2$ is a $U(2)$ bundle). The commutant of ${\rm SU}(7) \subset E_8$ is ${\rm SU}(2) \times U(1)$ and the underlying (fundamental) bundle geometry is ${{\cal L}^{\vee}} \oplus {\cal U}_2$ (the decomposition of ${\cal V}_2$ above is simply auxiliary information in this case).

To match the tuning in the heterotic and F-theory descriptions, the first step is to consider how such a decomposition is manifest when the bundle ${\cal V}={\cal V}_2 \oplus {\cal V}_3$ is described via a spectral cover as in \eqref{su6spec}. Can we match this decomposition to the enhancement given in \eqref{su7}? For a bundle described via a smooth spectral cover\footnote{{\it i.e.} For so-called ``regular" bundles \cite{Friedman:1997ih,lazaroiu}.}, decomposition of the vector bundle into a direct sum usually corresponds to factorization of the spectral cover ({\it i.e.} the spectral cover becomes reducible). In the case above, it is clear from \eqref{su7bun1} that the ${\rm SU}(2)$ portion of the bundle must split into a sum of two line bundles. As it turns out, this decomposition into Abelian components is a particularly subtle process from the point of view of a spectral cover description \cite{Friedman:1997ih,Friedman:1997yq,Choi:2012pr}. A well-behaved ({\it i.e.} smooth) spectral cover (corresponding to a rank $N$ vector bundle) intersects each fiber at exactly $N$ points. Thus, by definition, a smooth spectral cover associated to a line bundle must be a 1-sheeted cover of the base, intersecting each fiber exactly once. This however, is a familiar object in the fibration geometry: such a 1-sheeted cover is in fact \emph{a section of the elliptic fibration}. It is important to note that it is \emph{not} the case that any line bundle produces a section of the fibration under Fourier-Mukai transform. Instead, generic line bundles lead to singular/vertical spectral covers\footnote{In some cases these can correspond to so-called ``T-brane" solutions \cite{Anderson:2013rka}. See also Section 7.9 of \cite{Anderson:2014gla}.}. However, it is the case that any smooth 1-sheeted cover of the base (describing a line bundle) is also a section to the elliptic fibration.

Returning again to the expected geometry, it is clear from
\eqref{su7bun1} that upon tuning ${\rm SU}(6) \to SU(7)$  the
${\rm SU}(2)$ bundle should decompose as a sum of a line bundle and
its dual. As a result, if such a bundle can be described as a smooth
spectral cover, this should in turn \emph{correspond to an extra
  section appearing in the $K3$ geometry}. In this vein, a sum of the
form (${\cal L} \oplus {\cal L}^{\vee}$) should correspond to a
section and the ``inverse'' of that section (i.e, two sections leading
to marked points on each elliptic fiber with coordinates $[X',Y',Z]$
and $[X'',Y'',-Z]$ which sum to the zero section under the addition
law of the elliptic curve).

\subsubsection{The heterotic $K3$ geometry}\label{su7k3}
As described above, an SU(7) gauge symmetry must correspond to the reduction of an ${\rm SU}(2)$ bundle structure group to $S[U(1) \times U(1)]$, which after Fourier-Mukai transform will correspond to 1-sheeted spectral covers of the base $\mathbb{P}^1 \subset K3$. As a result, an SU(7) symmetry in the heterotic theory does not allow the form of the spectral cover and that of the $K3$ surface to be considered separately. They are intrinsically correlated and a generic SU(7) tuned F-theory geometry (with a heterotic dual) should lead to an enhanced Mordell-Weil group in the dual $K3$ geometry. To investigate this expectation we first consider the form of the Weierstrass coefficients $f_4$ and $g_6$ from \eqref{ffourgsix}. If this $K3$ geometry has an additional rational section ({\it i.e.} Mordell-Weil rank $1$), then as derived in \cite{Morrison:2012ei} it must be possible to write it in the following general form (see eq.(B.19) of \cite{Morrison:2012ei}):
\beq\label{morrisonpark}
Y^2= X^3 +\left(c_1c_3 - b^2 c_0 -\frac{{c_2}^2}{3} \right)X Z^4 + \left(c_0 {c_3}^2 - \frac{1}{3}c_1c_2c_3 + \frac{2}{27}{c_2}^3-\frac{2}{3}b^2 c_0 c_2+\frac{b^2{c_1}^2}{4} \right)Z^6
\eeq
for some functions $c_i, b$.

However, before comparing to \eqref{ffourgsix}, care must be taken with the stable degeneration limit. In matching the degrees of freedom in the heterotic and F-theory geometries, it is necessary to take the limit described above in \eqref{fib_prod} and \eqref{epsilons} in which the F-theory geometry undergoes stable degeneration. In this case we hope to compare the coefficients of $f_i$ with $0 \leq i \leq 3$ and $g_j$ with $0 \leq j \leq 5$ with the data of the spectral cover in \eqref{su6spec} (subject to the tuning in \eqref{su7}) and \eqref{het_wei} with the coefficients of $f_4$ and $g_6$ given in \eqref{ffourgsix}. However, as described in Section \ref{stab_degen}, we must consider  the powers of $\epsilon$ present in each term in the Weierstrass model in the limit that $\epsilon \to 0$. In fact, for the SU(7) solution, $g_6$ \eqref{ffourgsix} has been tuned in terms of $f_5$ and $f_6$, both of which carry additional powers of $\epsilon$. As a result, this dependence must be taken into account in the $\epsilon$ limits, where the terms dependent on $f_5$ and $f_6$ vanish.  Here the stable degeneration limit leads to a modified form for the ``middle'' coefficients in \eqref{ffourgsix}:
\begin{align}
&f_4 \to -6 \zeta_1 \lambda_1 -\frac{1}{12} \omega^2 + \psi_4 \delta^2 \\
&g_6 \to \frac{1}{108} \left(972 \delta^2 {\lambda_1}^2-108 \zeta_1^2 \psi_4-108 \zeta_1 \lambda_1 \omega-36 \delta^2 \psi_4 \omega-\omega^3 \right)\label{ffourgsix_sd}
\end{align}
It is these values that we must take as defining the heterotic base geometry \eqref{het_wei} in the stable degeneration limit. That is, this elliptically fibered $K3$ forms the divisor $D$ along which $Y_1$ and $Y_2$ are glued in \eqref{fib_prod}.

Remarkably, an inspection of \eqref{ffourgsix_sd} shows that it is of \emph{precisely the form required by \eqref{morrisonpark}}. The exact matching to the general two-section Weierstrass model is
\begin{align} \label{k3conds}
&c_0= \psi_4 \\
&c_1= (6 i) \lambda_1 \nonumber \\
&c_2= -(1/2) \omega \nonumber \\
&c_3=( i) \zeta_1 \nonumber \\
&b= \pm (i) \delta \nonumber
\end{align}

Choosing the sign identification of $b=- i\delta$, then the fiber coordinate of the new section in the $K3$ geometry is
\beq\label{newsection}
[X,Y,Z]=\left[-\frac{1}{3}\left(3 {\zeta_1}^2+\delta^2 \omega \right), i \left({\zeta_1}^3+\frac{1}{2} \zeta_1 \delta^2 \omega-3 \lambda_1 \delta^4\right), -i \delta \right]
\eeq

With these observations in hand, it is at last possible to compare the tuning given in \eqref{su7} to the spectral cover and the expected bundle geometry in \eqref{su7bun1} and \eqref{su7bun2}.

\subsubsection{SU(7) symmetry and the spectral cover}
We must begin by substituting the tuning (\ref{su7}) into the general
SU(6) spectral cover in (\ref{su6spec}). But first, it will be
convenient for to choose the constants of proportionality that were
left free in \eqref{dofmatch} through specific normalization (this
normalization choice is not physically significant, but will merely
serve to provide the clearest interpretation of the degrees of freedom
in the dual theories):
\begin{align}
&a_0 = \phi_2 \label{dofmatch_normalized} \\
&a_2  = -3\beta \nonumber \\
&b_0 =-3 \lambda \nonumber \\
&b_2  =  \nu \nonumber \\
&b_3 = -\alpha \nonumber
\end{align}
With these choices, the tuning of the complex structure in F-theory given by \eqref{su7} can be substituted into the heterotic spectral cover in \eqref{su6spec}, to yield a new spectral cover:
\begin{align}\label{su7spec1}
& \left( (3\zeta_1^2 + \delta^2 \omega)Z^2- 3 \delta^2 X) \right) \times \\
&\left(-3 \left(\frac{1}{3}{\zeta_1}^2\zeta_2 - \frac{1}{18}\zeta_1 \xi \omega+ \frac{1}{9}\delta^2 \zeta_2 \omega + \lambda_1\delta^2 \xi \right)Z^3 + (\zeta_2 \delta^2+ \zeta_1 \xi )XZ- \delta \xi Y \right)=0 \nonumber
\end{align}
Which can be re-written as
\begin{align}\label{rewrite}
&\left( (3\zeta_1^2 + \delta^2 \omega)Z^2- 3 \delta^2 X \right) \times \\
&\left(-\frac{\zeta_2 Z}{3}\left( (3 {\zeta_1}^2+\delta^2 \omega) Z^2 -3 \delta^2 X \right)+ \xi \left[-3(- \frac{1}{18}\zeta_1 \omega+\delta^2 \lambda_1)Z^3+ \zeta_1XZ - \delta Y \right] \right)=0 \nonumber
\end{align}
If the degrees of freedom have been paired correctly in the dual theories, this new spectral cover must be exactly the Fourier-Mukai transform of the reduced bundle geometry given in \eqref{su7bun1} and \eqref{su7bun2}.

To verify this, consider first the ${\rm SU}(2)$ component of the spectral cover. This vanishes on the locus
\beq\label{su2section}
X= \frac{1}{3\delta^2}(3\zeta_1^2 + \delta^2 \omega)Z^2
\eeq
But by inspection, this is exactly the constraint yielding the additional rational sections given in \eqref{newsection}! As expected, it is clear that over every point on the base, the two roots of the ${\rm SU}(2)$ spectral cover sweep out  precisely the new section and its ``inverse'' under the addition law of the elliptic curve (replacing $Z= - i\delta$ with $Z= + i \delta$). Now, since the new sections are ``horizontal'' in the K3 elliptic fibration, they intersect each fiber exactly once and at equal and opposite points on the elliptic curve relative to the zero section (located at $Z=0$). As a result, in the Fourier-Mukai transform this is exactly a pair of line bundles of the form
\beq
{\cal L} \oplus {{\cal L}^{\vee}}
\eeq
Thus, the tuning of \eqref{su7}, the enhanced rational sections of the
elliptically fibered $K3$ and the ${\rm SU}(2)$ component of the
spectral cover exactly match the expectation of the bundle
decomposition given in \eqref{su7bun1}. All that remains then, is to
examine the ${\rm SU}(3)$ component of the spectral cover.  

Not only must the ${\rm SU}(3)$ component also decompose, it is clear by inspection of \eqref{su7bun1} and \eqref{su7bun2} that one line bundle factor (${\cal L}^{\vee}$) is repeated in both the ${\rm SU}(2)$ and ${\rm SU}(3)$ components of the reducible bundle. In the spectral cover description, we have seen that the two roots of the ${\rm SU}(2)$ spectral cover have formed into a pair (corresponding to a line bundle and its dual) and now it must be verified that at least one of those roots has \emph{also become a root of the ${\rm SU}(3)$ spectral cover}. 

Inspecting \eqref{rewrite}, it is clear that the form of the ${\rm SU}(3)$ component on the locus where \eqref{su2section} is satisfied ({\it i.e.} the roots of the ${\rm SU}(2)$ factor) reduces to
\beq\label{su3remainder}
\xi \left[-3(- \frac{1}{18}\zeta_1 \omega+\delta^2 \lambda_1)Z^3+ \zeta_1XZ - \delta Y \right] =0
\eeq
We can now ask, does this vanish along either of the roots of
\eqref{su2section}? Substituting the new section of the $K3$
fibration \eqref{newsection} ({\it i.e.} the root with $Z$ coordinate
$Z=-i \delta$) into the remaining expression in \eqref{su3remainder},
we find that it vanishes identically. Thus, our expectations are fully
verified and a single ${\rm SU}(2)$ root is now overlapping with one
root of the ${\rm SU}(3)$ spectral cover. Moreover it is easy to
verify that this expression does not vanish on the other ${\rm SU}(2)$
root (with $Z=+ i \delta$) as required. The remaining two roots of the
${\rm SU}(3)$ component are distinct and correspond to the expected
rank $2$ bundle in the heterotic geometry given in \eqref{su7bun2}. 

Thus far, on this locus, the correspondence is perfect between the F-theory Weierstrass data, the heterotic bundle geometry, and the spectral cover description. There is only one remaining element to be considered and this is the presence of a Green-Schwarz massive U(1) in the heterotic effective theory.

From group theory alone, the reducible bundle ${\cal V}={\cal
  L}^{\vee} + {\cal U}_2$ with $c_1({\cal U}_2)=c_1({\cal L})$ given
in \eqref{su7bun2} has structure group ${\rm S}({\rm U}(2) \times {\rm U}(1))$ and
within $E_8$ this gives rise to a commutant ${\rm SU}(7) \times
{\rm U}(1)$. That is, at the level of group theory, any SU(7) symmetry
arising in a heterotic theory \emph{must} be accompanied by an
additional abelian gauge symmetry. Generically, by the Green-Schwarz
mechanism, this enhanced U(1) couples to the K\"ahler axions of the
base CY geometry (which transforms via shift symmetries) and becomes
massive (see
\cite{Dine:1987xk,Sharpe:1998zu,Lukas:1999nh,Blumenhagen:2005ga,Anderson:2009nt}). Since
the presence of Green-Schwarz massive U(1)s in F-theory has been the
topic of much recent interest (see for example 
\cite{Jockers:2005zy,Buican:2006sn,
Grimm:2010ez,Grimm:2011tb, Braun:2014nva,Anderson:2014yva,Garcia-Etxebarria:2014qua,Cvetic:2015uwu}), we turn
now to the appearance of this enhanced U(1) in heterotic/F-theory
pair described above. 

\subsubsection{SU(7) and Green-Schwarz massive ${\rm U}(1)$s}
As described in the previous subsection, the presence of an enhanced U(1) symmetry in the heterotic theory is unavoidable (by group theory within $E_8$). How then, are we to understand this U(1) from the point of view of F-theory? To explore this, consider the stable degeneration limit $ Y \to Y^{(1)} \cup_{D} Y^{(2)}$ as described in Section \ref{stab_degen} and \eqref{fib_prod}, and the Weierstrass model of $Y^{(1)}$ corresponding to the physics of a single $E_8$ factor of the heterotic effective theory. In this limit, the Weierstrass coefficients of $Y^{(1)}$ are given by
\begin{align}
f \sim & -\frac{\delta ^{12} \xi ^4}{48}+\sigma \left(-\frac{1}{6} \delta ^{10} \text{$\zeta_ 2$} \xi ^2-\frac{1}{6} \delta ^8 \text{$\zeta_1$} \xi ^3\right) \label{su7stabgen_f} \\
& +\frac{1}{6} \sigma^2 \left(-2 \delta ^8 \text{$\zeta_2$}^2-4 \delta ^6 \text{$\zeta_1$} \text{$\zeta_2$} \xi -\delta ^6 \xi ^2 \omega -5 \delta ^4 \text{$\zeta_1$}^2 \xi ^2\right) \nonumber\\
& +\sigma^3 \left(-\frac{2}{3} \delta ^4 \text{$\zeta_2$} \omega -3 \delta ^4 \text{$\lambda $1} \xi -2 \delta ^2 \text{$\zeta_1$}^2 \text{$\zeta_2$}-\frac{1}{6} \delta ^2 \text{$\zeta_1$} \xi  \omega -\text{$\zeta_1$}^3 \xi \right) \nonumber\\ 
& +\sigma^4 \left(\delta ^2 \text{$\psi_4 $}-6 \text{$\zeta_1$} \text{$\lambda _1$}-\frac{\omega ^2}{12}\right)\nonumber
\end{align}
and
\begin{align}
g \sim & \frac{\delta ^{18} \xi ^6}{864}+\frac{1}{72} \sigma \left(\delta ^{16}\zeta_2\xi ^4+\delta ^{14} {\zeta_1} \xi ^5\right) \label{su7stabgen_g} \\
& +\frac{1}{72} \sigma^2 \left(4 \delta ^{14} {\zeta_2}^2 \xi ^2+8 \delta ^{12} {\zeta_1}\zeta_2\xi ^3+\delta ^{12} \xi ^4 \omega +7 \delta ^{10} {\zeta_1}^2 \xi ^4\right) \nonumber\\
&+\frac{1}{216} \sigma^3 \left(16 \delta ^{12} {\zeta_2}^3+48 \delta ^{10} {\zeta_1} {\zeta_2}^2 \xi +24 \delta ^{10}\zeta_2\xi ^2 \omega +54 \delta ^{10} {\lambda_1} \xi ^3+120 \delta ^8 {\zeta_1}^2\zeta_2\xi ^2+15 \delta ^8 {\zeta_1} \xi ^3 \omega +70 \delta ^6 {\zeta_1}^3 \xi ^3\right) \nonumber\\
&+\frac{1}{144} \sigma^4 (32 \delta ^8 {\zeta_2}^2 \omega +144 \delta ^8\zeta_2{\lambda_1} \xi -12 \delta ^8 \xi ^2 {\psi_4}+96 \delta ^6 {\zeta_1}^2 {\zeta_2}^2+40 \delta ^6 {\zeta_1}\zeta_2\xi  \omega +216 \delta ^6 {\zeta_1} {\lambda_1} \xi ^2+5 \delta ^6 \xi ^2 \omega ^2 \nonumber\\
&+144 \delta ^4 {\zeta_1}^3\zeta_2\xi +32 \delta ^4 {\zeta_1}^2 \xi ^2 \omega +84 \delta ^2 {\zeta_1}^4 \xi ^2) \nonumber\\
&+\frac{1}{36} \sigma^5 (-12 \delta ^6\zeta_2{\psi_4}+72 \delta ^4 {\zeta_1}\zeta_2{\lambda_1}-12 \delta ^4 {\zeta_1} \xi  {\psi_4}+5 \delta ^4\zeta_2\omega ^2+36 \delta ^4 {\lambda_1} \xi  \omega +24 \delta ^2 {\zeta_1}^2\zeta_2\omega +180 \delta ^2 {\zeta_1}^2 {\lambda_1} \xi  \nonumber \\
&-\delta ^2 {\zeta_1} \xi  \omega ^2+36 {\zeta_1}^4 {\zeta_2}-6 {\zeta_1}^3 \xi  \omega) \nonumber\\
&+\frac{1}{108} \sigma^6 \left(972 \delta ^2 {\lambda_1}^2-36 \delta ^2 {\psi_4} \omega -108 {\zeta_1}^2 {\psi_4}-108 {\zeta_1} {\lambda_1} \omega -\omega ^3\right) \nonumber
\end{align}
where $\sigma=0$ defines a section of the $\mathbb{P}^1$ fiber of the
F-theory base geometry (recall $\sigma=0$ and $\sigma= \infty$ mark
the locations of the symmetry groups arising from each $E_8$ factor). 

By group theory alone, the U(1) accompanying the SU(7) in the
heterotic theory should be visible in this limit. How then do we see
it? Remarkably, we find that the U(1) is very much present in this
limit! Let us compare once more to the generic two-section Weierstrass
model given in \eqref{morrisonpark}.  We will see that in fact an
additional section, and hence U(1) symmetry has become visible in
the entire Weierstrass model of $Y^{(1)}$.

As in Section \ref{su7k3}, we must establish the dictionary that puts the coefficients given in \eqref{su7stabgen_f} and \eqref{su7stabgen_g} into the form necessary for a generic two-section model, \eqref{morrisonpark}. Here we find that this correspondence can be achieved by one of two choices. First,
\begin{align}
&c_0=-\frac{1}{4} \delta^6 \zeta_2^2 \sigma^2   + 
 (-\zeta_1^2 \zeta_2 - \frac{1}{2} \delta^2 \zeta_2 \omega) \sigma^3+  \psi_4 \sigma^4 \\
 &c_1= \frac{1}{2} i \delta^6 \zeta_2 \xi \sigma + 
  i( \delta^2 \zeta_1 \zeta_2 + \zeta_1^2 \xi +\frac{1}{2} \delta^2 \xi \omega)\sigma^2
     +6 i  \lambda_1 \sigma^3\\
    &c_2=\frac{1}{4}\delta^6 \xi^2 + 
 (-\frac{1}{2}\delta^4 \zeta_2 + \delta^2 \zeta_1 \xi)\sigma  - 
 \frac{1}{2} \omega \sigma^2 \\
 &c_3= \frac{i}{2} \delta^4 \xi+i \zeta_1 \sigma \\
 &b=-i\delta
\end{align}
The second choice for a variable change arises from the freedom associated to a Weierstrass model with Mordell-Weil rank $2$ ({\it i.e.} the freedom to define which, of the section $S_1 $ and it's inverse $-S_1$, we choose to call ``positive'' in the elliptic addition law). The other solution is
\begin{align}
&c_0=(-\frac{1}{4} \delta^6 \zeta_2^2 -\delta^4 \zeta_1 \zeta_2 \xi - \delta^2 \zeta_1^2 \xi^2)\sigma^2   + 
 (-\zeta_1^2 \zeta_2 - \frac{1}{2} \delta^2 \zeta_2 \omega-6\delta^2 \lambda_1 \xi) \sigma^3+  \psi_4 \sigma^4 \\
 &c_1= i(\frac{1}{2}  \delta^6 \zeta_2 \xi +\delta^4\zeta_1\xi^2)\sigma + 
  i( \delta^2 \zeta_1 \zeta_2 + \zeta_1^2 \xi -\frac{1}{2} \delta^2 \xi \omega)\sigma^2
     +6 i  \lambda_1 \sigma^3\\
    &c_2=\frac{1}{4}\delta^6 \xi^2 + 
 (-\frac{1}{2}\delta^4 \zeta_2 -2 \delta^2 \zeta_1 \xi)\sigma  - 
 \frac{1}{2} \omega \sigma^2 \\
 &c_3= -\frac{i}{2} \delta^4 \xi+i \zeta_1 \sigma \\
 &b=-i\delta
\end{align}
Either of these two solutions makes it clear that in the stable degeneration limit, $Y^{(1)}$ has a non-trivial Mordell-Weil group and a new rational section. Once again, in the stable degeneration limit and from the section addition law of the elliptic fiber of the $dP_9$ fiber of $Y^{(1)}$, this $U(1)$ symmetry was expected to arise and it is gratifying to see it manifestly appear. It only remains then to understand in the F-theory geometry how to understand the Green-Schwarz massive nature of this Abelian symmetry. We will return to this question in a following section.

\subsubsection{SU(7) matter transitions}\label{spec_su7_matter}
For SU(7) it is possible to once again compare the realization of F-theory matter transitions with heterotic small instanton transitions. From the F-theory geometry in Section \ref{ftrans}, a matter transition of the form \eqref{eq:su7transitionftheory}:
\beq
3\times{\tiny \yng(1,1)}+3\times{\tiny \yng(1)}+8\times\mathbf{1} \rightarrow \textbf{Superconformal Matter} \rightarrow {\tiny \yng(1,1,1)} + 8\times {\tiny \yng(1)} + \mathbf{1}, 
\eeq
is realized by first deforming
\begin{align}\label{su7f_tune1}
& \xi \rightarrow a^3 \xi' \\
&\zeta_2 \rightarrow a^4 \zeta_2' \nonumber \\
& \lambda_1 \rightarrow a \lambda_1' \nonumber \\
& \psi_4 \rightarrow a^2 \psi_4' \nonumber
\end{align}
and then tuning further so that the common factor is then absorbed into $\delta$ and $\zeta_1$:
\begin{align} \label{su7f_tune2}
&a \delta \rightarrow \delta' \\
& a\zeta_1 \rightarrow \zeta_1' \nonumber
\end{align}
In the heterotic geometry this should correspond to removing $m$ point-like instantons (taking $a$ to be a degree $m$ polynomial over $\mathbb{P}^1$ above) from either factor of the reducible bundle ${\cal L} \oplus {\cal U}_2$ in \eqref{su7bun2} and then smoothing them back into a new sum ${\cal L}' \oplus {\cal U}'_2$. Let us consider this from the point of view of the spectral cover given in \eqref{su7spec1}:
\begin{align}\label{su7spec1_again}
& \left( (3\zeta_1^2 + \delta^2 \omega)Z^2- 3 \delta^2 X) \right) \times \\
&\left(-3 \left(\frac{1}{3}{\zeta_1}^2\zeta_2 - \frac{1}{18}\zeta_1 \xi \omega+ \frac{1}{9}\delta^2 \zeta_2 \omega + \lambda_1\delta^2 \xi \right)Z^3 + (\zeta_2 \delta^2+ \zeta_1 \xi )XZ- \delta \xi Y \right)=0 \nonumber
\end{align}
Recall that this corresponds to the bundle geometry $({\cal L} \oplus {\cal L}^{\vee})\oplus ({\cal L}^{\vee} \oplus {\cal U}_2)$. Substituting the tuning of \eqref{su7f_tune1} into \eqref{su7spec1_again} above we find a transitional spectral cover
\begin{align}\label{su7_transit}
& \left( (3\zeta_1^2 + \delta^2 \omega)Z^2- 3 \delta^2 X) \right) \times (a^2) \times \\
&\left(-3 \left(\frac{1}{3}(a^2){\zeta_1}^2\zeta_2 - \frac{1}{18}(a)\zeta_1 \xi \omega+ \frac{1}{9}(a^2)\delta^2 \zeta_2 \omega + \lambda_1(a^2)\delta^2 \xi \right)Z^3 + (\zeta_2 (a^2)\delta^2+ (a)\zeta_1 \xi )XZ- (a)\delta \xi Y \right)=0 \nonumber
\end{align}
This is precisely of the form required to consistently allow $\delta$ and $\zeta_1$ to each absorb a factor of $a$ (as in \eref{su7f_tune2}) and return the spectral cover to its canonical form of \eqref{su7spec1} but with the degrees of the relevant functions shifted. 

The remarkable observation to be made is that it is clear from 
the quadratic terms in this spectral cover that the only transitions possible are ones which involve a perfect square vertical factor ($a^2$) as above. That is, the only consistent transitions must involve an \emph{even number of point-like instantons}. This exactly matches the observations made in Section \ref{hettransitions} (and under \eqref{mrtype}) on the smooth heterotic geometry, even-ness of Chern classes and the particle content of the SU(7) heterotic theory on $K3$. Finally it should be noted that the transition above clearly deforms the $dP_9$-fibered $3$-fold 
 $Y^{(1)}$ as well as
the class of the enhanced sections to the elliptic fibrations (with non-trivial Mordell-Weil group) of the elliptically fibered $K3$ surface. As a result, its impact on the dual heterotic/F-theory pair is more substantial than in the case of SU(6) theories.

\subsection{SU(8) Tuning}\label{su8_spec_tune}
In this section, we will repeat the symmetry enhancement analysis described above for the tuning of ${\rm SU}(7) \to {\rm SU}(8)$.
Once again, we return to the F-theory tuning described in Section \ref{sec:F-theory} for SU(8) gauge symmetry. 

Recall that from Section \ref{fsu8}, the tuning taking ${\rm SU}(7) \rightarrow SU(8)$ is given by
\begin{align}
&\xi=\delta^2 \tau \label{su8tune} \\
&\text{$\zeta_2$}=\delta ^2 \text{$\zeta_3$}+\frac{\text{$\zeta_4$} \tau ^2}{2} \nonumber\\
&\zeta_1=\zeta_4 \tau \nonumber\\
&\omega =\delta ^2 \text{$\omega_1$}+4 \text{$\zeta_3$} \text{$\zeta_4$} \nonumber\\
&\text{$\lambda_1$}=\delta ^2 \text{$\lambda_2$} \tau -\frac{\text{$\zeta_4$} \tau  \text{$\omega_1$}}{6} \nonumber\\
&\text{$\psi_4$}=\delta ^4 \text{$\phi_4$}-6 \delta ^2 \text{$\zeta_3$} \text{$\lambda_2$}-\frac{\delta ^2 \text{$\omega_1$}^2}{4}-3 \text{$\zeta_4$} \text{$\lambda_2$} \tau ^2 \nonumber\\
& f_5=\delta ^2 \text{$\psi_5$}+2 \text{$\zeta_4$} \text{$\phi_4$}\nonumber \\
&g_7=\frac{1}{12} (-72 \delta ^4 \text{$\lambda_2$} \text{$\phi_4$}-4 \delta ^4 \text{$\psi_5$} \text{$\omega_1$}-16 \delta ^2 \text{$\zeta_3$} \text{$\zeta_4$} \text{$\psi_5$}+16 \delta ^2 \text{$\zeta_4$} \text{$\omega_1$} \text{$\phi_4$}+16 \text{$\zeta_3$} \text{$\zeta_4$}^2 \text{$\phi_4$}-12 \text{$\zeta_4$}^2 \tau ^2 \text{$\psi_5$}\nonumber \\
&-4 \delta ^6 \text{$\zeta_3$} f_6-6 \delta ^4 \text{$\zeta_4$} f_6 \tau ^2+\delta ^{10} (-f_7) \tau ^2\nonumber)
\end{align}
and again the ``middle'' coefficients ($f_4,g_6$) will be related (in the stable degeneration limit) to the heterotic Weierstrass model.

As described in Section \ref{stab_degen}, it is clear that the SU(8) symmetry depends on a Weierstrass model with structure that is spread across both ``halves'' of the F-theory base geometry and not easily localized on a single patch of the $\mathbb{P}^1$ fiber. Unlike in ${\rm SU}(N)$ heterotic/F-theory dual pairs with $N \leq 5$, care must be taken in the stable degeneration limit. 

In order to take the stable degeneration limit described by \eqref{epsilons}, specifying the overall powers of $\epsilon$ in each Weierstrass coefficient $f_i,g_j$ is not sufficient. Since $f_5, g_7$ are determined by the functions $\xi, \zeta_2, \zeta_1, \ldots$ above, it it must further be specified how these functions are chosen to scale so that $f_5, g_7 \to 0$ in the $\epsilon \to 0$ limit. 

A Groebner-basis calculation (using \cite{Gray:2008zs}) demonstrates that there are three possible paths to the stable degeneration limit in this case corresponding to
\begin{enumerate}
\item $ \psi_5, ~\phi_4 \to 0$ 
\item $ \zeta_4, ~\delta \to 0$
\item $ \left(\text{$\zeta_3$} \text{$\psi_5$}^2 \text{$\phi_4$}-24 \text{$\lambda_1$} \text{$\phi_4$}^3-\tau ^2 \text{$\psi_5$}^3-4 \text{$\psi_5$} \text{$\omega_1$} \text{$\phi_4$}^2\right) \to 0$
\end{enumerate}
As argued in Appendix \ref{su8paths}, in fact only the first of these paths leads to a smooth $K3$ surface in the dual heterotic theory.

Thus, the appropriate limit to take which leads to a smooth, weakly coupled (perturbative) heterotic theory with SU(8) symmetry is
\beq
\psi_5\rightarrow 0~~,~~f_6 \rightarrow 0~~,~~ \phi_4 \rightarrow 0~~,~~f_7 \rightarrow 0
\eeq

In this case then, the form of the SU(8) Weierstrass model of $Y^{(1)}$ is given by

\begin{align}
&f \sim -\frac{\delta ^{20} \tau ^4}{48}+\sigma \left(-\frac{1}{6} \delta ^{16}\zeta_3 \tau ^2-\frac{1}{4} \delta ^{14}{\zeta_4} \tau ^4\right) \\
&+\frac{1}{12} \sigma^2 \left(-4 \delta ^{12}{\zeta_3}^2-2 \delta ^{12} \tau ^2 {\omega_1}-20 \delta ^{10}{\zeta_3}{\zeta_4} \tau ^2-15 \delta ^8{\zeta_4}^2 \tau ^4\right) \\
& +\sigma^3 \left(-\frac{2}{3} \delta ^8{\zeta_3} {\omega_1}-3 \delta ^8 {\lambda_2}\tau ^2-\frac{8}{3} \delta ^6{\zeta_3}^2{\zeta_4}-4 \delta ^4{\zeta_3}{\zeta_4}^2 \tau ^2-2 \delta ^2{\zeta_4}^3 \tau ^4\right) \\
& +\sigma^4 \left(-6 \delta ^4{\zeta_3}{\lambda_2}-\frac{\delta ^4 {\omega_1}^2}{3}-\frac{2}{3} \delta ^2{\zeta_3}{\zeta_4} {\omega_1}-9 \delta ^2{\zeta_4} {\lambda_2}\tau ^2-\frac{4{\zeta_3}^2{\zeta_4}^2}{3}+\text{$\zeta_4$}^2 \tau ^2 {\omega_1}\right)
\end{align}
and
\begin{align}
g \sim & \frac{\delta ^{30} \tau ^6}{864}+\sigma \left(\frac{1}{72} \delta ^{26} {\zeta_3} \tau ^4+\frac{1}{48} \delta ^{24}{\zeta_4} \tau ^6\right) +\frac{1}{72} \sigma^2 \left(4 \delta ^{22} {\zeta_3}^2 \tau ^2+\delta ^{22} \tau ^4 {\omega_1}+16 \delta ^{20} {\zeta_3}{\zeta_4} \tau ^4+12 \delta ^{18}{\zeta_4}^2 \tau ^6\right)  \nonumber\\
&+\frac{1}{108} \sigma^3 \left(8 \delta ^{18} {\zeta_3}^3+12 \delta ^{18} {\zeta_3} \tau ^2 {\omega_1}+27 \delta ^{18} {\lambda_2}\tau ^4+84 \delta ^{16} {\zeta_3}^2{\zeta_4} \tau ^2+9 \delta ^{16}{\zeta_4} \tau ^4 {\omega_1}+144 \delta ^{14} {\zeta_3}{\zeta_4}^2 \tau ^4+72 \delta ^{12}{\zeta_4}^3 \tau ^6\right) \nonumber \\
& \frac{1}{36} \sigma^4 (8 \delta ^{14} {\zeta_3}^2 {\omega_1}+54 \delta ^{14} {\zeta_3} {\lambda_2}\tau ^2+2 \delta ^{14} \tau ^2 {\omega_1}^2+32 \delta ^{12} {\zeta_3}^3{\zeta_4}+22 \delta ^{12} {\zeta_3}{\zeta_4} \tau ^2 {\omega_1}+81 \delta ^{12}{\zeta_4} {\lambda_2}\tau ^4 \nonumber\\
&+116 \delta ^{10} {\zeta_3}^2{\zeta_4}^2 \tau ^2+3 \delta ^{10}{\zeta_4}^2 \tau ^4 {\omega_1}+120 \delta ^8 {\zeta_3}{\zeta_4}^3 \tau ^4+45 \delta ^6{\zeta_4}^4 \tau ^6)  \nonumber\\
&+\frac{1}{36} \sigma^5 (72 \delta ^{10} {\zeta_3}^2{\lambda_2}+8 \delta ^{10} {\zeta_3} {\omega_1}^2+36 \delta ^{10} {\lambda_2}\tau ^2 {\omega_1}+40 \delta ^8 {\zeta_3}^2{\zeta_4} {\omega_1}+360 \delta ^8 {\zeta_3}{\zeta_4} {\lambda_2}\tau ^2+80 \delta ^6 {\zeta_3}^3{\zeta_4}^2 \nonumber \\
&+270 \delta ^6{\zeta_4}^2 {\lambda_2}\tau ^4+120 \delta ^4 {\zeta_3}^2{\zeta_4}^3 \tau ^2-30 \delta ^4{\zeta_4}^3 \tau ^4 {\omega_1}+60 \delta ^2 {\zeta_3}{\zeta_4}^4 \tau ^4+18{\zeta_4}^5 \tau ^6) \nonumber\\
&+\frac{1}{108} \sigma^6 (216 \delta ^6 {\zeta_3} {\lambda_2}{\omega_1}+972 \delta ^6{\lambda_2}^2 \tau ^2+8 \delta ^6 {\omega_1}^3+864 \delta ^4 {\zeta_3}^2{\zeta_4}{\lambda_2}+24 \delta ^4 {\zeta_3}{\zeta_4} {\omega_1}^2\nonumber \\
&-324 \delta ^4{\zeta_4} {\lambda_2}\tau ^2 {\omega_1}-48 \delta ^2 {\zeta_3}^2{\zeta_4}^2 {\omega_1}+648 \delta ^2 {\zeta_3}{\zeta_4}^2 {\lambda_2}\tau ^2+72 \delta ^2{\zeta_4}^2 \tau ^2 {\omega_1}^2-64 {\zeta_3}^3{\zeta_4}^3\nonumber \\
&+72 {\zeta_3}{\zeta_4}^3 \tau ^2 {\omega_1}+324{\zeta_4}^3 {\lambda_2}\tau ^4)
\end{align}
A remarkable observation can be made at this stage by considering the coefficients of $\sigma^4$ in $f$ and $\sigma^6$ in $g$ above. In this limit, the heterotic $K3$ surface is defined by
\begin{align}
f_4 = & \left(-6 \delta ^4{\zeta_3}{\lambda_2}-\frac{\delta ^4 {\omega_1}^2}{3}-\frac{2}{3} \delta ^2{\zeta_3}{\zeta_4} {\omega_1}-9 \delta ^2{\zeta_4} {\lambda_2}\tau ^2-\frac{4{\zeta_3}^2{\zeta_4}^2}{3}+\text{$\zeta_4$}^2 \tau ^2 {\omega_1}\right) \\
g_6 =& \frac{1}{108} (216 \delta ^6 {\zeta_3} {\lambda_2}{\omega_1}+972 \delta ^6{\lambda_2}^2 \tau ^2+8 \delta ^6 {\omega_1}^3+864 \delta ^4 {\zeta_3}^2{\zeta_4}{\lambda_2}+24 \delta ^4 {\zeta_3}{\zeta_4} {\omega_1}^2\nonumber \\
&-324 \delta ^4{\zeta_4} {\lambda_2}\tau ^2 {\omega_1}-48 \delta ^2 {\zeta_3}^2{\zeta_4}^2 {\omega_1}+648 \delta ^2 {\zeta_3}{\zeta_4}^2 {\lambda_2}\tau ^2+72 \delta ^2{\zeta_4}^2 \tau ^2 {\omega_1}^2-64 {\zeta_3}^3{\zeta_4}^3 \nonumber \\
&+72 {\zeta_3}{\zeta_4}^3 \tau ^2 {\omega_1}+324{\zeta_4}^3 {\lambda_2}\tau ^4)
\end{align}
which leads to a discriminant locus for the $K3$ of the form
\beq\label{su8disc_k3}
\Delta_{K3}=-\left(-3 \delta ^6 {\lambda_2}+\delta ^4{\zeta_4} {\omega_1}+2 \delta ^2{\zeta_3}{\zeta_4}^2+\text{$\zeta $4}^3 \tau ^2\right)^2\left(96{\zeta_3}^3 {\lambda_2}+4{\zeta_3}^2 {\omega_1}^2-108{\zeta_3} {\lambda_2} \tau ^2 {\omega_1}-243 {\lambda_2}^2 \tau ^4-4 \tau ^2 {\omega_1}^3\right) 
\eeq
Inspection of this discriminant leads us to an immediate and important
observation. As we have constructed this SU(8) solution thus far, it
is clear that the discriminant in \eqref{su8disc_k3} is
\emph{generically singular}. As a result, the generic heterotic dual
of the SU(8) F-theory Weierstrass model defined in \eqref{su8tune}
must be generically non-perturbative. But from Section \ref{su8het} it
is clear that \emph{perturbative} SU(8) heterotic theories do
exist. It is natural to inquire then, under what circumstances could
we find a perturbative heterotic dual to \eref{su8tune}? By inspection
of \eqref{su8disc_k3} it is clear that the $K3$ surface will be
singular unless the first quadratic factor in the determinant is in
fact a constant. Consulting Table \ref{tab:su8dofs}, the degrees of
the functions appearing in this term are given by
\begin{equation}
\begin{array}{|c|c|c|c|c|c|c|}
\hline
& \delta & \lambda_2 &\zeta_4 & \omega_1& \zeta_3 &\tau \\ \hline
{\rm degree} & \frac{r}{2} & 4+r-n & 3r-n & 4-r & n+4-3r & n+2-\frac{5r}{2} \\ \hline
\end{array}\label{degrees_su8} 
\end{equation}
Thus, the quadratic factor $\left(-3 \delta ^6 {\lambda_2}+\delta ^4{\zeta_4} {\omega_1}+2 \delta ^2{\zeta_3}{\zeta_4}^2+\text{$\zeta $4}^3 \tau ^2\right)^2$ in \eqref{su8disc_k3} will be a (generically non-zero) constant if $4(r+1)=n$. In this section $r=2r_8$ and thus the non-trivial condition on the spectrum is that
\beq\label{smooth_k3_cond}
4(2r_{8}+1)=n
\eeq
which is precisely the integer restriction seen to determine the topology of the bundles ${\cal V}={\cal L} \oplus {\cal L}^{\vee}$ in Section \ref{su8het}! (See \eqref{oops} and the following equations for the restricted spectrum.) As a result, we see a perfect correspondence between those SU(8) F-theory solutions which have a perturbative heterotic dual, as well as the origin of generically non-perturbative heterotic duals with singular $K3$ surfaces.

All that remains is to match the spectral cover associated to the tuned complex structure in \eqref{su8tune} to the bundle geometry given in Section \ref{su8het}. What do we expect to happen to the spectral cover in this case? From the bundle analysis in Section \ref{su8het}, we expect the $S[U(2) \times U(1)]$ bundle from \eqref{su7bun2} to decompose further into a sum of line bundles with structure group $S[U(1)\times U(1)]$ as
\beq\label{su8bund}
{\cal L} \oplus {\cal U}_2 \rightarrow {\cal L} \oplus ({\cal L} \oplus {{\cal L}^{\vee}}^{\otimes 2})
\eeq
In the spectral cover then, one must once again consider the possible
overlapping of roots for the order 2 component. Under the tuning given in
\eqref{su8tune} the SU(7) spectral cover of \eqref{rewrite}
specializes further to
\begin{align}\label{rewrite_su8}
&\left( (3(\zeta_4 \tau)^2 + \delta^2 (\delta ^2 \text{$\omega_1$}+4 \text{$\zeta_3$} \text{$\zeta_4$} ))Z^2- 3 \delta^2 X \right) \times \\
&(\frac{1}{3}{(\delta ^2 \text{$\zeta_3$}+\frac{\text{$\zeta_4$} \tau ^2}{2}) Z}\left( (3 {\zeta_4 \tau}^2+\delta^2 (\delta ^2 \text{$\omega_1$}+4 \text{$\zeta_3$} \text{$\zeta_4$} )) Z^2 -3 \delta^2 X \right) \nonumber \\
&+\delta ^2 \tau \left[-3(- \frac{1}{18}(\zeta_4 \tau) (\delta ^2 \text{$\omega_1$}+4 \text{$\zeta_3$} \text{$\zeta_4$} )+\delta^2 (\delta ^2 \text{$\lambda_2$} \tau -\frac{\text{$\zeta_4$} \tau  \text{$\omega_1$}}{6} ))Z^3+ \zeta_4\tau XZ - \delta Y \right] )=0 \nonumber
\end{align}

The essential structure of the enhanced Mordell-Weil group of both the heterotic $K3$ surface and $Y^{(1)}$ are both preserved in the present case. For the $K3$ geometry, on the patch $Z=1$, the rational section is now defined by
\beq\label{k3sec_su8}
\left[X\to \frac{\delta ^2 \left(\delta ^2 \text{$\omega_1$}+4 \text{$\zeta_3$} \text{$\zeta_4$}\right)+3 \text{$\zeta_4$}^2 \tau ^2}{3 \delta ^2},Y\to -\frac{\frac{1}{2} \delta ^2 \text{$\zeta_4$} \tau  \left(\delta ^2 \text{$\omega_1$}+4 \text{$\zeta_3$} \text{$\zeta4$}\right)-3 \delta ^4 \left(\delta ^2 \text{$\lambda_2$} \tau -\frac{\text{$\zeta_4$} \tau  \text{$\omega_1$}}{6}\right)+\text{$\zeta_4$}^3 \tau ^3}{\delta ^3},Z\to 1\right]
\eeq 
In order to make the correspondence between \eqref{su8bund} and \eqref{rewrite_su8}, we need to demonstrate that if the line bundle ${\cal L}$ corresponds to the rational section $S_1$, then $-2S_1$ (corresponding to ${{\cal L}^{\vee}}^{\otimes 2}$) is now a root of \eqref{su8bund}. Given the rational section, $S_1$ in \eqref{k3sec_su8}, we need first to determine the coordinates of $-2 S_1$ under the addition law of the elliptic fiber. This addition is reviewed in Appendix \ref{elliptic_addition}. Using these standard techniques, $-2S_1$ can be found to correspond to the points on the fiber given by
\begin{align}
&[X\to \frac{\delta ^2 \text{$\zeta_3$}^2}{\tau ^2}+\frac{\text{$\zeta_4$}^2 \tau ^2}{4 \delta ^2}-\frac{1}{3} 2 \delta ^2 \text{$\omega_1$}+\frac{\text{$\zeta_3$} \text{$\zeta_4$}}{3},  \\
&Y\to \frac{8 \delta ^6 \left(\text{$\zeta_3$}^3-\text{$\zeta_3$} \tau ^2 \text{$\omega_1$}-3 \text{$\lambda_2$} \tau ^4\right)+4 \delta ^4 \text{$\zeta_4$} \tau ^2 \left(\text{$\zeta_3$}^2-\tau ^2 \text{$\omega_1$}\right)-2 \delta ^2 \text{$\zeta_3$} \text{$\zeta_4$}^2 \tau ^4-\text{$\zeta_4$}^3 \tau ^6}{8 \delta ^3 \tau ^3},\nonumber \\
& Z\to 1 ]\nonumber
\end{align}
Finally, then it is possible to substitute this point into the spectral cover given by \eqref{rewrite_su8} and verify that it vanishes identically. Thus, as expected, the vector bundle has reduced to two copies of the rational section $S_1$ and one of $-2 S_1$ which corresponds exactly to the required bundle geometry: ${\cal L} \oplus ({\cal L} \oplus {{\cal L}^{\vee}}^{\otimes 2})$.

As pointed out in Section \ref{hettransitions}, due to the restricted
spectrum imposed by the condition for a smooth $K3$ manifold,
\eqref{smooth_k3_cond} matter transitions are not possible in this
stable degeneration limit. If we allow for singular $K3$ surfaces the
analysis is of the same form as those for SU(7) given in Section
\ref{spec_su7_matter} above.

We turn now to other Higgsing transitions in the language of spectral covers.

\subsection{Higgsing on antisymmetric matter}
In this section we briefly review the deformations of spectral covers corresponding to Higgsing on antisymmetric matter in the ${\rm SU}(6), {\rm SU}(7)$ and SU(8) theories described in the previous sections. Such Higgsing chains are surprisingly simple in the language of spectral covers

Consider a heterotic gauge bundle ${\cal V}={\cal V}_1 \oplus {\cal
  V}_2$ with reducible structure group $G_1 \times G_2$. As discussed
above, this can be described via a (possibly further reducible)
spectral cover with structure group ${\rm SU}(n) \times {\rm SU}(m)$:
\beq {\cal S}_1 \cup {\cal S}_2=(a_0Z^{n} + \ldots a_n
X^{\frac{n}{2}})(b_0 Z^{m} + \ldots b_m X^{\frac{m-3}{2}}Y) \eeq
(illustrating here the case where $n$ is even and $m$ is odd).
\\ Then Higgsing on an antisymmetric tensor field corresponds to
non-trivial deformations of either \beq {\cal V}_1 \oplus {\cal
  O}_{K3} \to {\cal V}' \eeq or \beq {\cal V}_2 \oplus {\cal O}_{K3}
\to V' \eeq In the language of spectral covers, these deformations
correspond to \beq {\cal S}_1 \to (a_0 Z^{n+1} \oplus \ldots a_n Y +
a_{n+1} X^{\frac{n-2}{2}}Y)~~~~\text{or}~~~~{\cal S}_2 \to (b_0
Z^{m+1} \oplus \ldots b_m Y + b_{m+1} X^{\frac{m+1}{2}}) \eeq
controlled by the coefficients $a_{n+1}$ and $b_{m+1}$ respectively.

\subsection{Comments on Green-Schwarz massive U(1)s}
The nature of the Green-Schwarz massive U(1) symmetries of the previous section provides an intriguing puzzle in heterotic/F-theory duality. In the heterotic theory, the U(1) symmetries are required by the group theory of $E_8$ subgroups alone. The fact that they generically become massive arises from a separate field theory mechanism: namely the transformation of K\"ahler axions under U(1) shift symmetries. Since the U(1) symmetry is clearly visible in the Weierstrass model of $Y^{(1)}$ in the stable degeneration limit \eqref{fib_prod} it is natural to ask: how can we understand the origin of its mass term in F-theory? While we leave a systematic study of this question to future work, for now we simply raise two possibilities:
\begin{enumerate}
\item In $6$-dimensional compactifications of F-theory the presence of
  massless U(1) symmetries is controlled by the structure of the
  Mordell-Weil group of $Y$, the CY $3$-fold. Since in all the cases
  studied here the U(1) is present only in ``half'' the geometry
  ($Y^{(1)}$), one should not view this as generating a massless
  U(1). This agrees with the analysis of \cite{Cvetic:2015uwu}. In
  particular only in the limit where $\epsilon=0$ would this become
  truly massless. To some extent this agrees with the expectation in
  the heterotic theory since the non-trivial mass terms scale as
  $1/$(Kahler modulus) \cite{Lukas:1999nh} and in the limit of
  strictly infinite volume these would vanish.
\item The geometric origin of the heterotic U(1) mass term is separate from the group theory of the elliptic fibration. Thus, it is possible that in F-theory the mass originates (even in $6$-dimensions) from a source entirely separate to the holomorphic geometry of $Y$ ({\it i.e.} a ``Stuckelberg'' mechanism \cite{Jockers:2005zy,Grimm:2011tb,Grimm:2010ez,Braun:2014nva}).
\end{enumerate}
The clarification of these possibilities is an intriguing area to study further. For now, though we make one final observation: it should be noted that the presence of U(1) symmetries in the stable degeneration limit is itself a subtle thing\footnote{While this paper was in the final stages of preparation we became aware of the work of \cite{Cvetic:2015uwu} which is also focused on the question of massless and Green-Schwarz massive U(1) symmetries in heterotic/F-theory duality and has some overlap with the content of this section and Appendix \ref{stab_degenu1}.}. We give a brief illustration of some of the uncertainty that may arise in Appendix \ref{stab_degenu1}.

\section{Matter transitions in other gauge groups}
\label{sec:other}

In the preceding sections we have focused on matter transitions
involving gauge groups SU($N$) with $N= 6, 7, 8$ and 3-antisymmetric
($\Lambda^3$) representations.  There are a variety of other
situations where similar transitions can occur.  Here we explore a few
such cases, particularly those related to the SU($N$) $\Lambda^3$ matter
transitions.  These include models with Sp(3) and SO(12) gauge groups,
and a class of SU(3) transitions involving matter in the symmetric
representation, where there is an intricate structure to the
Weierstrass model similar to that found recently in \cite{ckpt}.  The
existence of distinct families of Sp(3), SO(12), and SU(3) models with
varying matter content was recognized in \cite{Bershadsky-all}; we
explore here the explicit connection of these models through matter
transitions and  comment on generalizations to other related models.

\subsection{Sp($N$) matter transitions}

\subsubsection{Field theory}

While breaking an SU($N$) theory on a pair of $k$-index antisymmetric
representations gives a theory with gauge group SU($N -k$) $\times$
SU($k$),  it is also possible to Higgs SU($2k$) on a single
$\Lambda^2$ representation, giving the breaking
\begin{equation}
{\rm SU}(2k) \rightarrow {\rm Sp}(k) \,.
\end{equation}
In this case, the Higgs field takes the VEV $\Phi = J$, where $J$ is
the antisymmetric matrix defining Sp($N$) through $[h, J] = 0$ for $h
\in \gsu(N)$.
For SU(6), this breaking gives a branching of representations ${\bf
  20} \rightarrow  {\bf 14'}+ {\bf 6}$ and
${\bf 15} \rightarrow {\bf 14} + {\bf 1}$.  The SU(6) blocks
(\ref{eq:model-6}) thus are Higgsed to Sp(3) blocks
\begin{equation}
(16 + 2n + 3r/2) \times {\bf 6} +
(r/2) \times {\bf 14'} + (1 + n-r) \times {\bf 14} \,.
\end{equation}
The transition (\ref{eq:equivalence-6})  becomes a transition between
Sp(3) representations
\begin{equation} 
\frac{1}{2}{\bf 14}' +\frac{3}{2} {\bf 6}
\leftrightarrow
 {\bf 14} + 2 \times {\bf 1} \,.
\label{eq:sptrans1}
\end{equation}

Similarly, for SU(8) there is a breaking
\begin{eqnarray}
SU(8) & \rightarrow & Sp(4) \\
{\bf 56}  &  \rightarrow &  {\bf 48} + {\bf 8}\\
{\bf  28}  &  \rightarrow &  {\bf 27} + {\bf 1} \\
{\bf  8}  &  \rightarrow &  {\bf  8} {\bf 1} \,,
\end{eqnarray}
from which we expect an Sp(4) transition that follows from
(\ref{eq:equivalence-81})
\begin{equation}
 {\bf 48} + 10 \times {\bf 8} \leftrightarrow
4 \times {\bf 27} + 20 \times {\bf 1} \,.
\end{equation}
Note that the {\bf 70} of SU(8) branches to 
${\bf 42} + {\bf 27} + {\bf 1}$, so we would expect SU(8) with a
$\Lambda^4$ representation to break to an Sp(4) with a matter field in
the {\bf 42} representation.

\subsubsection{F-theory}

From the F-theory point of view, the Sp($k$) and SU($2k$) models are
very closely related.  Both come from a Kodaira type $I_k$
singularity, the only difference is whether it is a ``split'' type
singularity or not.  At the level of the Weierstrass model, the
Higgsing is achieved by allowing $\phi_0^2$ to deform into an
irreducible polynomial $\phi$ in the tunings of
\S\ref{subsec:ftheorytunings}.  This deformation must be associated
with an analogous deformation of $\alpha, \beta$ in (\ref{eq:6-0}). Specifically, $\alpha^2$ is allowed to deform into the irreducible polynomial $h$, while $\beta$ is unchanged. With this modification, all the tuning, transition, and Higgsing
analysis for SU(6) goes through unchanged for Sp(3).  The story would
be more complicated for SU(8), where $\alpha$ and $\beta$ themselves
are decomposed into further components.

\subsubsection{Heterotic description}

From the heterotic point of view the Sp($N$) matter transitions can be computed using techniques similar to those in Section \ref{sec:heterotic}. Looking at the Sp($3$) case as an example, the relevant group theory is as follows.
\begin{eqnarray}
E_8 &\supset& {\rm Sp}(3) \times {\rm SU}(2) \times G_2 \\
{\bf 248} &=& ({\bf 1},{\bf 3},{\bf 1}) + ({\bf 21},{\bf 1},{\bf 1})+({\bf 14}',{\bf 2},{\bf 1})+({\bf 6},{\bf 2},{\bf 7})+({\bf 14},{\bf 1},{\bf 7})+({\bf 1},{\bf 1},{\bf 14})
\end{eqnarray}
We denote the ${\rm SU}(2)$ gauge bundle by ${\cal V}_2$ and the $G_2$
gauge bundle by ${\cal V}'_2$. We then have the correspondence between
representations and cohomologies, and derive from this the
multiplicities of matter representations, as given in Table
\ref{tab5}.

\begin{table}[!ht]
\begin{center}
\begin{tabular*}{14.1cm}{|c|c|c|}
\hline
Representation & Cohomology & Multiplicity \\ \hline
${\bf 14}'$ & $H^1({\cal V}_2)$  & $c_2({\cal V}_2) -4$ \\
${\bf 6}$ & $H^1({\cal V}_2 \otimes {\cal V}'_2)$ & $7 c_2({\cal V}_2)+2 c_2({\cal V}'_2) -28$\\ 
${\bf 14}$ & $H^1({\cal V}'_2)$ & $c_2({\cal V}'_2) -14$\\
${\bf 1}$ &$ H^1(\textnormal{End}_0({\cal V}_2)) \oplus H^1(\textnormal{End}_0({\cal V}'_2)) $ &$(4 c_2({\cal V}_2) -6) +(4 c_2({\cal V}'_2) -28)$\\\hline
\end{tabular*}
\caption{{\it The cohomology associated to each representation of the low-energy gauge group $Sp(3)$.}}
\label{tab5}
\end{center}
\end{table}

Computing the total number of hyper multiplets from the data in Table \ref{tab5} we obtain the familiar anomaly cancelation condition.
\begin{eqnarray} \label{oopssp}
n_H+29 n_T-n_V =273 \\ \nonumber
\Rightarrow  c_2({\cal V}_2) + \frac{1}{2} c_2({\cal V}'_2) + \frac{1}{60} c_2(\textnormal{End}_0({\cal V}_{E_8})) =24
\end{eqnarray}

From Equation (\ref{oopssp}) and Table \ref{tab5} we see that the following matter transition between full hypermultiplets is possible upon small instanton transition (recalling that all of the representations involved are real).
\begin{eqnarray}
\frac{1}{2} \times {\bf 14}'+ \frac{3}{2} \times {\bf 6} \leftrightarrow  \times {\bf 14} + 2 \times {\bf 1} 
\end{eqnarray}
This is precisely of the form (\ref{eq:sptrans1}) expected from field theory considerations.

\subsection{SO($N$) models}

There are two ways in which SO($N$) models may exhibit transitions
like those described for SU($N$) groups in the earlier sections.  One
is for SO($N$) models with matter in the analogue of the $\Lambda^3$
representation.  For example, for SO(12), which has the self-conjugate spinorial representations $\mathbf{32}$ and $\mathbf{32}^\prime$, there are blocks 
\begin{equation}
(r/2) \times {\bf 32} + \frac{4 + n-r}{2}  \times {\bf 32'} + (n + 8)
  \times {\bf 12} \,,
\end{equation}
and associated transitions
\begin{equation}
\frac{1}{2}{\bf 32} \leftrightarrow \frac{1}{2}{\bf 32'} \,.
\end{equation}
These representations
branch to the representations in the SU(6) transition
(\ref{eq:equivalence-6}) under the embedding SU(6) $\subset$
SO(12).  In the F-theory picture this transition follows much like
the SU(6) transition, and can be constructed by taking $\alpha
\rightarrow 0$ in the general SU(6) tuning (\ref{eq:tuning-f-6}),
(\ref{eq:gsu6}). The $\frac{1}{2}\mathbf{32}$ and $\frac{1}{2}\mathbf{32}^\prime$ representations are respectively given by the loci $\beta=\sigma=0$ and $\nu=\sigma=0$, and the transition involves transferring factors between $\nu,\lambda$ and $\beta,\phi_2$. Like the previous transitions considered, the $\gso(12)$ transition involves passing through a superconformal point. In the heterotic picture the SO(12) models can be
constructed by taking an $E_8$ bundle with structure group
$\gsu(2)\times\gsu(2)$, as explained in Section \ref{sec:spectransition}.
These models, and similar constructions for SO(14), {\it etc.} are
similar in principle to the models already considered, with similar
features on both the  F-theory and heterotic sides when applicable.

The heterotic picture, however, suggests another kind of SO($N$)
transition that may occur when we construct bundles with a product
structure group for the SO(32) theory.  For example, consider a
SO(32) heterotic compactification on K3 in the presence of a gauge
bundle with SO(6) $\times$ SU(2) structure group. First we need
the appropriate group theory.
\begin{eqnarray} \label{branching}
{\rm SO}(32) &\supset& {\rm SO}(22) \times {\rm SU}(2) \times {\rm SO}(6) \times {\rm SU}(2) \\ \nonumber
{\bf 496} &=& ({\bf 231},{\bf 1},{\bf 1},{\bf 1}) + ({\bf 22},{\bf 1},{\bf 6},{\bf 1})+({\bf 1},{\bf 1},{\bf 15},{\bf 1})+({\bf 22},{\bf 2},{\bf 1},{\bf 2})\\ \nonumber &&+({\bf 1},{\bf 2},{\bf 6},{\bf 2})+({\bf 1},{\bf 3},{\bf 1},{\bf 1})+({\bf 1},{\bf 1},{\bf 1},{\bf 3})
\end{eqnarray}
We will take the last two factors above to be those associated to the gauge bundle. In particular we will describe the situation in terms of a ${\rm SU}(4)$ vector bundle ${\cal V}_1$ and an ${\rm SU}(2)$ bundle ${\cal V}_2$. We are using an ${\rm SU}(4)$ rather than SO(6) structure group for ease of description. We must account for this, however, in matching cohomologies to representations in the decomposition (\ref{branching}) with the relevant cohomologies for determining matter multiplicities, which are presented in Table \ref{tabso321}.

\vspace{0.2cm}

\begin{table}[!ht]
\begin{center}
\begin{tabular*}{13.5cm}{|c|c|c|}
\hline
Representation & Cohomology & Multiplicity \\ \hline
$({\bf 22},{\bf 1})$ & $H^1(\wedge^2 {\cal V}_1)$  & $2 c_2({\cal V}_1) -12$ \\
$({\bf 22},{\bf 2})$ & $H^1({\cal V}_2)$ & $c_2({\cal V}_2) -4$\\ 
$({\bf 1},{\bf 2})$ & $H^1(\wedge^2 {\cal V}_1 \otimes {\cal V}_2)$ & $4 c_2({\cal V}_1) + 6 c_2({\cal V}_2) -24$\\
$({\bf 1},{\bf 1})$ & $H^1(\textnormal{End}_0({\cal V}_1)) \oplus H^1(\textnormal{End}_0({\cal V}_2))$ & $8 c_2({\cal V}_1) -30 + 4 c_2({\cal V}_2) -6$\\ \hline
\end{tabular*}
\caption{{\it The cohomology associated to each representation of the low-energy gauge group $SO(22) \times {\rm SU}(2)$.}}
\label{tabso321}
\end{center}
\end{table}

The anomaly cancellation condition $n_H+29n_T-n_V=273$ results in the following condition in this case
\begin{eqnarray} \label{10anom}
 c_2({\cal V}_1) +c_2({\cal V}_2) =24\;.
\end{eqnarray}
Equation (\ref{10anom}) is of course simply the 10D anomaly cancellation condition as it should be.

Using (\ref{10anom}) and Table \ref{tabso321}, we see that, in terms of full multiplets, the following kind of matter change can be implemented by small instanton transitions.
\begin{eqnarray}
({\bf 22},{\bf 1}) + 2 \times ({\bf 1},{\bf 1}) \leftrightarrow \frac{1}{2}({\bf 22},{\bf 2}) +({\bf 1},{\bf 2}) \label{eq:so22nettrans}
\end{eqnarray}
But small instantons behave differently in $\gso(32)$ theories than in $\text{E}_8\times \text{E}_8$ theories. Unlike the superconformal points of the $\text{E}_8\times \text{E}_8$ models discussed earlier, the $\gso(32)$ small instanton point leads to a new $\gsu(2)$ symmetry that can be analyzed with field theory \cite{Witten:1995gx}. The $\gso(22)\times \gsu(2)$ transition can therefore be understood completely in terms of Higgsing and unHiggsing transitions (see also \cite{Douglas:2004yv,Cicoli:2013zha} for matter transitions of a similar nature), as can be seen in its F-theory realization.

The matter content of Table \ref{tabso321} suggests an F-theory compactification with base $\mathbb{F}_4$ where the $\gso(22)$ symmetry is tuned on a curve $u=0$ of divisor class $S$. In fact, F-theory models dual to $\gso(32)$ heterotic string theory can only be realized on $\mathbb{F}_4$ \cite{Aspinwall:1996nk}.  Meanwhile, the $\gsu(2)$ symmetry should be tuned on a curve $\sigma = 0 $ of divisor class $S+(4+r)F$, with $r=c_2(\mathcal{V}_2) - 4$ denoting the number of $(\mathbf{22},\mathbf{2})$ half-multiplets. The global Weierstrass model is then 
\begin{equation}
y^2 = x^3 + \Biggparen{-\frac{1}{48}\Phi^2 u^2 + F_1 \sigma u^6}x + \Biggparen{\frac{1}{864}\Phi^3 u^3 - \frac{1}{12} F_1 \Phi \sigma u^7 + \gamma_2^2 \sigma^2 u^{10}}.
\end{equation}
$\Phi$, $F_1$, and $\gamma_2$ are respectively sections of $\mathcal{O}(3S + 12 F)$, $\mathcal{O}(S+(20-r)F)$, and $\mathcal{O}(14-r)F$. 

For a transition where $r$ increases, $F_1$ and $\gamma_2$ develop a
common factor $a$ that is a section of $\mathcal{O}(F)$ ($F_1
\rightarrow a F_1, \gamma_2 \rightarrow a \gamma_2$). $a$ is then
absorbed into $\sigma$ ($a\sigma \rightarrow \sigma'$), and the divisor
class of $\sigma=0$ changes from $\tilde{S}+rF$ to $\tilde{S} +
(r+1)F$. Immediately after absorbing $a$, $\sigma$ is a reducible
curve. $\sigma$ can then be deformed into a non-reducible curve,
completing the transition. To perform the transition in the reverse
direction, we let $\sigma$ become $a\sigma$ and reabsorb $a$ into
$F_1$ and $\gamma_2$ ({\it i.e.} $a F_1 \rightarrow {F_1}'$ and
$a\gamma_2 \rightarrow {\gamma_2}'$).

At the transition point, there is an additional $I_2$ singularity on the $a=0$ locus, signaling the expected appearance of a new $\gsu(2)$ symmetry. We will refer to this new symmetry as $\gsu(2)_a$ to distinguish it from the original $\gsu(2)$ tuned on $\sigma=0$. There is also matter charged under $\gsu(2)_a$, as $a=0$ intersects the curves $u=0$ and $\sigma=0$ once. In terms of the $\gso(22)\times\gsu(2)\times\gsu(2)_a$ representations, this charged matter consists of a half-multiplet of $(\mathbf{22},\mathbf{1},\mathbf{2}_a)$ matter and a full $(\mathbf{1},\mathbf{2},\mathbf{2}_a)$ multiplet; the $a$ subscript is used to identify the $\gsu(2)_a$ representations. The two sides of the transition correspond to the two ways of Higgsing $\gsu(2)_a$. Giving a VEV to the $(\mathbf{1},\mathbf{2},\mathbf{2}_a)$ multiplet merges to the two $\gsu(2)$ symmetries into a single, diagonal $\gsu(2)$, and the $(\mathbf{22},\mathbf{1},\mathbf{2}_a)$ half-multiplet reduces to a half-multiplet of bifundamental $(\mathbf{22},\mathbf{2})$ matter. But when the $(\mathbf{22},\mathbf{1},\mathbf{2}_a)$ half-multiplet is given a VEV, $\gsu(2)_a$ is Higgsed separately from the original $\gsu(2)$. As a result, the  $(\mathbf{22},\mathbf{1},\mathbf{2}_a)$ half-multiplet is left as a full $(\mathbf{22},\mathbf{1})$ multiplet. Tracking the matter fully, the total transition can be summarized as 
\begin{equation}
(\mathbf{22},\mathbf{1}) + 2\times (\mathbf{1},\mathbf{2}) + 3\times (\mathbf{1},\mathbf{1}) \leftrightarrow \frac{1}{2}(\mathbf{22},\mathbf{2}) + 3\times (\mathbf{1},\mathbf{2}) + 1\times(\mathbf{1},\mathbf{1}),
\end{equation}
reproducing the net matter change of Equation
\eqref{eq:so22nettrans}. Even though the transition consists only of
Higgsing and unHiggsing transitions, the fact that 29 multiplets
participate in the transition
suggests a parallel structure to the matter transitions mediated
through superconformal points, which may reflect the underlying small-instanton
behavior.

A plethora of exotic transitions of this type are possible in different compactifications of the SO($32$) heterotic string. As in the $\gso(22)\times \gsu(2)$ model, these transitions should be described by phenomena accessible from field theory.

\subsection{SU(3) with symmetric matter}
\label{sec:SU(3)}

A particularly interesting set of exotic matter representations are
the representations of SU($N$) that have Young diagrams with more than
one column.  As described in \cite{kpt}, there is a natural quantity
$g$ associated with any representation $R$ of a simple Lie group $G$
that plays the role of a ``genus'' of the representation.  The
geometric interpretation of this quantity is conjectured to be that
when $g > 0$, for any representation other than the adjoint,
the representation $R$ is realized in F-theory through a Kodaira
singularity on a divisor $D$ that is itself singular, where $g$
represents the arithmetic genus contribution of the singularity to the
curve $D$ in the 6D case.  This correspondence works most simply for
the symmetric representation of SU($N$), which has $g = 1$, and which
can be realized on a double point singularity of $D$ as first
suggested by Sadov \cite{Sadov}, described further in \cite{mt-singularities}, and recently confirmed through an
explicit F-theory construction \cite{ckpt}.  The explicit F-theory
construction of the symmetric matter representation of SU(3) has the
unusual feature that the Weierstrass model cannot be built from a
standard generic Tate SU(3) construction; rather, the vanishing of the
discriminant to order 3 follows from a nontrivial cancellation that
involves the explicit algebraic structure of the singular divisor
locus carrying the SU(3) gauge group.  Understanding matter
transitions in this context gives further insight into this story.

We can realize a symmetric representation of SU(3) by breaking Sp(3)
into SU(3), or more directly by breaking SU(6) into
SU(3) $\times$ SU(3) and then breaking
\begin{equation}
{\rm SU}(3) \times {\rm SU}(3) \rightarrow {\rm SU}(3)
\end{equation}
by Higgsing a bifundamental field.
The anomaly equivalent matter representations in the resulting
theories exchange an adjoint (plus a singlet) with a symmetric and an
antisymmetric matter representation
\begin{equation}
{\bf 8} + {\bf 1} \leftrightarrow {\bf 6} + {\bf 3} \,.
\label{eq:su3trans1}
\end{equation}
Thus, the generic SU(3) model with $g$ adjoints and $b \cdot b = n$
(\ref{eq:model-general}) has anomaly equivalent variations
\begin{equation}
(g-r) {\bf 8} + r {\bf 6} + (18 + 6n -18 g + r) {\bf 3} \,.
\label{eq:su3matter}
\end{equation}

\begin{figure}[tbp]
\centering
\begingroup
\setlength{\unitlength}{1cm}
\begin{picture}(12,7)(-6,-3)

\put(-4.4,2.25){$\gsu(6)$}
\put(-3,2){\circle*{.15}}
\put(-3.5,2){\circle*{.15}}
\put(-4,2){\circle*{.15}}
\put(-4.5,2){\circle*{.15}}
\put(-5,2){\circle*{.15}}
\multiput(-5,2)(.5,0){4}{\line(1,0){.5}}

\put(3.6,2.25){$\gsp(3)$}
\put(3,2){\circle*{.15}}
\put(3.5,2){\circle*{.15}}
\put(4,2){\circle*{.15}}
\put(4.5,2){\circle*{.15}}
\put(5,2){\circle*{.15}}
\multiput(3,2)(.5,0){4}{\line(1,0){.5}}

\qbezier(3.5006,2.02453)(3.5,2.)(3.5006,1.97547)
\qbezier(3.50541,1.92663)(3.50961,1.90245)(3.51498,1.87851)
\qbezier(3.52923,1.83156)(3.53806,1.80866)(3.54801,1.78622)
\qbezier(3.57114,1.74295)(3.58427,1.72221)(3.5984,1.70215)
\qbezier(3.62952,1.66422)(3.64645,1.64645)(3.66422,1.62952)
\qbezier(3.70215,1.5984)(3.72221,1.58427)(3.74295,1.57114)
\qbezier(3.78622,1.54801)(3.80866,1.53806)(3.83156,1.52923)
\qbezier(3.87851,1.51498)(3.90245,1.50961)(3.92663,1.50541)
\qbezier(3.97547,1.5006)(4.,1.5)(4.02453,1.5006)
\qbezier(4.07337,1.50541)(4.09755,1.50961)(4.12149,1.51498)
\qbezier(4.16844,1.52923)(4.19134,1.53806)(4.21378,1.54801)
\qbezier(4.25705,1.57114)(4.27779,1.58427)(4.29785,1.5984)
\qbezier(4.33578,1.62952)(4.35355,1.64645)(4.37048,1.66422)
\qbezier(4.4016,1.70215)(4.41573,1.72221)(4.42886,1.74295)
\qbezier(4.45199,1.78622)(4.46194,1.80866)(4.47077,1.83156)
\qbezier(4.48502,1.87851)(4.49039,1.90245)(4.49459,1.92663)
\qbezier(4.4994,1.97547)(4.5,2.)(4.4994,2.02453)

\qbezier(3.0012,2.04907)(3.,2.)(3.0012,1.95093)
\qbezier(3.01082,1.85327)(3.01921,1.80491)(3.02997,1.75702)
\qbezier(3.05846,1.66311)(3.07612,1.61732)(3.09601,1.57244)
\qbezier(3.14227,1.4859)(3.16853,1.44443)(3.19679,1.4043)
\qbezier(3.25905,1.32844)(3.29289,1.29289)(3.32844,1.25905)
\qbezier(3.4043,1.19679)(3.44443,1.16853)(3.4859,1.14227)
\qbezier(3.57244,1.09601)(3.61732,1.07612)(3.66311,1.05846)
\qbezier(3.75702,1.02997)(3.80491,1.01921)(3.85327,1.01082)
\qbezier(3.95093,1.0012)(4.,1.)(4.04907,1.0012)
\qbezier(4.14673,1.01082)(4.19509,1.01921)(4.24298,1.02997)
\qbezier(4.33689,1.05846)(4.38268,1.07612)(4.42756,1.09601)
\qbezier(4.5141,1.14227)(4.55557,1.16853)(4.5957,1.19679)
\qbezier(4.67156,1.25905)(4.70711,1.29289)(4.74095,1.32844)
\qbezier(4.80321,1.4043)(4.83147,1.44443)(4.85773,1.4859)
\qbezier(4.90399,1.57244)(4.92388,1.61732)(4.94154,1.66311)
\qbezier(4.97003,1.75702)(4.98079,1.80491)(4.98918,1.85327)
\qbezier(4.9988,1.95093)(5.,2.)(4.9988,2.04907)

\put(-5.1,-1.75){$\gsu(3)\times\gsu(3)$}
\put(-3,-2){\circle*{.15}}
\put(-3.5,-2){\circle*{.15}}
\put(-4.5,-2){\circle*{.15}}
\put(-5,-2){\circle*{.15}}
\multiput(-5,-2)(1.5,0){2}{\line(1,0){.5}}

\put(3.5,-1.75){$\suthree$}
\put(3,-2){\circle*{.15}}
\put(3.5,-2){\circle*{.15}}
\put(4.5,-2){\circle*{.15}}
\put(5,-2){\circle*{.15}}
\multiput(3,-2)(1.5,0){2}{\line(1,0){.5}}
\qbezier(3.5006,-1.97547)(3.5,-2.)(3.5006,-2.02453)
\qbezier(3.50541,-2.07337)(3.50961,-2.09755)(3.51498,-2.12149)
\qbezier(3.52923,-2.16844)(3.53806,-2.19134)(3.54801,-2.21378)
\qbezier(3.57114,-2.25705)(3.58427,-2.27779)(3.5984,-2.29785)
\qbezier(3.62952,-2.33578)(3.64645,-2.35355)(3.66422,-2.37048)
\qbezier(3.70215,-2.4016)(3.72221,-2.41573)(3.74295,-2.42886)
\qbezier(3.78622,-2.45199)(3.80866,-2.46194)(3.83156,-2.47077)
\qbezier(3.87851,-2.48502)(3.90245,-2.49039)(3.92663,-2.49459)
\qbezier(3.97547,-2.4994)(4.,-2.5)(4.02453,-2.4994)
\qbezier(4.07337,-2.49459)(4.09755,-2.49039)(4.12149,-2.48502)
\qbezier(4.16844,-2.47077)(4.19134,-2.46194)(4.21378,-2.45199)
\qbezier(4.25705,-2.42886)(4.27779,-2.41573)(4.29785,-2.4016)
\qbezier(4.33578,-2.37048)(4.35355,-2.35355)(4.37048,-2.33578)
\qbezier(4.4016,-2.29785)(4.41573,-2.27779)(4.42886,-2.25705)
\qbezier(4.45199,-2.21378)(4.46194,-2.19134)(4.47077,-2.16844)
\qbezier(4.48502,-2.12149)(4.49039,-2.09755)(4.49459,-2.07337)
\qbezier(4.4994,-2.02453)(4.5,-2.)(4.4994,-1.97547)

\qbezier(3.0012,-1.95093)(3.,-2.)(3.0012,-2.04907)
\qbezier(3.01082,-2.14673)(3.01921,-2.19509)(3.02997,-2.24298)
\qbezier(3.05846,-2.33689)(3.07612,-2.38268)(3.09601,-2.42756)
\qbezier(3.14227,-2.5141)(3.16853,-2.55557)(3.19679,-2.5957)
\qbezier(3.25905,-2.67156)(3.29289,-2.70711)(3.32844,-2.74095)
\qbezier(3.4043,-2.80321)(3.44443,-2.83147)(3.4859,-2.85773)
\qbezier(3.57244,-2.90399)(3.61732,-2.92388)(3.66311,-2.94154)
\qbezier(3.75702,-2.97003)(3.80491,-2.98079)(3.85327,-2.98918)
\qbezier(3.95093,-2.9988)(4.,-3.)(4.04907,-2.9988)
\qbezier(4.14673,-2.98918)(4.19509,-2.98079)(4.24298,-2.97003)
\qbezier(4.33689,-2.94154)(4.38268,-2.92388)(4.42756,-2.90399)
\qbezier(4.5141,-2.85773)(4.55557,-2.83147)(4.5957,-2.80321)
\qbezier(4.67156,-2.74095)(4.70711,-2.70711)(4.74095,-2.67156)
\qbezier(4.80321,-2.5957)(4.83147,-2.55557)(4.85773,-2.5141)
\qbezier(4.90399,-2.42756)(4.92388,-2.38268)(4.94154,-2.33689)
\qbezier(4.97003,-2.24298)(4.98079,-2.19509)(4.98918,-2.14673)
\qbezier(4.9988,-2.04907)(5.,-2.)(4.9988,-1.95093)

\thicklines
\put(-1,2.25){Monodromy}
\put(-5.7,0.2){Remove}
\put(-5.5,-0.2){Node}
\put(-2,2){\vector(1,0){4}}
\put(-2,-2){\vector(1,0){4}}
\put(-4,1.5){\vector(0,-1){2.8}}
\put(4,0.8){\vector(0,-1){2.1}}
\end{picture}
\endgroup
\caption{Higgsing chain involving $\gsu(6)$, $\gsp(3)$, $\gsu(3)\times\gsu(3)$, and $\suthree$ along with associated Dynkin diagrams. Dotted lines indicate nodes exchanged under monodromy. The Higgsing chain involves two types of F-theory deformations that either introduce monodromy or remove the central node in the Dynkin diagrams. The $\suthree$ singularity occurs when both types of deformations are performed.}
\label{fig:su3leveltwoHiggschain}
\end{figure}
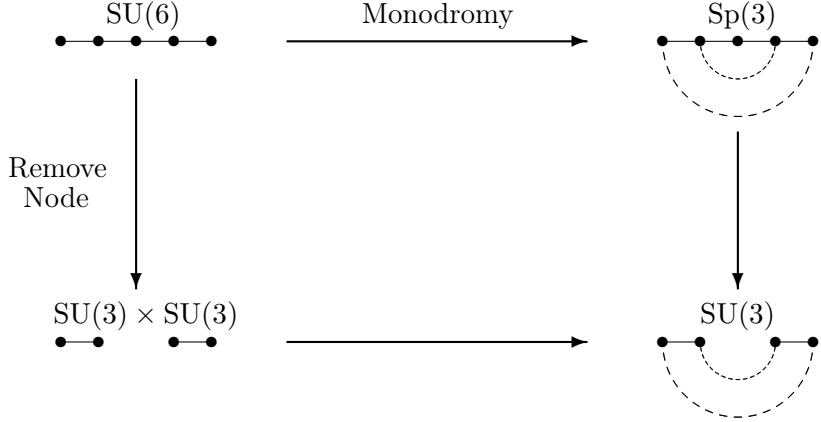

In F-theory\footnote{Here, we essentially describe the F-theory
  realization of the level-two $\suthree$ (or $\suthree_2$) discussed
  in \cite{Bershadsky-all}. However, our notation differs slightly. We
  take $n$ to refer to $b\cdot b$ or the self-intersection number of
  the curve with the $\gsu(3)$ singularity; in \cite{Bershadsky-all},
  $n$ refers to the base $\mathbb{F}_n$. $r$ is also smaller by 2 in
  our conventions.}, the Higgsing chain for $\gsu(6)$, $\gsp(3)$, and
$\gsu(3)\times\gsu(3)$, illustrated in Figure
\ref{fig:su3leveltwoHiggschain}, involves two distinct
deformations. $\gsu(6)$ is Higgsed to $\gsu(3)\times\gsu(3)$ by
removing the central node in the $A_5$ Dynkin diagram. Meanwhile, the
$\gsp(3)$ model is produced by introducing monodromy effects in the
$\gsu(6)$ model that cause a $\mathbb{Z}_2$ folding of the $A_5$
Dynkin diagram. The $\suthree$ symmetry is produced by applying both
deformations to $\gsu(6)$. One can think of the $\suthree$ singularity
as consisting of two $A_2$ singularities that are mapped onto each
other through the monodromy-induced $\mathbb{Z}_2$ folding. Thus, the
two $\gsu(3)$ algebras in the $\gsu(3)\times\gsu(3)$ product model
reduce to a single $\suthree$ algebra.

For an explicit construction of $\suthree$, consider an F-theory compactification with base $\mathbb{F}_m$ and an $\gsu(6)$ singularity tuned on a curve $\sigma=0$ of divisor class $\tilde{S}$. Using a strategy similar to that of section \ref{subsubsec:ftheory3asHiggs}, the $\gsu(6)$ symmetry can then be Higgsed to $\gsu(3)\times\gsu(3)$, with the two $\gsu(3)$ singularities tuned on the curves
\begin{displaymath}
\sigma \pm \frac{\alpha\epsilon_1}{2}=0.
\end{displaymath}
This situation could be thought of as a single $A_2$ singularity tuned on the reducible curve 
\begin{displaymath}
\sigma^2-\frac{\alpha^2\epsilon_1^2}{4}=0;
\end{displaymath}
since this curve can be reduced into the product of two components, it
actually represents two distinct $\gsu(3)$ algebras. The reducible
curve can then be smoothed into a non-factorizable quadratic
polynomial in $\sigma$, thereby reducing the product group to a single
$\gsu(3)$. Specifically, $\alpha$ always appears in even powers in the
$\gsu(3)\times\gsu(3)$ Weierstrass model, so $\alpha^2$ can be
consistently replaced with a parameter $h$. Note that this deformation
is the same as that for the $\gsu(6)\rightarrow\gsp(3)$ Higgsing
process described earlier.  
Assuming that neither $h$ nor $\epsilon_1$ is constant,
the $A_2$ singularity is now tuned on the non-factorizable curve
\begin{equation}
\sigma^2-\frac{h\epsilon_1^2}{4}=0,
\end{equation}
so the resulting gauge algebra is a single $\gsu(3)$ tuned on the divisor class $2\tilde{S}$. The corresponding Weierstrass model has
\begin{multline}
f= -\frac{1}{48}\Bigparen{h\beta^2 + \epsilon_1^2 \nu^2 + 4 \beta \nu \sigma}^2 - \frac{1}{6}\Bigparen{9 \epsilon_1^2 \lambda \nu + h \beta \phi_2 + (18\beta\lambda + 2\nu\phi_2)\sigma}\Bigparen{\sigma^2-\frac{h\epsilon_1^2}{4}} \\+ \Bigparen{f_4 + f_5 \sigma}\Bigparen{\sigma^2-\frac{h\epsilon_1^2}{4}}^2+\mathcal{O}\Bigparen{\sigma^2-\frac{h\epsilon_1^2}{4}}^3 \label{eq:su32f}
\end{multline}
and
\begin{multline}
g = \frac{1}{864}\Bigparen{h\beta^2 + \epsilon_1^2 \nu^2 + 4 \beta \nu \sigma}^3+\frac{1}{72}\Bigparen{h\beta^2 + \epsilon_1^2 \nu^2 + 4 \beta \nu \sigma}\Bigparen{9 \epsilon_1^2 \lambda \nu + h \beta \phi_2 + (18\beta\lambda + 2\nu\phi_2)\sigma}\Bigparen{\sigma^2-\frac{h\epsilon_1^2}{4}} \\
+\frac{1}{36}\Bigparen{81 \epsilon_1^2 \lambda^2 + h \phi_2^2 + 36\lambda \phi_2 \sigma -3 (f_4 + f_5 \sigma)(h\beta^2+\epsilon_1^2\nu^2+4 \beta\nu\sigma)}\Bigparen{\sigma^2-\frac{h\epsilon_1^2}{4}}^2+\mathcal{O}\Bigparen{\sigma^2-\frac{h\epsilon_1^2}{4}}^3\label{eq:su32g}
\end{multline}

Taking the $\epsilon_1\rightarrow 0$ limit recovers the $\gsp(3)$
model, even though the $\gsp(3)$ model was never directly used to find
the $\suthree$ Weierstrass tuning. This fact confirms that our ${\rm SU}(3)$
tuning agrees with the Higgsing chain given in Figure \ref{fig:su3leveltwoHiggschain}, as we could reach the $\gsp(3)$ model indirectly via $\gsu(3)\times\gsu(3)$ and $\suthree$. Moreover, we can directly see that the monodromy ``inherited'' from the $\gsp(3)$ model is crucial in the $\suthree$ tuning. The $\suthree$ singularity could alternatively be thought of as two $A_2$ singularities tuned on the two curves
\begin{displaymath}
\sigma \pm \frac{h^{1/2}\epsilon_1}{2} = 0.
\end{displaymath}
However, the curves would interchange under the transformation $h\rightarrow e^{2\pi i}h$, the same transformation involved in the $\gsp(3)$ monodromy, and the two $A_2$ subdiagrams should be identified with one another. We are therefore left with a single $\suthree$ algebra, with the $h\rightarrow e^{2\pi i}h$ transformation providing the $\mathbb{Z}_2$ folding depicted in Figure \ref{fig:su3leveltwoHiggschain}.

Turning to the matter content, a curve with divisor class $2\tilde{S}$ has a genus $g$ given by
\begin{equation}
g = 1 + \frac{1}{2}\left(2\tilde{S}\right)\cdot\left(K_B + 2\tilde{S}\right) = m-1.
\end{equation}
The $\suthree$ model therefore has a total of $m-1$ charged multiplets
in either the adjoint or symmetric representation. Distinguishing
between the two representations requires examining the two possible
sources of double point singularities in the $2\tilde{S}$ curve:
double point singularities that can be deformed away contribute
adjoint ($\mathbf{8}$) matter, whereas non-deformable double point
singularities gives one ${\tiny \yng(2)}$ ($\mathbf{6}$) multiplet and
one fundamental ($\mathbf{3}$) multiplet. The situation is the Higgsed
version of a similar feature in the $\gsu(3)\times\gsu(3)$ model,
where there  are subtle differences between the
$(\mathbf{3},\bar{\mathbf{3}})$ and $(\mathbf{3},\bar{\mathbf{3}})$
bifundamental representations \cite{mt-singularities}. There, the distinction between the two
bifundamental representations does not have significant physical
implications. But upon Higgsing to $\suthree$, the two bifundamentals
branch to dramatically different representations, turning into either
$\mathbf{8}+\mathbf{1}$ or $\mathbf{6}+\mathbf{3}$. From an F-theory
perspective, monodromy identifies the nodes of the two $A_2$
subdiagrams in a particular way. With this identification, the
redefinitions of the gauge algebras that made the bifundamental
representations essentially equivalent are no longer valid.

We can now find the specific double-point singularities that
contribute symmetric matter. $h$ can have perfect square factors that
lead to double point singularities. But since $h$ is not required to
have any perfect square factors, such double points can be deformed
away by modifying the form of $h$. These double point singularities
therefore contribute localized adjoints. The zeroes of $\epsilon_1$
lead to double point singularities as well, but these double points
cannot be removed by simply deforming one of the free
parameters. Letting $r$ be the order of $\epsilon_1$, $\epsilon_1$ contributes $r$ multiplets of ${\tiny
  \yng(2)}$ $(\mathbf{6})$ matter and $r$ fundamental multiplets. To
find the additional fundamentals provided by the discriminant loci, it
is easiest to expand the discriminant around $\sigma \pm \frac{1}{2}
h^{1/2} \epsilon_1$. When expanded around $\sigma + \frac{1}{2}
h^{1/2} \epsilon_1$, there are $3m+18$ codimension two loci where the
discriminant vanishes to order 4; there are the same number of loci
when the discriminant is expanded around $\sigma - \frac{1}{2} h^{1/2}
\epsilon_1$. A total of $6m+36$ fundamentals therefore come from the
discriminant. Considering all of these contributions, the $\suthree$
models have $r$ $\mathbf{6}$ multiplets, $m-1-r$ $\mathbf{8}$
multiplets, and $6m + 36 +r$ $\mathbf{3}$ multiplets. Noting that $2\tilde{S}$ has self-intersection number $n=4m$ and genus $g=m-1$, this is in agreement with \eqref{eq:su3matter}.

It is tempting to transfer factors from $h$ to $\epsilon_1$ in order to systematically introduce non-deformable double-point singularities. If this were possible, we could transform adjoint matter into symmetric matter. However, there are terms in the Weierstrass model of Equations \eqref{eq:su32f} and \eqref{eq:su32g} that depend on $\epsilon_1$ and not on $h$, or vice versa. The only consistent way to transfer factors from $h$ to $\epsilon_1$ is to use a matter transition similar to that of $\gsu(6)$, which requires moving through a superconformal point. Suppose we wish to convert an adjoint $\mathbf{8}$ multiplet to a $\mathbf{6}$ multiplet (along with other fundamentals and singlets). Just as in the $\gsu(3)\times\gsu(3)$ transition, $h$,$\nu$, and $\lambda$ develop common factors:
\begin{align}
h &\rightarrow a^2 h',\\
\nu &\rightarrow a \nu',\\
\lambda &\rightarrow a\lambda.'
\end{align}
Note that there is now superconformal point at the locus $a=\sigma=0$. $\epsilon_1$, $\beta$, and $\phi_2$ then absorb the common factor:
\begin{align}
a \epsilon_1 &\rightarrow \epsilon_1',\\
a \beta &\rightarrow \beta,'\\
a \phi_2 &\rightarrow \phi_2'.
\end{align}
We have thus introduced a non-deformable double point singularity through the matter transition. Overall, the transition is summarized as
\begin{equation}
\mathbf{Adj} + 6\times {\tiny \yng(1)} + 3\times\mathbf{1} \rightarrow
\textbf{Superconformal Matter} \rightarrow {\tiny \yng(2)} +
7\times{\tiny \yng(1)} + 2 \times \mathbf{1},
\label{eq:adjoint-transition}
\end{equation}
giving a net matter change of
\begin{equation}
\mathbf{Adj} + \mathbf{1} \rightarrow {\tiny \yng(2)} + {\tiny \yng(1)}.
\end{equation}
Note the usual appearance of 29 matter fields in the spectrum on each
side of the full transition (\ref{eq:adjoint-transition})

Finally, we note that the Weierstrass model of Equations
\eqref{eq:su32f} and \eqref{eq:su32g} has a ``non-Tate''
structure. The $\suthree$ tuning cannot be written in the generic
$\gsu(3)$ form given in \cite{mt-singularities}. Instead, the curve
with the $A_2$ singularity has a specific structure that depends on
variables used in the tuned Weierstrass coefficients. The coefficients
then conspire to ensure all terms in the discriminant are proportional
$(\sigma^2-\frac{h\epsilon_1^2}{4})^3$. But for the special case with
no symmetric matter, the $\suthree$ tuning can be written in the
standard form. If $r=0$, $\epsilon_1$ is a non-zero constant, and we
can set $\epsilon_1$ to 1 without loss of generality. With this
simplification, $f$ and $g$ can be rewritten as
\begin{align}
f&= \frac{-1}{48}\Phi_0^4 + \frac{1}{2}\Phi_0 \Psi_0 \Bigparen{\sigma^2 - \frac{h}{4}} + F_2 \Bigparen{\sigma^2-\frac{h}{4}}^2 + \mathcal{O}\Bigparen{\sigma^2 - \frac{h}{4}}^3,\label{eq:su3adjf}\\
g&= \frac{1}{864}\Phi_0^6 -\frac{1}{24}\Phi_0^3 \Psi_1 \Bigparen{\sigma^2 - \frac{h}{4}} + \left(\frac{\Psi_1^2}{4}-\frac{1}{12}F_2 \Phi_0^2\right)\Bigparen{\sigma^2 - \frac{h}{4}}^2 +\mathcal{O}\Bigparen{\sigma^2 - \frac{h}{4}}^3,\label{eq:su3adjg}
\end{align}
where
\begin{align}
\Phi_0 &= \nu + 2 \beta\sigma,\\
\Psi_1 &=  -3\lambda - \frac{2}{3}\phi_2\sigma +\frac{1}{3} \beta^2\Phi_0,\\
F_2 &= f_4 + f_5 \sigma -\frac{1}{3}\beta^4+\frac{2}{3}\beta \phi_2.
\end{align}
In fact, Equations \eqref{eq:su3adjf} and \eqref{eq:su3adjg} are in the standard $\suthree$ forms given in \cite{mt-singularities}.

This behavior parallels that observed in the $\gsu(3)$ models derived
in \cite{ckpt}. There, all of the higher-genus $\suthree$ models with
symmetric matter had non-Tate structures. The construction presented
here further supports the idea that non-Tate structures are necessary
for symmetric matter; indeed, our models can only be expressed in
standard forms for exactly those cases without symmetric matter.
However, our $\suthree$ tuning seems to be different from the tuning
given in \cite{ckpt}. 
The connection between these classes of models is left as a question
for future investigations.

\vspace{0.2cm}

Matter transitions in ${\rm SU}(3)$ models with symmetric matter can also be realized within the heterotic context. The relevant group theory in this instance is as follows.
\begin{eqnarray} 
E_8 &\supset& G_2 \times {\rm SU}(3) \times {\rm SU}(3) \\ \nonumber
{\bf 248} &=& ({\bf 14},{\bf 1},{\bf 1})+({\bf 7},{\bf 8},{\bf 1}) + ({\bf 1},{\bf 8},{\bf 1}) + ({\bf 1},{\bf 1},{\bf 8})+ ({\bf 7},\overline{{\bf 3}},{\bf 3}) + ({\bf 1},{\bf 6},{\bf 3})+ ({\bf 7},{\bf 3},\overline{{\bf 3}}) +({\bf 1},\overline{{\bf 6}},\overline{{\bf 3}})
\end{eqnarray}

We denote the $G_2$ gauge bundle by ${\cal V}_2$ and the ${\rm SU}(3)$ gauge bundle by ${\cal V}_3$. It should be noted that we choose the second of the two ${\rm SU}(3)$'s to be associated with the bundle structure group. We then have the multiplicities of representations in the six-dimensional theory given in  Table \ref{tabsu3222} .

\begin{table}[!h]
\begin{center}
\begin{tabular*}{14.285cm}{|c|c|c|}
\hline
Representation & Cohomology & Multiplicity \\ \hline
${\bf 1}$ & $H^1(\textnormal{End}_0({\cal V}_2)) \oplus H^1(\textnormal{End}_0({\cal V}_3))$  & $(4 c_2({\cal V}_2) -28)+(6 c_2({\cal V}_3) -16)$ \\
${\bf 3}$ &  $ H^1({\cal V}_2 \otimes {\cal V}^{\vee}_3)$  & $(3 c_2({\cal V}_2)+ 7 c_2({\cal V}_3)-42)$\\  
$\overline{{\bf 3}}$ &$ H^1({\cal V}_2 \otimes {\cal V}_3)$&$(3 c_2({\cal V}_2)+ 7 c_2({\cal V}_3)-42)$\\
${\bf 6}$ &$H^1({\cal V}_3)$& $c_2({\cal V}_3) -6$\\
$\overline{{\bf 6}}$&$H^1({\cal V}_3^{\vee})$&$c_2({\cal V}_3) -6$\\
${\bf 8}$&$H^1({\cal V}_2)$& $c_2({\cal V}_2)-14$\\\hline 
\end{tabular*}
\caption{{\it The cohomology associated to each representation of the low-energy gauge group SU(3).}}
\label{tabsu3222}
\end{center}
\end{table}
 
The anomaly cancelation condition in this case gives the relations
\begin{eqnarray}
n_H + 29 n_T -n_V =273 \\ \label{10danom22}
\Rightarrow  \frac{1}{2}c_2({\cal V}_2) + c_2({\cal V}_3) + \frac{1}{60} c_2(\textnormal{End}_0({\cal V}_{E_8}))= 24 \;.
\end{eqnarray}
Equation (\ref{10danom22}) is of course simply the 10D anomaly cancelation condition as one would expect.

Given Equation (\ref{10danom22}) and the matter multiplicities given in Table \ref{tabsu3222} we arrive at the following matter transition induced by
small instanton transitions in ten dimensions.
\begin{eqnarray}
{\bf 1} + {\bf 8} \leftrightarrow {\bf 3} +{\bf 6}
\end{eqnarray}
This is exactly of the form given in Equation (\ref{eq:su3trans1}) in the six-dimensional field theory discussion.

\section{Conclusions}
\label{sec:conclusions}

\subsection{Summary of results}

\noindent
{\bf Novel matter transitions}

\begin{figure}
\centering
\begin{picture}(200,100)(- 100,- 40)
\put(-120,55){\color{Blue} \vector(4, -1){240}}
\put(120,55){\color{Purple} \vector(-4, -1){240}}
\put(0,25){\circle*{5}}
\multiput(0,25)(0,10){4}{\line(0,1){5}}
\put(0,65){\vector(0,1){5}}
{\color{Red}
\put(0,21){\vector(4, -1){80}}
\put(0,21){\vector(-4, -1){80}}
\put(0,21){\circle*{3}}
\put(0,10){\makebox(0,0){$\hat{\varepsilon} = 0$}}
\put(-70,-7){\makebox(0,0){$\hat{\varepsilon} < 0$}}
\put(70,-7){\makebox(0,0){$\hat{\varepsilon} > 0$}}
}
\put(140, 10){\makebox(0,0){\color{Blue} ${\bf  15}\left(\;{\tiny\yng(1,1)}\;\right)
+{\bf  1}$}}
\put(-140, 10){\makebox(0,0) {\color{Purple} $\frac{1}{2} {\bf 20} \;\left(\;\frac{1}{2}{\tiny\yng(1,1,1)}\;\right)+
{\bf  6}\left({\tiny\yng(1)}\right)$}}
\put(0,82){\makebox(0,0){\small tensor branch}}
\end{picture}
\caption[x]{\footnotesize The matter transitions studied in this paper
  can be seen as arising along one-parameter families of theories,
  with the transition point at $\hat{\varepsilon} = 0$, and field
  theories with the same gauge group but distinct matter
  representation content at $\hat{\varepsilon} > 0$ and
$\hat{\varepsilon} < 0$.  For these matter transitions in 6D theories,
  the transition point at $\hat{\varepsilon} = 0$ is a superconformal
  field theory, from which an additional tensor branch generally
  extends, as well as separate branches for each of the field theories
  with distinct matter contents, though the tensor branch is
  incidental to the matter transition.  The matter representations
  shown are for the simplest (SU(6)) matter transition.}
\label{f:transitions}
\end{figure}
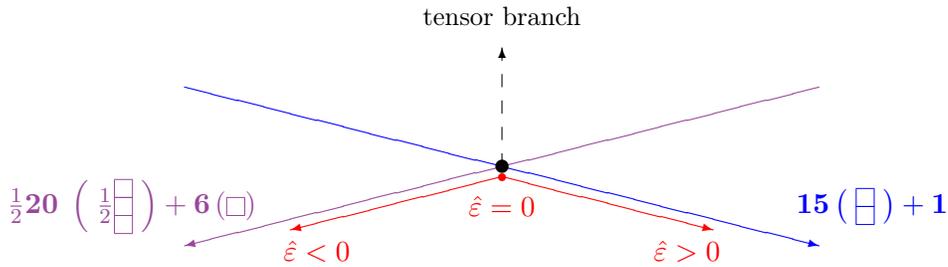

We have identified a new class of field theory transitions through
which matter fields in one set of representations transform into
matter fields in another set of representations without changing the
gauge group.  We have explicitly described these transitions in 6D
models where the field theory is coupled to gravity, from both the
heterotic and F-theory perspectives.  These pictures match, and agree
with constraints from gauge and gravitational anomaly conditions in
6D.  
In both pictures these transitions involve passing through a
superconformal fixed point (Figure~\ref{f:transitions}),  at which an infinite family of light
fields appear and the simple perturbative low-energy field theory
picture breaks down.  
Similar transitions should be possible in 4D
theories, though in some cases may be obstructed by a superpotential
(see \S\ref{sec:further}).  The simplest example we have studied is
the transition between a matter field in the three-index antisymmetric
($\Lambda^{3}$) representation of SU(6), along with a fundamental
($\frac{1}{2}${\bf 20} + {\bf 6}) to a two-index antisymmetric representation and a
singlet ({\bf 15} + {\bf 1}).  Similar transitions occur for other
groups such as SU(7),  Sp(3), SO(12), and SU(3). 
\vspace*{0.1in}

\noindent
{\bf Heterotic picture}

In the heterotic picture, these transitions occur by moving an
instanton between different simple factors in the structure group of a
single $E_8$ bundle, such as SU(3) $\times$ SU(2) in the SU(6)
$\Lambda^3$ case.  
Such transitions had not been previously explored systematically in heterotic theories.
Several novel features arise in these heterotic
models.  For the SU(7) and SU(8) models with $\Lambda^3$ matter, the
necessary bundles are only possible for a fixed subset of the moduli
space of K3 compact spaces, giving a nontrivial coupling between
bundle and complex structure moduli of the K3 surface.  We find a new form of the stable degeneration limit in these cases. The SU(7)
construction also leads to an increased Mordell-Weil rank on one side
of the stable degeneration limit, associated with a massive U(1) field
in the low-energy theory.
\vspace*{0.1in}

\noindent
{\bf F-theory picture}

In the F-theory picture, we have an explicit description of these
transitions in terms of Weierstrass models.  At the transition points
between models with different representations, codimension two
singular loci $L$ develop in the Weierstrass model where $f$ and $g$
vanish to orders 4 and 6, signaling SCFTs in the low-energy theory.
Matter transitions occur when the Weierstrass model moves from one
branch to another without changing the geometry of the
compactification base $B$.  Tensionless string transitions, on the
other hand, are
associated with a blowup of the locus $L$ and a corresponding change
in the topology of the base $B$
that in 6D changes the number of tensor multiplets.  The Weierstrass models for all
theories with non-generic matter representations ({\it e.g.}, for
SU($N$) anything but the fundamental, $\Lambda^2$ and adjoint
representations) are realized in a way that is not captured by the
simple general Tate form for the associated gauge group.
\vspace*{0.1in}

\noindent
{\bf Heterotic/F-theory duality}

In our analysis we have carried out a careful matching of the degrees
of freedom in the spectral cover construction of the heterotic bundles
with the parameters in the F-theory Weierstrass model.  Matching these
parameters gives a clear picture of the duality between these classes
of models and illuminates many features such as the appearance of
SCFTs at the transition points and the matching of constraints between
the two pictures. We view this work as important first step into extending heterotic/F-theory duality to include more complex and phenomenologically relevant Calabi-Yau geometries and vector bundles (including, for example the geometries in \cite{Anderson:2014hia,Anderson:2013xka,Anderson:2012yf,Anderson:2011ns} which involve bundles with reducible structure groups and Green-Schwarz massive $U(1)$ symmetries).
\vspace*{0.1in}

\noindent
{\bf Higgsing deformations}

In the process of constructing models with transitions, we have
discussed the F-theory and heterotic manifestations of a variety of
Higgsing processes. On the F-theory side, this analysis involved
identifying explicit Weierstrass deformations for particular Higgsing
processes, For instance, our analysis has described the F-theory
deformations that correspond to giving VEVs to antisymmetric
representations of $\gsu(N)$. Examining the Higgsing connections
between different models clarified how the transitions are related by
Higgsing and allowed us to investigate transitions in product
group models. Moreover, the explicit F-theory deformations illuminate
how the F-theory degrees of freedom correspond to those of the
low-energy theory, at least in six dimensions. Further investigations into Higgsing deformations
could help to develop a more explicit dictionary between F-theory, the heterotic effective theory, and
their low-energy limits.

\vspace*{0.1in}

\noindent
{\bf Higher genus matter}

An interesting subject of study in recent work is the appearance of
``higher genus'' matter in 6D supergravity theories.  Each
representation $R$ of a group $G$ can be associated with a genus
contribution $g$, which in F-theory should have the interpretation of
an arithmetic genus contribution from a singularity in the divisor
supporting the gauge factor $G$.  We have found an explicit example of
such a realization in models with the symmetric ({\bf  6})
representation of SU(3).  These models are connected through a matter
transition to other models with an adjoint representation, but are
realized, as in \cite{ckpt}, by non-Tate Weierstrass models that
exhibit a highly nontrivial cancellation in the vanishing of the
discriminant at higher orders.  The role of SCFTs in matter
transitions that we have uncovered here suggests a resolution of a
question regarding under what circumstances a Weierstrass model can
exhibit such an exotic higher genus matter representation other than
the adjoint: it seems in particular that a model formed by starting
with a Tate model for a group $G$ on a smooth divisor $D$ and then
deforming the divisor to a singular geometry will not develop an exotic
matter representation in the singular limit without moving through a
superconformal fixed point to a different branch of the set of
Weierstrass models
where the
algebraic structure of the model is changed, modifying
the cancellation mechanism in the discriminant.

\subsection{Further directions}
\label{sec:further}

\vspace*{0.1in}

\noindent
{\bf 4D realizations}

While we have focused here on 6D models, over which we have the
greatest level of control, similar transitions should be possible in
4D ${\cal N} = 1$ field theories, particularly in the context of
global models coupled to supergravity.  From the heterotic and
F-theory points of view, the 4D constructions are almost identical to
the 6D constructions.  In the heterotic picture, bundles with
structure groups within $E_8$ that are products like SU(3) $\times$
SU(2) can arise on Calabi-Yau threefolds as well as on K3.  And in the
F-theory picture, the algebraic structure of Weierstrass models giving
these matter transitions is formally identical in 4D or in 6D.  One
issue that may arise, however, is that while in 6D the geometric
moduli space precisely matches the moduli space of flat directions in
the low-energy theory, in four dimensions there is a superpotential
that lifts some of the flat directions.  Such a superpotential could
in principle either obstruct the passage between branches with
different matter content, or render unstable one of the branches.  In
4D there are fewer constraints from anomaly cancellation in the ${\cal
  N} = 1$ low-energy theory, so it is less clear why the specific
combinations of matter representations that admit matter transitions
in the 6D theories should be singled out particularly.  Recent work
\cite{Anderson:2014gla} has given a description of a very broad class
of F-theory/heterotic dual pairs for 4D F-theory compactifications.
Further study of 4D versions of the matter transitions explored here
in some of these dual geometries
would provide an interesting direction for further investigations.

\vspace*{0.1in}

\noindent
{\bf Geometry of transitions}

The matter transitions described here should correspond geometrically
to geometric transitions between distinct elliptically fibered
Calabi-Yau manifolds with the same $h^{1, 1}(X)$ but different
intersection structure.  Such transitions would be interesting to
study further from the purely geometric perspective, or in the context
of other types of string compactification.

\vspace*{0.1in}

\noindent
{\bf Other groups and representations}

We have focused here on a set of gauge groups and matter transitions
for the classical groups SU($N$), SO($N$) and Sp($N$)  that are related to
the basic SU(6) $\Lambda^3$ example by simple Higgsing transitions.
It would be interesting to explore further other possibilities,
including exceptional groups and higher representations, such as
tri-fundamental representations
of SU(2) $\times$ SU(2) $\times$ SU(2), {\it etc.}

\vspace*{0.1in}

\noindent
{\bf Global models and string universality}

The models with unusual matter representations that we have considered
here provide an interesting test of the 6D string universality
conjecture \cite{universality}, which suggests that all consistent
low-energy theories of matter fields, gauge fields and tensor fields
coupled to supergravity in six dimensions should have a  UV
description in string theory or F-theory.  The low-energy models with
SU(6) and SU(7) gauge groups and $\Lambda^3$ representations that are
acceptable from low-energy anomaly cancellation conditions we have
identified here in both heterotic and F-theory constructions as models
with good UV completions in string theory.  The SU(8) theory with a
$\Lambda^4$ matter representation, however, seems acceptable from the
low-energy point of view but does not seem to have a clear realization
in either the heterotic or F-theory pictures.  This raises the
question of whether this theory suffers from some as-yet-undiscovered
inconsistency, or can be realized in some new way in F-theory or
another string construction, or whether it actually represents a
counterexample to the string universality conjecture.  Similar
questions can be asked of the SU(9) $n_T = 1$ model with $\Lambda^3$
representations.  
An interesting feature of these models that seem acceptable from the
low-energy point of view but for which we cannot identify a consistent
F-theory or heterotic construction is that on the F-theory side they
would involve in principle an embedding of a Dynkin diagram of the
gauge group into $\hat{E}_8$ and not $E_8$ at the singular point.
Resolving whether there is some low-energy problem with such
configurations or some novel F-theory mechanism for realizing such
constructions is an interesting open problem.
It seems promising that models with larger $N$
and/or higher-index antisymmetric representations, such as SU(10) with
$\Lambda^3$ or SU(9) with $\Lambda^4$ seem to violate anomaly
cancellation so there is a close connection between what is allowed in
the low-energy theory and what can be realized through F-theory, with
the difference between these conditions giving only a small
intermediate zone of uncertainty.  One
possibly surprising feature here is that the constraints from the
low-energy theory seem to depend on the gravitational anomaly
cancellation condition, while the F-theory constraints on gauge groups
that admit a $\Lambda^3$ representation seem to come from local
considerations.  This should be understood better.

\vspace*{0.1in}

\noindent
{\bf  Higher genus matter}

The explicit F-theory construction found here of matter in the
symmetric representation of SU(3) complements another family of
Weierstrass models constructed recently \cite{ckpt} that realize the
same kind of matter fields.  These constructions, however, seem to
give slightly different classes of models.  It would be desirable to
have a more general understanding of how such Weierstrass models are
constructed and a general framework that would encompass both of these
classes of models.  It would also be interesting to construct more
general types of matter, such as a 3-symmetric representation of
SU($N$), using the kinds of analysis developed here.

\vspace*{0.1in}

\noindent
{\bf  Transitions between conjugate representations}

An interesting question, which we have not explored here, is the
extent to which matter transitions can occur for smaller groups like
SU(5).  While in 6D supergravity theories, the $\Lambda^3$
representation of SU(5) is the conjugate of the $\Lambda^2$
representation and therefore lies in the same hypermultiplets, a
Higgsing of the SU(6) $\Lambda^3$ matter transition appears to give a
class of SU(5) Weierstrass models where there is a transition between
$\Lambda^3$ and $\Lambda^2$ representations through an SCFT.  For 6D
theories, both branches of the theory seem to represent special cases
of the general SU(5) construction, so that there is no obstacle to
deforming the theory from one branch to the other without passing
through the SCFT.  In the F-theory picture this follows by taking a
general form of the parameter $\phi_0$ from (\ref{eq:fsu5}),
(\ref{eq:gsu5}) without any factorization as in
(\ref{eq:6-0})-(\ref{eq:6-2}).
In the heterotic picture the transition can be realized by building an
SU(5) bundle with a specialized structure group S(U(2)$\times$ U(3)),
where there can be small instanton transitions between the factors,
but for compactifications on K3 there is no obstruction to deformation
to a general SU(5) bundle (so long as the required matter is present for higgsing).
For 4D theories, on the other hand, there may be obstructions to
moving between these branches, which would stem physically from
the fact that in 4D ${\cal N} = 1$ theories there is a distinction
between chiral multiplets in the $\Lambda^2$ and conjugate
representations.  This would be interesting to investigate
further.

\vspace*{0.1in}

\noindent
{\bf   Massive U(1)'s in heterotic/F-theory duality}

We have found that some models such as the SU(7) and SU(8) models with exotic
matter seem to always give rise to massive U(1)'s in the heterotic description, which
correspond to an enhanced Mordell-Weil group on one side of the stable
degeneration limit in the corresponding F-theory picture, but not
both.  It would be interesting to further explore the physical consequences of these massive abelian symmetries in the low-energy theory away from the stable degeneration limit. Particularly in 4D compactifications, it is expected that the discrete remnants of such massive symmetries can significantly affect the structure of the theory --including Yukawa couplings, K\"ahler potentials and the vacuum structure of the theory \cite{Blumenhagen:2005ga,Blumenhagen:2006ux,Kuriyama:2008pv,Anderson:2010tc,Anderson:2010ty}.
\vspace*{0.1in}

\noindent
{\bf   Superconformal points}

We have found that the transitions between different matter fields
occur at points in the heterotic moduli space where instantons have
shrunk to a point, corresponding to points in the F-theory Weierstrass
moduli space where (4, 6) singularities have arisen at codimension two
loci in the base.  These are the same transition points that give
superconformal fixed points in the low-energy theory, and which lead
to tensionless string transitions to models with more tensor
multiplets in the 6D framework.  A large class of 6D SCFT's were
constructed from the F-theory point of view in
\cite{Heckman-Morrison-Vafa} and couplings of these theories
to supergravity were
explored in \cite{DelZotto-Heckman-Morrison-Park}.
It would be interesting to investigate further how the SCFT's that
play the role of mediating the matter transitions considered here fit
into this picture.
In particular, it would be nice to find some clear way of following
the transition from one field theory to another through the
superconformal point strictly in the language of the low-energy theory.

\vspace{.5in}

{\bf Acknowledgements}: We would like to thank Mirjam Cvetic, Jim Halverson, Jonathan Heckman, Antonella Grassi,
Denis Klevers, and David Morrison for helpful
discussions.  LA and JG would like to thank the University of
Pennsylvania and MIT's Center for Theoretical Physics for hospitality during the completion of this work. The
work of
NR and WT was supported by
the DOE under contract
\#DE-SC00012567, and was also supported in
part by the National Science Foundation under Grant No. PHY-1066293. The work of LA  is supported by NSF grant PHY-1417337 and that of JG is supported by NSF grant PHY-1417316.

\appendix

\section{Addition of sections}\label{elliptic_addition}
In this section we review briefly the addition of points on the elliptic fiber (see \cite{Morrison:2012ei,hartshorn} for reviews). Given a Weierstrass model in $\mathbb{P}[1,2,3]$ 
\beq
y^2=x^3+fxz^4 +gz^6
\eeq 
the zero section can be defined to be $(x,y,z)=(1,1,0)$. On the affine patch where $z=1$, the addition (denoted $[+]$) of two points $(a,b)$ and $(A,B)$ on the elliptic fiber can be defined as follows. On this patch the Weierstrass equation can be written as
\begin{align}
y^2=x^3+fx +g & =(x-a)(x^2+ax+c)+b^2 \\
&=(x-A)(x^2+Ax +C)+B^2
\end{align}
Then, ${\cal P}=p [+] P=({\cal A}, {\cal B})$ can be defined by demanding that $({\cal A},{\cal -B})$ is the third intersection point of the line that joins the points $p$ and $P$. Then, in the notation of \cite{Morrison:2012ei},
\beq
{\cal P}=\left( \left(\frac{B-b}{A-a}\right)^2-(a+A),-\left(\frac{B-b}{A-a}\right)^3+(2a+A)\left(\frac{B-b}{A-a} \right)-b \right)
\eeq
Likewise, the point $2P=P [+] P$ is defined via
\beq
2P=\left( \left(\frac{C+2A^2}{2B}\right)^2-2A),-\left(\frac{C+2A^2}{2B}\right)^3+(3A)\left(\frac{C+2A^2}{2B} \right)-B \right)
\eeq

\section{Three stable degeneration limits for SU(8)}\label{su8paths}
As described in Section \ref{su8_spec_tune}, there are three possible paths that lead to stable degeneration limits for the SU(8) F-theory geometry. Of these, only one leads to a smooth $K3$ manifold in the heterotic dual theory. In order to take the stable degeneration limit given by equations \eqref{fib_prod} and \eqref{epsilons}, it is necessary to decide on a limit which takes $f_i, g_j \to 0$ for $i>4$ and $j>6$. A Groebner basis calculation \cite{Gray:2008zs} demonstrates that the most general path to such a solution can be described by the following equations (the primary decomposition of the stable degeneration locus):
\begin{align}\label{branches}
4 \text{$\zeta_3$} \text{$\psi_5$}^2 \text{$\phi_4$}-24 \text{$\lambda_2$} \text{$\phi_4$}^3-\tau ^2 \text{$\psi_5$}^3-4 \text{$\psi_5$} \text{$\omega_1$} \text{$\phi_4$}^2   = 0\\
\delta ^2 \text{$\psi_5$}+2 \text{$\zeta_4$} \text{$\phi_4$}=0 \label{eq:branchtwo}\\
12 \delta ^2 \text{$\lambda_2$} \text{$\phi_4$}^2+4 \text{$\zeta_3$}\text{$\zeta_4$} \text{$\psi_5$} \text{$\phi_4$}-\text{$\zeta_4$} \tau ^2 \text{$\psi_5$}^2-4 \text{$\zeta_4$} \text{$\omega_1$} \text{$\phi_4$}^2=0\\
6 \delta ^4 \text{$\lambda_2$} \text{$\phi_4$}-2 \delta ^2 \text{$\zeta_4$} \text{$\omega_1$} \text{$\phi_4$}-4 \text{$\zeta_3$}\text{$\zeta $4}^2 \text{$\phi_4$}+\text{$\zeta $4}^2 \tau ^2 \text{$\psi_5$}=0\\
3 \delta ^6 \text{$\lambda_2$}-\delta ^4 \text{$\zeta_4$} \text{$\omega_1$}-2 \delta ^2 \text{$\zeta_3$}\text{$\zeta $4}^2-\text{$\zeta $4}^3 \tau ^2=0
\end{align}
Note that if either $\zeta_4$ or $\tau$ share factors with $\delta$, $(4,6)$ singularities are unavoidable on the shared factor. 

We will first consider \eqref{eq:branchtwo}. This equation would imply that $\zeta_4 \phi_4$ is proportional to to $\delta^2$, and every zero of $\delta$ must be in either $\zeta_4$ or $\phi_4$. But $\delta$ and $\zeta_4$ cannot share any zeroes if we wish to avoid (4,6) singularities. As a result, $\phi_4$ must be proportional to $\delta^2$, or
\begin{equation}
\phi_4 = \phi_4^\prime \delta^2.
\end{equation}
This in turn implies that
\begin{equation}
\psi_5 = -2 \zeta_4 \phi_4^\prime
\end{equation}
If the expression for $\psi_5$ is substituted into \eref{branches}, it leads to
\begin{equation}
4{\phi_4^\prime}^3 \left(2\zeta_4^3 \tau^2 + \delta^2\left(\zeta_3 \zeta_4^2 + 2 \delta^2 \zeta_4 \omega_1 - 6 \delta^4 \lambda_2\right)\right) = 0
\end{equation}
Unless $\phi_4^\prime=0$, we must have that $\zeta_4^3 \tau^2$ is proportional to $\delta^2$. However, this possibility would introduce $(4,6)$ singularities and will therefore not be a valid geometry. Thus, the only option is to take $\phi_4^\prime=0$, and both $\phi_4$ and $\psi_5$ must go to zero. In fact, once \eqref{eq:branchtwo} is solved as above, all of the other equations similarly lead to singular geometries. As a result the $\phi_4=\psi_5=0$ branch is the unique smooth solution. 

\section{Stable Degeneration Limits and U(1) Symmetries}\label{stab_degenu1}
In this section, we briefly explore the compatibility of the stable degeneration limit and the existence of non-zero rank Mordell-Weil group in the Calabi-Yau geometry. The existence of a log semi-stable degeneration limit \cite{Donagi:2012ts,Aspinwall:1998bw} depends globally on being able to consistently define $dP_9$-fibered $n$-folds $Y^{(1)}$ and $Y^{(2)}$ and their ability to share a common divisor $D=X_n$ a CY manifold of one dimension lower. Moreover, the limit requires that the fibrations of $Y^{(i)}$ and $X_n$ be compatible -- that is, the elliptic fiber of $X_n$ should be the same form as the elliptic fiber of $Y^{(i)}$, etc. In what follows, we will demonstrate that the process of stable degeneration and presentation of a manifold in Weierstrass form do not necessarily commute. When this paper was in the final stages of preparation, \cite{Cvetic:2015uwu} appeared which comprehensively studies the above questions using a different approach. For concreteness, here we will illustrate the relevant ambiguities in stable degeneration with elliptically fibered threefold, $Y$, with base $\mathbb{F}_n$.

To begin, we briefly review the ``standard" stable degeneration limit in the case that the elliptic fibration of $Y$ admits a single section (i.e. the rank of $MW(Y)$ is vanishing). That is, $Y$ is a generic Weierstrass model over $\mathbb{F}_n$. We can realize this torically as a hypersurface with $P_{123}[6]$ fiber. That is, the charge matrix
\beq
 Y= \begin{tabular}{c|ccccccc}
     & $y$  & $x$  & $z$  & $x_0$  & $x_1$ & $y_0$  & $y_1$      \\ 
    \hline 
    6  & 3 & 2 & 1 & 0 & 0 & 0 & 0 \\
   0 & 0  & 0  & -2 & 0 & 0 & 1 & 1\\
   0 & 0 & 0 &  -2-n & 1 & 1 & n & 0
  \end{tabular}
  \label{standardcase}
\eeq
where the first column denotes a degree $(6,0,0)$ hypersurface in the toric ambient space. For this choice of global description of the manifold, the stable degeneration limit corresponds to choosing
\beq
Y^{(1)}= \begin{tabular}{c|ccccccc}
     & $y$  & $x$  & $z$  & $x_0$  & $x_1$ & $y_0$  & $y_1$      \\ 
    \hline 
    6  & 3 & 2 & 1 & 0 & 0 & 0 & 0 \\
   0 & 0  & 0  & -1 & 0 & 0 & 1 & 1\\
   0 & 0 & 0 &  -2-n & 1 & 1 & n & 0
  \end{tabular} \label{halfk3_standard}
\eeq
(with $Y^{(2)}$ similar). The defining equation of $Y^{(1)}$ is of the same form as that of $Y$ (i.e. of Weierstrass form: $y^2=x^3+f x z^4+ g z^6$) but with $f,g$ truncated at degree $\leq 4,6$ in the coordinate ($\sigma=y_0$). This global description of $Y^{(1)}$ is equivalent to the scaling limit defined in \eqref{epsilons}. Briefly, for the Weierstrass coefficients of $Y$ above, we choose
\beq
f\sim \sum_{i=0}^{8} f_i \sigma^i~~~~~,~~~~g \sim  \sum_{j=0}^{12} g_j \sigma^j
\eeq
where $\sigma=0$ and $\sigma=\infty$ define the poles of the $\mathbb{P}^1$ fiber of $\mathbb{F}_n$. To take the stable degeneration limit, a scaling is chosen in which
\begin{align}\label{epsilons2}
& f_i ~~\text{scales as}~~ \epsilon^{(i-4)} \\
& g_j ~~\text{scales as} ~~\epsilon^{(j-6)}~~. \nonumber
\end{align}
which in the limit that $\epsilon \to 0$ ``separates" the $dP_9$-fibered halves of the $Y$. Note that in this scaling, the divisor $D=K3$ along which $Y^{(1)}$ and $Y^{(2)}$ are glued is defined by $f_4,g_6$ (the coefficients of weight zero in $\epsilon$). These "middle" coefficients define the moduli of an elliptically fibered $K3$ surface with a compatible ({\it i.e.} $P_{123}[6]$) fibration.

Now, in constrast, when the F-theory geometry \emph{has a higher rank Mordell-Weil group} then the stable degeneration limit may differ with the global description of $Y$. The following example provides a simple example of a geometry where the process of presenting a manifold in Weierstrass form and the stable degeneration limit do not commute.

Following \cite{Morrison:2012ei}, let us realize the elliptic fiber of a generic $rk(MW)=1$ F-theory geometry, $Y$, as a toric blow up of $P_{112}[4]$. For concreteness, consider the global geometry 
\beq
 Y= \begin{tabular}{c|ccccccccc}
     & $u$  & $v$  & $w$  & $t$ & $s$ & $x_0$  & $x_1$ & $y_0$  & $y_1$      \\ \hline 
    1 & 1 & 0 & 1& -1 & 0 &0 & 0 & 0 & 0 \\
   4 & 1  & 1  & 2 & 0 & 0&  0 & 0 & 0 & 0\\
   2 & 0 & 0 &   1 & 0 & 1 & 0 & 0 & 0 & 0 \\
    0 & 0  & -2  & 0 &0 &0 & 0 & 0 & 1 & 1\\
   0 & 0 & -n-2 &  0 &0 & 0& 1 & 1 & n & 0
  \end{tabular}
  \label{2secmodel}
\eeq
This leads to generic defining equation of the form 
\beq\label{def2sec}
w^2s+b_0 u^2 w s^2 t+b_1 u v w s t + b_2 v^2 w t=c_0 u^4 t^2 s^3 +c_1 u^3 v t^2 s^2+ c_2 u^2 v^2 t^2 s + c_3 u v^3 t^2
\eeq
where $b_i, c_j$ are functions of the base coordinates $(x,y)$. By shifting $w$ by a multiple of $u$ it is possible to set $b_0=b_1=0$. Labeling $b_2$ simply as $b$, and letting subscripts denote degree in $x_i$, we have explicitly
\begin{align}\label{explicit_coeffs}
&b = y_0^4 b_{4-2n}(x)+ y_0^3 y_1 b_{4-n}(x)+y_0^2 y_1^2b_4(x)+ y_0 y_1^3 b_{n+1}(x)+y_1^4b_{2n+4}(x) \\
&c_1=  y_0^2{(c_1)}_{2-n}(x)+y_0y_1 {(c_1)}_2(x) + y_1^2 {(c_1)}_{n+2}(x) \nonumber \\
&c_2= y_0^2 {(c_2)}_{4-2n}(x)+ y_0^3y_1 {(c_2)}_{4-n}(x)+ y_0^2y_1^2 {(c_2)}_{4}(x)+y_0 y_1^3{(c_2)}_{n+4}(x)+y_1^4 {(c_2)}_{2n+4}(x) \nonumber \\
&c_3 = y_0^6 {(c_3)}_{6-3n}(x)+ y_0^5 y_1 {(c_3)}_{6-2n}(x) + y_0^4y_1^2 {(c_3)}_{6-n}(x) +y_0^3y_1^3{(c_3)}_6(x) \nonumber \\&+y_0^2y_1^4 {(c_3)}_{n+6}(x)+y_0y_1^5{(c_3)}_{2n+6}(x) + y_1^6{(c_3)}_{3n+6}(x) \nonumber
\end{align}

Moreover, $Y$ in (\ref{2secmodel}) has a compatible $K3$ fibration given by

\beq
K3= \begin{tabular}{c|cccccccc}
     & $u$  & $v$  & $w$  & $t$ & $s$  & $y_0$  & $y_1$      \\ \hline 
    1 & 1 & 0 & 1& -1 & 0 & 0 & 0 \\
   4 & 1  & 1  & 2 & 0 & 0& 0 & 0\\
   2 & 0 & 0 &   1 & 0 & 1 & 0 & 0 \\
    0 & 0  & -2  & 0 &0 &0 & 1 & 1\\
  \end{tabular}
  \label{K3_2secmodel}
\eeq
Here once again we have the defining relation (\ref{def2sec}) where the degrees of the coefficients in the $K3$ case in terms of the base variables ($y$) are given by
\beq
deg(c_0)=0~,~ deg(c_1)=2~,~deg(b)=deg(c_2)=4~,~deg(c_3)=6
\eeq

Now to take the stable degeneration limit of the threefold given in (\ref{2secmodel}) we need to identify the ``middle" $K3$ along which $Y^{(i)}$ will be glued. Here the "middle" $K3$ coefficients inside of $X$ are given by 
\begin{align}\label{goodK3}
&b \sim y_0^2 y_1^2b_4 \\
&c_1 \sim  y_0y_1 {(c_1)}_2 \\
&c_2 \sim  y_0^2y_1^2 {(c_2)}_{4} \\
&c_3 \sim y_0^3y_1^3{(c_3)}_6  
\end{align}
More precisely, as in the standard case above, the stable generation limit can be defined via a scaling of the coefficients of $b, c_i$. To take the stable degeneration limit $Y \to Y^{(1)} \cup_{D} Y^{(2)}$ with $Y$ defined by \eqref{2secmodel} and $D$ defined via \eqref{K3_2secmodel} it is possible to choose
\begin{align}\label{epsilons3}
& b_i ~~~~~~\text{scales as}~~ \epsilon^{(i-4)} \\
& (c_1)_j ~~\text{scales as} ~~\epsilon^{(j-2)} \nonumber \\
& (c_2)_k~~\text{scales as} ~~\epsilon^{(k-4)} \nonumber \\
& (c_3)_l~~\text{scales as} ~~\epsilon^{(l-6)}~~~. \nonumber
\end{align}
In the limit that $\epsilon \to 0$ this separates $Y$ into two $dP_9$ fibered (non-CY) $3$-folds. Now we come to the central observation to be made from the above geometry: In this stable degeneration limit, it is straightforward to verify that \emph{the presence of a non-trivial Mordell-Weil group is fully preserved} not only in each of $Y^{(1)}, Y^{(2)}$ but also in the "middle" $K3$ surface \eqref{K3_2secmodel} (defined by the $\epsilon$ weight zero terms above).

Having come thus far, it should now be recalled \cite{Morrison:2012ei} that it is always possible via coordinate redefinitions (i.e. the Jacobian procedure) to put a two-section model such as $Y$ in \eref{2secmodel} explicitly into Weierstrass form. The dictionary to Weierstrass form defines the coefficients

\begin{align}\label{morrison_park}
&f=c_1c_3-b^2c_0 -\frac{1}{3}c_2^2 \\
&g=c_0 c_3^2 - \frac{1}{3} c_1c_2c_3 + \frac{2}{27}c_2^3 -\frac{2}{3}b^2c_0
   c_2 + \frac{1}{4}b^2 c_1^2
\end{align}
With this explicit defining relation we are once again considering a manifold with $\mathbb{P}_{123}[6]$ fiber type of the form given in \eqref{standardcase}. For such a geometry, the usual stable degeneration limit (i.e. the splitting given in \eqref{halfk3_standard}) and the scalings defined in \eqref{epsilons2} can be employed. 

To begin, the ``middle" $K3$ coefficients can be readily identified as those of $\epsilon$ weight zero. For example combining \eqref{explicit_coeffs} with \eqref{morrison_park} the coefficients of ${y_0}^4{y_1}^4$ in $f$ include the following terms from the product $c_1c_3$
\beq\label{part_of_f}
c_1 c_3 \sim ({y_0}^4{y_1}^4)\left({(c_1)}_2{(c_3)}_6+{(c_1)}_{n+2}{(c_3)}_{-n+6}+{(c_1)}_{-n+2} {(c_3)}_{n+6} \right)
\eeq
(with many further terms of this order arising from $b^2 c_0$ and ${c_2}^2$). These coefficients (and the coefficients of $y_0^6 y_1^6$ in $g$) determine a "middle" $K3$ surface along which the standard stable degeneration limit would glue the new $\mathbb{P}_{123}[6]$-fibered manifolds $Y^{(i)}$. Note that this same result could be obtained simply by performing the scaling rule given in \eqref{epsilons3} since according to that rule, the epsilon factors cancel in terms like ${(c_1)}_{n+2}{(c_3)}_{-n+6}$ above.

It is clear by inspection of the Weierstrass coefficients of the $K3$ surface (including the terms in \eqref{part_of_f}) that it is not of the form required for a higher rank Mordell-Weil group. Moreover it \emph{does not} define the same $K3$ surface as we obtained in \eqref{K3_2secmodel} and \eqref{goodK3}. Although the Weirerstrass form for $Y$ given by \eqref{morrison_park} initially has a non-trivial Mordell-Weil group, that structure is \emph{not in general preserved in either $Y^{(1)},Y^{(2)}$ or the gluing divisor $D=K3$}. In this case, unlike in the stable degeneration limit of \eqref{2secmodel}, the rank of Mordell-Weil is reduced in stable degeneration. Thus finally, we have reached our central observation: the procedures of stable degeneration of a global elliptically fibered geometry and putting that fibration into explicit Weierstrass form, do not in general commute. That is, varying the order of these operations leads to different weakly coupled limits.


\section{Details of heterotic Higgsing analysis} \label{hetHiggsapp}

In this appendix, we provide some of the details of the computations which underly the heterotic description of the Higgsing processes as described in Tables \ref{tab2trans}, \ref{tab3trans} and \ref{tab4trans}. The cases of Higgsing SU(6) on fundamental, {\rm SU}(7) on two-index antisymmetric and SU(8) on three-index antisymmetric matter are described in the main text.

\subsection{Higgsing on fundamental matter}

\subsubsection{SU(7)}

Let us consider what happens to bundle topology as we Higgs from SU(7) to SU(6). The relevant group theory is
\begin{eqnarray} \label{agroupbreak1}
{\rm SU}(7) &\to& {\rm SU}(6) \times {\rm U}(1) \\ \nonumber
{\bf 7} &=&{\bf 1}_{-6} + {\bf 6}_1 \\\nonumber
{\bf 21} &=& {\bf 6}_{-5} + {\bf 15}_2 \\\nonumber
{\bf 35} &=& {\bf 15}_{-4} + {\bf 20}_{3}
\end{eqnarray}
Clearly, we wish to turn on the singlet of SU(6) inside SU(7) to achieve the Higgsing. However, we see from Table \ref{tab10} that there are two types of ${\bf 7}$ (and indeed $\overline{{\bf 7}}$) one associated to $H^1({\cal V}_2^{\vee} \otimes {\cal L})$ and another to $H^1({\cal L}^{\vee 2})$. It turns out that, at an arbitrary point in the moduli space of the base K3, we can not simply choose which of these to give a VEV too. A combination of both must be turned on simultaneously.

We can see this structure by examining the change in bundle geometry in going from an SU(7) to SU(6) visible symmetry. Turning on the VEV associated to $H^1({\cal V}_2^{\vee} \otimes {\cal L})$ corresponds to forming the following extension.
\begin{eqnarray} \label{avsu3}
0 \to {\cal L} \to {\cal V}_{SU(3)} \to  {\cal V}_2 \to 0
\end{eqnarray}
Turning on the field associated to $H^1({\cal L}^{\vee 2})$ corresponds to forming this extension.
\begin{eqnarray} \label{vsu2}
0 \to {\cal L}^{\vee } \to {\cal V}_{SU(2)} \to {\cal L} \to 0
\end{eqnarray}
The conjugate representations that must also be
given VEVs correspond to the dual to the above two sequences and the actual SU(3) and SU(2) bundles are formed from the combination of these mutually dual extensions in the usual manner \cite{Li:2004hx}.

We see now that the need to turn on both types of ${\bf 7}$ corresponds to the need to form a bundle with structure group SU(3) $\times$ SU(2), which is the relevant case to arrive at SU(6). At special loci in the moduli space of K3 one can leave one of the two extensions (\ref{avsu3}) and (\ref{vsu2}) split and still maintain bundle poly-stability (this is actually the same locus on which the original S(U(2) $\times$ U(1)) bundle is poly-stable). Splitting one of these two bundles in this fashion would induce an extra U(1) factor in the commutant of the bundle structure group inside $E_8$. This additional abelian factor will be Green-Schwarz anomalous however \cite{Dine:1987xk,Blumenhagen:2005ga,Lukas:1999nh,Sharpe:1998zu,Anderson:2009nt,Anderson:2009sw}.

One can check that the matter one achieves in the SU(6) case obtained by Higgsing SU(7) in this fashion agrees with what one would get by plugging in the topology of the above bundles into a direct computation of the spectrum of an SU(6) model, as given in Table \ref{tab9}, a calculation to which we now turn.

Equation (\ref{agroupbreak1}) describes how the SU(7) representations that we have in our initial theory branch to SU(6) representations under the breaking. We start with a number of each of these representations, determined by the topology of the bundle ${\cal V} = {\cal V}_2  \oplus {\cal L}$, as detailed in Table \ref{tab10}. This information is enough to determine the spectrum after breaking.

Alternatively, from equations (\ref{avsu3}) and (\ref{vsu2}) we can determine the topology of the bundle ${\cal V}_{SU(3)} \oplus {\cal V}_{SU(2)}$ that we transition to and from there, using Table \ref{tab9}, the matter content after the Higgsing. The second Chern classes of the two bundles can easily be determined to be,
\begin{eqnarray} \label{ac2c2}
c_2 ({\cal V}_{SU(3)}) &=& c_2({\cal V}_2) -  c_1({\cal L})^2 \\ \nonumber
c_2 ({\cal V}_{SU(2)}) &=& - c_1({\cal L})^2\;,
\end{eqnarray}
which can be then used to determine the matter content of the resulting SU(6) theory.

In Table \ref{atranstab11} we present the matter content of an SU(6) bundle with the topology just described, together with the matter content that would be naively expected under a transition from the SU(7) theory to SU(6) using Table \ref{tab10} and the branching rules (\ref{agroupbreak1}).

\begin{table}[!ht]
\begin{center}
\begin{tabular*}{15.7cm}{|c|c|c|}
\hline
SU(6) Representation & \# from SU(7) multiplet decomposition & \# found after transition \\ \hline
${\bf 1}$ & $6 c_2({\cal V}_2) -10 c_1({\cal L})^2 -18$  & $6 c_2({\cal V}_2) -10 c_1({\cal L})^2 -22$ \\
${\bf 6}$ & $2 c_2({\cal V}_2) - 10 c_1({\cal L})^2 - 10$  & $2 c_2({\cal V}_2) - 5 c_1({\cal L})^2 - 12$ \\
$\overline{{\bf 6}}$ & $2 c_2({\cal V}_2) - 10 c_1({\cal L})^2 - 10$& $2 c_2({\cal V}_2) - 5 c_1({\cal L})^2 - 12$\\
${\bf 15}$ & $c_2({\cal V}_2)- c_1({\cal L})^2 - 6$ &  $c_2({\cal V}_2)- c_1({\cal L})^2 - 6$\\ 
$\overline{{\bf 15}}$ & $c_2({\cal V}_2)- c_1({\cal L})^2 - 6$ &  $c_2({\cal V}_2)- c_1({\cal L})^2 - 6$\\
${\bf 20}$ & $- c_1({\cal L})^2 -4$&  $- c_1({\cal L})^2 -4$\\ 
\hline
\end{tabular*}
\caption{{\it Matter content after Higgsing an SU(7) to an SU(6) theory, both via a naive decomposition of the initial SU(7) multiplets and via a direct computation from the resulting SU(6) bundle.}}
\label{atranstab11}
\end{center}
\end{table}

We can now observe that this result matches the field theory analysis of the Higgs mechanism given in Section \ref{higsfund}. 

\subsubsection{SU(8)}\label{su8app}

The relevant group theory in this case is as follows.
\begin{eqnarray} \label{agroupbreak2}
{\rm SU}(8) &\to& {\rm SU}(7) \times {\rm U}(1)\\
{\bf 8} &=& {\bf 7}_1 + {\bf 1}_{-7} \\
{\bf 28} &=& {\bf 7}_{-6} + {\bf 21}_2\\
{\bf 56} &=& {\bf 21}_{-5} + {\bf 35}_3
\end{eqnarray} 
Given this, giving a VEV to the singlet of SU(7) inside the $\overline{{\bf 8}}$ of SU(8) will Higgs SU(8) to SU(7). Looking at Table \ref{tab11}, we see that the $\overline{{\bf 8}}$'s correspond to elements of $H^1({\cal L}^{\vee 3})$. Giving such a field a VEV corresponds to forming the following extension (and its dual via the associated ${\bf 8}$ VEV).
\begin{eqnarray} \label{asu8trans1}
0 \to {\cal L}^{\vee 2} \to {\cal V}_2 \to {\cal L} \to 0
\end{eqnarray}
This is a U(2) bundle which can form part of an S(U(2)$\times$ U(1)) object, in order to break to SU(7) (with, in addition, a Green-Schwarz anomalous U(1)), as follows.
\begin{eqnarray} \label{asu8trans2}
{\cal V} =  {\cal V}_2\oplus {\cal L}
\end{eqnarray}

As in previous cases, we now compare the SU(7) spectrum that is achieved by a decomposition of the parent SU(8) theory to the spectrum associated to the SU(7) theory defined by (\ref{asu8trans1}) and (\ref{asu8trans2}).

\begin{table}[!h]
\begin{center}
\begin{tabular*}{15.7cm}{|c|c|c|}
\hline
SU(7) Representation & \# from SU(8) multiplet decomposition & \# found after transition \\ \hline
${\bf 1}$ & $-9c_1({\cal L})^2-4$  & $-9 c_1({\cal L})^2 -6$ \\
${\bf 7}$ & $-\frac{13}{2}c_1({\cal L})^2 -4$  & $-\frac{13}{2} c_1({\cal L})^2 - 6$ \\
$\overline{{\bf 7}}$ & $-\frac{13}{2}c_1({\cal L})^2 -4$& $-\frac{13}{2} c_1({\cal L})^2 - 6$\\
${\bf 21}$ & $-\frac{5}{2} c_1({\cal L})^2 - 4$ &  $-\frac{5}{2} c_1({\cal L})^2 - 4$\\ 
$\overline{{\bf 21}}$ & $-\frac{5}{2} c_1({\cal L})^2 - 4$ &  $-\frac{5}{2} c_1({\cal L})^2 - 4$\\ 
${\bf 35}$ & $-\frac{1}{2} c_1({\cal L})^2 - 2$ &  $-\frac{1}{2} c_1({\cal L})^2 - 2$\\
$\overline{{\bf 35}}$ & $-\frac{1}{2} c_1({\cal L})^2 - 2$&  $-\frac{1}{2} c_1({\cal L})^2 - 2$\\ 
\hline
\end{tabular*}
\caption{{\it Matter content after Higgsing an SU(8) to an SU(7) theory, both via a naive decomposition of the initial SU(8) multiplets and via a direct computation from the resulting SU(7) bundle.}}
\label{atranstab2}
\end{center}
\end{table}

In Table \ref{atranstab2} we have used the fact that, for the bundle given in (\ref{asu8trans1}), 
\begin{eqnarray}
c_2({\cal V}_2) &=& -2 c_1({\cal L})^2\;.
\end{eqnarray}

As in previous cases, the differences between the second and third columns in Table \ref{atranstab2} precisely match what we would expect from an analysis of the Higgs mechanism in such a situation. This Higgsing is precisely of the form described in a field theory context in Section \ref{antihigs}.

It should also be mentioned that, during this transition, the ray in the moduli space of K3 where the S(U(1)$\times$ U(1)) bundle and S(U(2)$\times$ U(1)) bundle are slope poly-stable is the same. Therefore the K3 moduli expectation values need not change during this process.

\subsection{Higgsing on two-index antisymmetric matter}

\subsubsection{SU(6)}

The relevant group theory in this case is as follows.

\begin{eqnarray} \label{agroupbreaksu62}
{\rm SU}(6) &\to& {\rm SU}(4) \times {\rm SU}(2) \times {\rm U}(1) \\
{\bf 6} &=& ({\bf 1},{\bf 2})_{-2}+({\bf 4},{\bf 1})_1 \\ 
{\bf 15} &=& ({\bf 1},{\bf 1})_{-4} + ({\bf 4},{\bf 2})_{-1} + ({\bf 6},{\bf 1})_2 \\
{\bf 20} &=& ({\bf 4},{\bf 1})_{-3}+( \overline{{\bf 4}},{\bf 1})_3 + ({\bf 6},{\bf 2})_0
\end{eqnarray}

The breaking pattern in (\ref{agroupbreaksu62}) corresponds to giving a VEV to a ${\bf 15}$, $\overline{{\bf 15}}$ pair. The ${\bf 15}$'s, according to Table \ref{tab9}, lie in the cohomology $H^1({\cal V}_3^{\vee})$. In terms of bundle topology, giving an expectation value to such a field corresponds to forming the following bundle.
\begin{eqnarray} \label{currentguy}
&&{\cal V}= {\cal V}_2 \oplus {\cal V}_4  \\ \nonumber
\textnormal{where} && 0 \to {\cal V}_3 \to {\cal V}_4 \to {\cal O} \to 0
\end{eqnarray}
The conjugate representation which must also be given a
VEV corresponds to the dual to the above sequence and the actual SU(3) bundle is formed from the combination of these mutually dual extensions in the usual manner \cite{Li:2004hx}.

The group theory for a SU(4) $\times$ SU(2) compactification of heterotic is as follows.
\begin{eqnarray}
E_8 &\supset& {\rm SU}(4) \times {\rm SU}(2) \times {\rm SU}(4) \times {\rm SU}(2) \\ \nonumber
{\bf 248} &=& ({\bf 1},{\bf 1},{\bf 1},{\bf 3})+({\bf 6},{\bf 2},{\bf 1},{\bf 2})+({\bf 1},{\bf 2},{\bf 6},{\bf 2})+({\bf 4},{\bf 1},\overline{{\bf 4}},{\bf 2})+(\overline{{\bf 4}},{\bf 1},{\bf 4},{\bf 2}) \\ \nonumber&&+({\bf 1},{\bf 3},{\bf 1},{\bf 1})+(\overline{{\bf 4}},{\bf 2},{\bf 4},{\bf 1}) + ({\bf 4},{\bf 2},\overline{{\bf 4}},{\bf 1})+({\bf 6},{\bf 1},{\bf 6},{\bf 1}) \\ \nonumber &&+({\bf 15},{\bf 1},{\bf 1},{\bf 1})+({\bf 1},{\bf 1},{\bf 15},{\bf 1})
\end{eqnarray}
This leads to the low-energy spectrum given in Table \ref{atabastrans1zero}. 
\begin{table}[!h]
\begin{center}
\begin{tabular*}{12.8cm}{|c|c|c|}
\hline
Representation & Cohomology & Multiplicity \\ \hline
$({\bf 1},{\bf 1})$ & $H^1 (\textnormal{End}_0({\cal V}_2)) \oplus H^1 (\textnormal{End}_0({\cal V}_4))  $ &  $4 c_2({\cal V}_2)+8 c_2({\cal V}_4) -36$\\ 
$({\bf 6},{\bf 2})$ & $H^1 ({\cal V}_2) $ &  $c_2({\cal V}_2) -4$\\ 
$({\bf 1},{\bf 2})$ & $H^1( {\cal V}_2 \otimes \wedge^2 {\cal V}_4)$&$6 c_2({\cal V}_2) + 4 c_2({\cal V}_4) -24$\\
$({\bf 4},{\bf 1})$ & $ H^1( {\cal V}_4^{\vee} \otimes {\cal V}_2)$&$4 c_2({\cal V}_2) + 2 c_2({\cal V}_4) -16 $\\
$({\bf 4},{\bf 2})$&$ H^1({\cal V}_4^{\vee})$&$c_2({\cal V}_4)  -8$\\
$({\bf 6},{\bf 1})$&$H^1(\wedge^2{\cal V}_4) $&$2c_2({\cal V}_4) - 12$\\
\hline
\end{tabular*}
\caption{{\it The cohomology associated to each representation of the low-energy gauge group SU(4) $\times$ SU(2). }}
\label{atabastrans1zero}
\end{center}
\end{table}
The number of massless hypermultiplets can be read off from Table \ref{atabastrans1zero} and leads to the following anomaly cancelation condition.
\begin{eqnarray} \nonumber
n_H +29 n_T -n_V =273\\
\Rightarrow c_2({\cal V}_2) + c_2({\cal V}_4)+ \frac{1}{60} c_2(\textnormal{End}_0({\cal V}_{E_8}) =24
\end{eqnarray}
We can now specialize this result to the particular SU(4) $\times$ SU(2) bundle that we found in (\ref{currentguy}). In this instance we have,
\begin{eqnarray}
c_2({\cal V}_4) = c_2({\cal V}_3)
\end{eqnarray}
We can now construct the equivalent table to those that have been formed in the proceeding cases.  We compare the spectrum which is obtained by decomposing the initial SU(6) multiples with that obtained by direct computation from the bundle after transition. The results of this comparison are in Table \ref{atranstab52}.
\begin{table}[!h]
\begin{center}
\begin{tabular*}{17.05cm}{|c|c|c|}
\hline
SU(4) $\times$ SU(2) Representation & \# from SU(6) multiplet decomposition & \# found after transition \\ \hline
$({\bf 1},{\bf 1})$ & $4 c_2({\cal V}_2) + 8 c_2({\cal V}_3)-34$  & $4 c_2({\cal V}_2) + 8 c_2({\cal V}_3)-36$ \\
$({\bf 6},{\bf 2})$ & $c_2({\cal V}_2)-4$  & $c_2({\cal V}_2)-4$ \\
$({\bf 1},{\bf 2})$ & $6 c_2({\cal V}_2)+4 c_2({\cal V}_3) -24$& $6 c_2({\cal V}_2)+4 c_2({\cal V}_3) -24$\\
$({\bf 4},{\bf 1})$ & $4 c_2({\cal V}_2) + 2 c_2({\cal V}_3)-16$ &  $4 c_2({\cal V}_2) + 2 c_2({\cal V}_3)-16$\\ 
$({\bf 4},{\bf 2})$ & $ c_2({\cal V}_3)-6$ &  $ c_2({\cal V}_3)-8$\\
$({\bf 6},{\bf 1})$ & $2 c_2({\cal V}_3)-12$&  $2 c_2({\cal V}_3)-12$\\ 
\hline
\end{tabular*}
\caption{{\it Matter content after Higgsing an SU(6) to an SU(4) $\times$ SU(2) theory, both via a naive decomposition of the initial SU(6) multiplets and via a direct computation from the resulting SU(4) $\times$ SU(2) bundle. }}
\label{atranstab52}
\end{center}
\end{table}

As in all of the cases in this section, the result is fully consistent with the field theory analysis given in Section \ref{antihigs}.

\subsubsection{SU(8)}

The relevant group theory in this case is the following.
\begin{eqnarray} \label{aSU8asHiggsing1}
{\rm SU}(8) &\supset& {\rm SU}(6) \times {\rm SU}(2) \times {\rm U}(1) \\ \nonumber
{\bf 28} &=& ({\bf 1},{\bf 1})_{-6} +({\bf 6},{\bf 2})_{-2}+({\bf 15},{\bf 1})_2 \\\nonumber
{\bf 8} &=& ({\bf 1},{\bf 2})_{-3} + ({\bf 6},{\bf 1})_1 \\\nonumber
{\bf 56} &=& ({\bf 6},{\bf 1})_{-5} + ({\bf 20},{\bf 1})_3 +({\bf 15},{\bf 2})_{-1} 
\end{eqnarray}

This breaking pattern of interest in (\ref{aSU8asHiggsing1})  corresponds to giving a VEV to a ${\bf 28}$, $\overline{{\bf 28}}$ pair. The ${\bf 28}$'s, according to Table \ref{tab11}, lie in the cohomology $H^1({\cal L}^{\vee 2})$. In terms of bundle topology, giving an expectation value to such a field corresponds to forming the following bundle.
\begin{eqnarray} \label{athischappy}
&& {\cal V}= {\cal V}_2 \oplus  {\cal L} \oplus {\cal L}^{\vee} \\
\textnormal{where} && 0 \to {\cal L}^{\vee } \to {\cal V}_2 \to {\cal L} \to 0 
\end{eqnarray}

As in previous cases one should really think of ${\cal V}_2$ as being a deformation of this extension and its dual \cite{Li:2004hx}.  Here ${\cal V}$ is an SU(2) $\times$ S(U(1) $\times$ U(1)) bundle. The correct embedding of SU(2) $\times$ U(1) does indeed break $E_8$ to SU(6) $\times$ SU(2), with an additional Green-Schwarz massive U(1) being present.

The group theory for a SU(6) $\times$ SU(2) compactification of heterotic is as follows.
\begin{eqnarray} \label{asu6group}
E_8 &\supset& {\rm SU}(6) \times {\rm SU}(2) \times {\rm SU}(2) \times {\rm U}(1) \\ \nonumber
{\bf 248} &=& ({\bf 1},{\bf 1},{\bf 3})_0+ ({\bf 1},{\bf 1},{\bf 1})_0+({\bf 1},{\bf 2},{\bf 1})_3+({\bf 1},{\bf 2},{\bf 1})_{-3}+({\bf 1},{\bf 3},{\bf 1})_0 \\ \nonumber
&&+({\bf 35},{\bf 1},{\bf 1})_0+(\overline{{\bf 15}},{\bf 1},{\bf 1})_{-2}+(\overline{{\bf 15}},{\bf 2},{\bf 1})_1+({\bf 15},{\bf 1},{\bf 1})_2+({\bf 15},{\bf 2},{\bf 1})_{-1} \\ \nonumber
&&+({\bf 6},{\bf 1},{\bf 2})_{-2}+({\bf 6},{\bf 2},{\bf 2})_1+(\overline{{\bf 6}},{\bf 1},{\bf 2})_2+(\overline{{\bf 6}},{\bf 2},{\bf 2})_{-1} + ({\bf 20},{\bf 1},{\bf 2})_0
\end{eqnarray}
This leads to the low-energy spectrum given in Table \ref{atabastrans1one}.
\begin{table}[!h]
\begin{center}
\begin{tabular*}{10.35cm}{|c|c|c|}
\hline
Representation & Cohomology & Multiplicity \\ \hline
$({\bf 1},{\bf 1})$ & $H^1 (\textnormal{End}_0({\cal V}_2))  $ &  $4 c_2({\cal V}_2) -6$\\ 
$({\bf 1},{\bf 2})$ & $H^1 ({\cal L}^3) \oplus H^1 ({\cal L}^{\vee 3}) $ &  $2(-\frac{9}{2} c_1({\cal L})^2-2)$\\ 
$({\bf 15},{\bf 1})$ & $H^1( {\cal L}^2)$&$-2 c_1({\cal L})^2 -2 $\\
$({\bf 15},{\bf 2})$ & $ H^1( {\cal L}^{\vee})$&$-\frac{1}{2} c_1({\cal L})^2 -2 $\\
$({\bf 6},{\bf 1})$&$ H^1({\cal V}_2 \otimes  {\cal L}^{\vee 2})$&$c_2({\cal V}_2)  -4 c_1({\cal L})^2 -4$\\
$({\bf 6},{\bf 2})$&$H^1({\cal V}_2 \otimes {\cal L})$&$c_2({\cal V}_2) - c_1({\cal L})^2 -4$\\
$({\bf 20},{\bf 1})$&$H^1({\cal V}_2)$&$c_2({\cal V}_2) -4$\\
\hline
\end{tabular*}
\caption{{\it The cohomology associated to each representation of the low-energy gauge group ${\rm SU}(6)\times {\rm SU}(2) $. }}
\label{atabastrans1one}
\end{center}
\end{table}
The number of massless hypermultiplets can be read off from this table leading to the following anomaly cancelation condition.
\begin{eqnarray} \label{aoops2}
n_H+29 n_T-n_V =273 \\ \nonumber
\Rightarrow  c_2({\cal V}_2) -3 c_1({\cal L})^2+ \frac{1}{60} c_2(\textnormal{End}_0({\cal V}_{E_8}))=24
\end{eqnarray}

We can now specialize this result to the particular ${\rm SU}(2) \times {\rm S}({\rm U}(1) \times {\rm U}(1))$ bundle that we obtained in (\ref{athischappy}). In this instance we have,
\begin{eqnarray}
c_2({\cal V}_2)&=&-c_1({\cal L})^2 \;.
\end{eqnarray}

We are now in a position to construct the equivalent table to those that have been formed in the other cases. We compare the spectrum which is obtained by a decomposition of the initial SU(8) multiplets with that obtained by direct computation from the bundle after transition. The results of this comparison are in Table \ref{atranstab5}.
\begin{table}[!h]
\begin{center}
\begin{tabular*}{17.05cm}{|c|c|c|}
\hline
${\rm SU}(6)\times {\rm SU}(2)$ Representation & \# from SU(8) multiplet decomposition & \# found after transition \\ \hline
$({\bf 1},{\bf 1})$ & $-4 c_1({\cal L})^2 -4$  & $ -4 c_1({\cal L})^2 -6$ \\
$({\bf 1},{\bf 2})$ & $ - 9 c_1({\cal L})^2 - 4$  & $-9 c_1({\cal L})^2 - 4$ \\
$({\bf 15},{\bf 1})$ & $-2 c_1({\cal L})^2 - 2$& $-2c_1({\cal L})^2 - 2$\\
$({\bf 15},{\bf 2})$ & $-\frac{1}{2} c_1({\cal L})^2 - 2$ &  $-\frac{1}{2} c_1({\cal L})^2 - 2$\\ 
$({\bf 6},{\bf 1})$ & $-5c_1({\cal L})^2 - 4$ &  $-5 c_1({\cal L})^2 - 4$\\
$({\bf 6},{\bf 2})$ & $- 2c_1({\cal L})^2 -2$&  $- 2 c_1({\cal L})^2 -4$\\ 
$({\bf 20},{\bf 1})$ & $-c_1({\cal L})^2 -4$&  $-  c_1({\cal L})^2 -4$\\ 
\hline
\end{tabular*}
\caption{{\it Matter content after Higgsing an SU(8) to an ${\rm SU}(6) \times {\rm SU}(2)$ theory, both via a naive decomposition of the initial SU(8) multiplets and via a direct computation from the resulting ${\rm SU}(2) \times S(U(1) \times U(1))$ bundle. }}
\label{atranstab5}
\end{center}
\end{table}

As in all of the cases in this section, the result is fully consistent with the field theory analysis given in Section \ref{antihigs}.

\subsection{Higgsing on three-index antisymmetric matter}

\subsubsection{SU(6) } \label{asu6Higgs}

The group theory relevant to this case is as follows.
\begin{eqnarray} \label{agroupbreaksu63}
{\rm SU}(6) &\to& {\rm SU}(3) \times {\rm SU}(3) \times {\rm U}(1) \\
{\bf 6} &=& ({\bf 3},{\bf 1})_1 + ({\bf 1},{\bf 3})_{-1}\\
{\bf 15} &=& (\overline{{\bf 3}},{\bf 1})_{2} + ({\bf 1},\overline{{\bf 3}})_{-2} + ({\bf 3},{\bf 3})_0\\
{\bf 20} &=& ({\bf 1},{\bf 1})_3+({\bf 1},{\bf 1})_{-3}+({\bf 3},\overline{{\bf 3}})_{-1}+(\overline{{\bf 3}},{\bf 3})_1
\end{eqnarray}
The breaking pattern we are interested in thus corresponds to giving a VEV to matter in the ${\bf 20}$ representation. The ${\bf 20}$, according to Table \ref{tab9}, is associated with the cohomology $H^1({\cal V}_2)$. In terms of bundle geometry, therefore, giving a VEV to matter in this representation corresponds to forming the following bundle.
\begin{eqnarray} \label{mdelasu3}
&& {\cal V} ={\cal V}_3 + \tilde{\cal V}_3 \\  \nonumber
\textnormal{where} &&  0 \to {\cal V}_2 \to \tilde{\cal V}_3 \to {\cal O} \to 0
\end{eqnarray}
As in the other cases in this appendix, one should really think of ${\cal V}_3$ as being a deformation of this extension and its dual \cite{Li:2004hx}.

The group theory associated to a SU(3) $\times$ SU(3) compactification of heterotic string theory is as follows.
\begin{eqnarray}
E_8 &\supset& {\rm SU}(3) \times {\rm SU}(3) \times {\rm SU}(3) \times {\rm SU}(3) \\ \nonumber
{\bf 248} &=& ({\bf 1},{\bf 1},{\bf 1},{\bf 8})+({\bf 3},{\bf 1},{\bf 3},{\bf 3})+({\bf 1},{\bf 3}, \overline{{\bf 3}},{\bf 3})+(\overline{{\bf 3}} ,\overline{{\bf 3}} ,{\bf 1},{\bf 3}) + (\overline{{\bf 3}},{\bf 1},\overline{{\bf 3}},\overline{{\bf 3}}) \\ \nonumber
&&+({\bf 1},\overline{{\bf 3}},{\bf 3},\overline{{\bf 3}}) +({\bf 3},{\bf 3},{\bf 1},\overline{{\bf 3}})+(\overline{{\bf 3}},{\bf 3},{\bf 3},{\bf 1})+({\bf 3},\overline{{\bf 3}},\overline{{\bf 3}},{\bf 1}) \\ && +({\bf 8},{\bf 1},{\bf 1},{\bf 1})+({\bf 1},{\bf 8},{\bf 1},{\bf 1})+({\bf 1},{\bf 1},{\bf 8},{\bf 1})
\end{eqnarray}
This leads to the matter content given in Table \ref{atabastrans22222}.
\begin{table}[!ht]
\begin{center}
\begin{tabular*}{12.79cm}{|c|c|c|}
\hline
Representation & Cohomology & Multiplicity \\ \hline
$({\bf 1},{\bf 1})$ & $H^1(\textnormal{End}_0({\cal V}_3))\oplus H^1(\textnormal{End}_0(\tilde{\cal V}_3))$& $6c_2({\cal V}_3) + 6c_2(\tilde{\cal V}_3) -32 $\\
$({\bf 3},{\bf 1})$ & $H^1 ({\cal V}_3 \otimes \tilde{\cal V}_3)$ &  $3c_2({\cal V}_3)+3 c_2(\tilde{\cal V}_3)-18$\\ 
$({\bf 1},{\bf 3})$ & $H^1 ({\cal V}_3 \otimes \tilde{\cal V}_3^{\vee})$& $3c_2({\cal V}_3)+3 c_2(\tilde{\cal V}_3)-18$\\
$({\bf 3},{\bf 3})$ & $H^1({\cal V}_3^{\vee})$&  $c_2({\cal V}_3) -6$\\ 
$({\bf 3},\overline{\bf 3})$ &$ H^1(\tilde{\cal V}_3^{\vee})$ & $c_2(\tilde{\cal V}_3) -6$\\ 
\hline
\end{tabular*}
\caption{{\it The cohomology associated to each representation of the low-energy gauge group SU(3) $\times$ SU(3). }}
\label{atabastrans22222}
\end{center}
\end{table}

The matter content in Table \ref{atabastrans22222} leads to the following anomaly cancelation condition.
\begin{eqnarray}
n_H+29n_T-n_V=273 \\
\Rightarrow c_2({\cal V}_3)+c_2(\tilde{\cal V}_3)+\frac{1}{60} c_2(\textnormal{End}_0({\cal V}_{E_8})) = 24
\end{eqnarray}
As in previous cases, we can now specialize this result to the particular SU(3) $\times$ SU(3) bundle which we obtain after transition, as given in equation (\ref{mdelasu3}). We have that,
\begin{eqnarray}
c_2(\tilde{\cal V}_3) = c_2({\cal V}_2) \;.
\end{eqnarray}
Using this, we can compare the spectrum which is obtained by a decomposition of the initial SU(6) multiplets to that obtained by a direct computation from the bundle after transition. The results of this comparison are in Table \ref{atranstab5aaa}.
\begin{table}[!h]
\begin{center}
\begin{tabular*}{17.05cm}{|c|c|c|}
\hline
${\rm SU}(3)\times {\rm SU}(3)$ Representation & \# from SU(6) multiplet decomposition & \# found after transition \\ \hline
$({\bf 1},{\bf 1})$ & $6c_2({\cal V}_3) + 6c_2({\cal V}_2) -30$  & $ 6c_2({\cal V}_3) + 6c_2({\cal V}_2) -32$ \\
$({\bf 3},{\bf 1})$ & $ 3c_2({\cal V}_3)+3 c_2({\cal V}_2)-18$  & $3c_2({\cal V}_3)+3 c_2({\cal V}_2)-18$ \\
$({\bf 1},{\bf 3})$ & $3c_2({\cal V}_3)+3 c_2({\cal V}_2)-18$& $3c_2({\cal V}_3)+3 c_2({\cal V}_2)-18$\\
$({\bf 3},{\bf 3})$ & $c_2({\cal V}_3) -6$ &  $c_2({\cal V}_3) -6$\\ 
$({\bf 3},\overline{\bf 3})$ & $c_2({\cal V}_2) -4$ &  $c_2({\cal V}_2) -6$\\
\hline
\end{tabular*}
\caption{{\it Matter content after Higgsing an SU(6) to an SU(3) $\times$ SU(3) theory, both via a naive decomposition of the initial SU(6) multiplets and via a direct computation from the resulting SU(3) $\times$ SU(3) bundle. }}
\label{atranstab5aaa}
\end{center}
\end{table}

As in all of the cases we look at, the result is fully consistent with the field theory analysis given in Section \ref{antihigs}

\subsubsection{SU(7)}

The analysis for Higgsing SU(7) on triple antisymmetrics is extremely similar. We have,
\begin{eqnarray}
{\rm SU}(7) &\supset& {\rm SU}(4) \times {\rm SU}(3) \times {\rm U}(1) \\\nonumber
{\bf 35} &=& ({\bf 1},{\bf 1})_{-12} +(\overline{{\bf 4}},{\bf 1})_9 + ({\bf 4},\overline{{\bf 3}})_{-5}+({\bf 6},{\bf 3})_2 \\\nonumber
{\bf 7} &=& ({\bf 1},{\bf 3})_{-4} + ({\bf 4},{\bf 1})_3 \\\nonumber
{\bf 21} &=& ({\bf 1},\overline{{\bf 3}})_{-8} + ({\bf 4},{\bf 3})_{-1} + ({\bf 6},{\bf 1})_6 \;.
\end{eqnarray}
The breaking pattern we are interested in thus corresponds to giving a VEV to a ${\bf 35}$, $\overline{{\bf 35}}$ pair. The ${\bf 35}$'s, according to Table \ref{tab10}, lie in the cohomology $H^1({\cal L})$. In terms of bundle topology, giving an expectation value to such a field corresponds to forming the following bundle.
\begin{eqnarray} \label{ahigtrans2}
{\cal V} &=& {\cal V}_2  \oplus \tilde{\cal V}_2 \\ \nonumber
\textnormal{where}  &&0 \to {\cal L} \to \tilde{\cal V}_2 \to {\cal O} \to 0
\end{eqnarray}
As in previous cases one should really think of $\tilde{\cal V}_2$ as being a deformation of this extension and its dual \cite{Li:2004hx}.  Here $\tilde{\cal V}_2$ is an U(2) bundle and the bundle ${\cal V}_2$ is unaffected by the transition. The overall structure group is ${\rm S}({\rm U}(2) \times {\rm U}(2))$ which does indeed break $E_8$ to ${\rm SU}(4) \times {\rm SU}(3) \times {\rm U}(1)$ (where the last, U(1), factor is  Green-Schwarz anomalous).

The group theory for a ${\rm SU}(4) \times {\rm SU}(3) \times {\rm U}(1)$ heterotic compactification is as follows,
\begin{eqnarray} \label{asu54subgroup}
E_8 &\supset& {\rm SU}(4) \times {\rm SU}(3) \times {\rm U}(1) \times {\rm SU}(2) \times {\rm SU}(2)  \\
{\bf 248} &=& ({\bf 1},{\bf 1},{\bf 1},{\bf 3})_{0} + ({\bf 1},{\bf 3},{\bf 2},{\bf 2})_{-2} + ({\bf 4},{\bf 3},{\bf 1},{\bf 2})_{1} + ({\bf 1},\overline{{\bf 3}},{\bf 2},{\bf 2})_{2} +(\overline{{\bf 4}},\overline{{\bf 3}},{\bf 1},{\bf 2})_{-1} \\ \nonumber
&&+ ({\bf 4},{\bf 1},{\bf 1},{\bf 2})_{-3} + (\overline{{\bf 4}},{\bf 1},{\bf 1},{\bf 2})_{3} + ({\bf 6},{\bf 1},{\bf 2},{\bf 2})_{0}+({\bf 1},{\bf 8},{\bf 1},{\bf 1})_{0} + ({\bf 1},\overline{{\bf 3}},{\bf 1},{\bf 1})_{-4} \\ \nonumber
&& +({\bf 4},\overline{{\bf 3}},{\bf 2},{\bf 1})_{-1} + ({\bf 6},\overline{{\bf 3}},{\bf 1},{\bf 1})_2+({\bf 1},{\bf 3},{\bf 1},{\bf 1})_4+(\overline{{\bf 4}},{\bf 3},{\bf 2},{\bf 1})_1+(\overline{{\bf 6}},{\bf 3},{\bf 1},{\bf 1})_{-2} + ({\bf 1},{\bf 1},{\bf 1},{\bf 1})_0 \\ \nonumber
&& +({\bf 1},{\bf 1},{\bf 3},{\bf 1})_0+(\overline{{\bf 4}},{\bf 1},{\bf 2},{\bf 1})_{-3}+({\bf 4},{\bf 1},{\bf 2},{\bf 1})_3+({\bf 15},{\bf 1},{\bf 1},{\bf 1})_0\;,
\end{eqnarray}
which leads to the matter content given in Table \ref{atabastrans2}.
\begin{table}[!ht]
\begin{center}
\begin{tabular*}{15.12cm}{|c|c|c|}
\hline
Representation & Cohomology & Multiplicity \\ \hline
$({\bf 1},{\bf 1})$ & $H^1(\textnormal{End}_0({\cal V}_2))\oplus H^1(\textnormal{End}_0(\tilde{\cal V}_2))$& $4c_2({\cal V}_2)+4c_2(\tilde{\cal V}_2)-2c_1(\tilde{\cal V}_2)^2-12$\\
$({\bf 1},{\bf 3})$ & $H^1 ({\cal V}_2 \otimes \tilde{\cal V}_2^{\vee}) \oplus H^1(\wedge^2{\cal V}_2^{\vee}\otimes \wedge^2 \tilde{\cal V}_2)$ &  $2c_2(\tilde{\cal V}_2) +2 c_2({\cal V}_2)-5 c_1(\tilde{\cal V}_2)^2-10$\\ 
$({\bf 4},{\bf 3})$ & $H^1({\cal V}_2^{\vee})$ & $c_2({\cal V}_2) - \frac{1}{2}c_1(\tilde{\cal V}_2)^2 -4$\\
$({\bf 4},{\bf 1})$ & $H^1(\tilde{\cal V}_2 \otimes \wedge^2 {\cal V}_2^{\vee}) \oplus H^1({\cal V}_2 \otimes \wedge^2 \tilde{\cal V}_2^{\vee})$&  $c_2(\tilde{\cal V}_2)+c_2({\cal V}_2)-5c_1(\tilde{\cal V}_2)^2-8$\\ 
$({\bf 6},{\bf 1})$ &$ H^1(\tilde{\cal V}_2 \otimes {\cal V}_2) $ & $2c_2({\cal V}_2) +2 c_2(\tilde{\cal V}_2)-c_1(\tilde{\cal V}_2)^2 -8$\\ 
$({\bf 6},\overline{{\bf 3}})$ & $H^1(\wedge^2{\cal V}_2^{\vee})$&$-\frac{1}{2} c_1(\tilde{\cal V}_2)^2 -2 $\\
$(\overline{{\bf 4}},{\bf 3})$&$H^1(\tilde{\cal V}_2)$&$c_2(\tilde{\cal V}_2) - \frac{1}{2} c_1(\tilde{\cal V}_2)^2-4$\\
\hline
\end{tabular*}
\caption{{\it The cohomology associated to each representation of the low-energy gauge group ${\rm SU}(4)\times {\rm SU}(3) \times U(1)$. }}
\label{atabastrans2}
\end{center}
\end{table}

This matter content leads to the following anomaly cancellation condition.
\begin{eqnarray} \label{aoops2b}
n_H+29 n_T-n_V =273 \\ \nonumber
\Rightarrow  c_2(\tilde{\cal V}_2)+c_2({\cal V}_1) -2 c_1(\tilde{\cal V}_2)^2+ \frac{1}{60} c_2(\textnormal{End}_0({\cal V}_{E_8}))=24
\end{eqnarray}
For the case of a ${\rm S}({\rm U}(2) \times {\rm U}(2))$ bundle obtained by a transition of the form (\ref{ahigtrans2}) we have the following topology.
\begin{eqnarray}
c_2(\tilde{\cal V}_2)&=&0 \\
c_1(\tilde{\cal V}_2) &=& c_1({\cal L})
\end{eqnarray}
Given this, and noting that the ${\bf 6}$ of ${\rm SU}(4)$ is real, we obtain Table \ref{atranstab3}, which gives the number of ${\rm SU}(4) \times {\rm SU}(3)$ multiplets obtained by decomposing the SU(7) matter content, compared to a direct computation of the spectrum of the four-dimensional theory given the bundle topology after transition.
\begin{table}[!h]
\begin{center}
\begin{tabular*}{17.2cm}{|c|c|c|}
\hline
${\rm SU}(4)\times {\rm SU}(3)$ Representation & \# from SU(7) multiplet decomposition & \# found after transition \\ \hline
$({\bf 1},{\bf 1})$ & $4 c_2({\cal V}_2) -2 c_1({\cal L})^2 -10$  & $ 4c_2({\cal V}_2) -2 c_1({\cal L})^2 -12$ \\
$({\bf 1},{\bf 3})$ & $2 c_2({\cal V}_2) - 5 c_1({\cal L})^2 - 10$  & $2 c_2({\cal V}_2) - 5 c_1({\cal L})^2 - 10$ \\
$({\bf 4},{\bf 3})$ & $ c_2({\cal V}_2) - \frac{1}{2} c_1({\cal L})^2 - 4$& $ c_2({\cal V}_2) - \frac{1}{2} c_1({\cal L})^2 - 4$\\
$({\bf 4},{\bf 1})$ & $c_2({\cal V}_2)- 5c_1({\cal L})^2 - 8$ &  $c_2({\cal V}_2)- 5 c_1({\cal L})^2 - 8$\\ 
$({\bf 6},{\bf 1})$ & $2 c_2({\cal V}_2)- c_1({\cal L})^2 - 8$ &  $2c_2({\cal V}_2)- c_1({\cal L})^2 - 8$\\
$({\bf 6},\overline{{\bf 3}})$ & $- \frac{1}{2}c_1({\cal L})^2 -2$&  $- \frac{1}{2} c_1({\cal L})^2 -2$\\ 
$(\overline{{\bf 4}},{\bf 3})$ & $- \frac{1}{2}c_1({\cal L})^2 -2$&  $- \frac{1}{2} c_1({\cal L})^2 -4$\\ 
\hline
\end{tabular*}
\caption{{\it Matter content after Higgsing an SU(7) to an ${\rm SU}(4)\times {\rm SU}(3)$ theory, both via a naive decomposition of the initial SU(7) multiplets and via a direct computation from the resulting $S(U(2) \times U(2))$ bundle.}}
\label{atranstab3}
\end{center}
\end{table}
Once more these results are consistent with the usual understanding of such a Higgsing process and with the field theory analysis in Section \ref{antihigs}

\end{document}